\documentclass[aps,preprint,superscriptaddress,groupedaddress,nofootinbib]{revtex4}
\usepackage{amssymb}
\usepackage{amsmath}
\usepackage{graphicx}
\usepackage{nicefrac}
\usepackage{slashed}
\usepackage{hyperref}

\usepackage{amsmath}

  \def\rt#1{\sqrt{#1}}

    \font\thin=cmbx8 
  \def\clnl{\thin \raise1.2pt\hbox to 0.3pt{: \hss}}
  \def\clnr{\thin \raise1.2pt\hbox to 0.3pt{\hss :}}
  \def\bm#1{\text{\boldmath$#1$}}

  \def\rl#1{\rightline{#1}}

  \def\pd{\partial} 

  \def\gsim{\mathrel{\rlap{\lower0.25em\hbox{$\sim$}}\raise0.2em\hbox{$>$}}} 
  \def\lsim{\mathrel{\rlap{\lower0.25em\hbox{$\sim$}}\raise0.2em\hbox{$<$}}}
  \def\lg{\mathrel{\rlap{\lower0.25em\hbox{$>$}}\raise0.25em\hbox{$<$}}}
  \def\gl{\mathrel{\rlap{\lower0.25em\hbox{$<$}}\raise0.25em\hbox{$>$}}}

         \let\b=\beta      \let\d=\delta     \let\g=\gamma
  \let\lam=\lambda    \let\o=\omega     \let\s=\sigma     \let\z=\zeta
       
  \let\D=\Delta       \let\G=\Gamma         
  \let\Th=\Theta      
  \def\mn{_{\mu\nu}}  \def\omn{^{\mu\nu}}

  \def\be{\begin{eqnarray}}       \def\ee{\end{eqnarray}}
  \def\bea{\begin{eqnarray}}      \def\eea{\end{eqnarray}}
  \def\bean{\begin{eqnarray*}}    \def\eean{\end{eqnarray*}}
  \def\bfig{\begin{figure}}       \def\efig{\end{figure}}
  \def\benu{\begin{enumerate}}    \def\eenu{\end{enumerate}}
  \def\bse{\begin{subequations}}  \def\ese{\end{subequations}}
             
  \def\eq#1{(\ref{#1})}        \def\nonu{\nonumber}

  \def\u#1{\underline{#1}}
  \def\Ang{$\buildrel{\lower0.25em\hbox{\tiny$\circ\,$}}\over{\rm A}$}

   \def\Im{\hbox{Im}}      \def\Re{\hbox{Re}}
  \def\sgn{\hbox{sgn}}    
        
  \def\Li{\hbox{Li}}

  \def\pd{\partial} 

  \def\l{\left}			\def\r{\right}

  \def\eg{e.\,g.\ }	
  \def\ie{i.\,e.\ }	

  \def\o3{\textsl{O}(3)}

  \def\gl2c{\textsl{GL}(2, \mathbb{C})}
  \def\lasu2{\mathfrak{su}(2)}
  
  \newcommand{\sumint}[1]{{\hbox{$\textstyle\sum$}\!\!\!\!\!\!\!\int\,}_{\!\!\!\!\raise-0.5ex\hbox{$\scriptstyle{#1}$}}}

  \def\cf{C_{\mbox{\rm \scriptsize F}}}

  \newcommand{\msbar}{{\overline{\mbox{\rm MS}}}}

\graphicspath{{}}
\newcommand{\figscale}{0.9} 

\renewcommand{\theequation}{\arabic{section}.\arabic{equation}}

\renewcommand{\thesubsection}{\arabic{section}.\arabic{subsection}}

\makeatletter 
\@addtoreset{equation}{section} 
\def\p@section{}
\def\p@subsection{}
\makeatother

\begin{document}

\rl{December 2019}
\vspace{.5cm}

\title{ 
  Two-loop thermal spectral functions with general kinematics
}

\author{G.~Jackson} 
\email{jackson@itp.unibe.ch}

\affiliation{
Albert Einstein Center, 
Institute for Theoretical Physics, 
University of Bern,  
Sidlerstrasse 5, CH-3012 Bern, Switzerland} 

\begin{abstract}
  Spectral functions 
  at finite temperature and two-loop order are investigated,
  for a medium consisting of massless particles.
  We consider them 
  in the timelike and spacelike domains, allowing
  the propagating particles to be any valid combination of bosons and fermions.
  Divergences (if present) are analytically derived and set aside
  for the remaining finite part to be calculated numerically.
  To illustrate the utility of these `master' functions,
  we consider transverse and longitudinal parts of the 
  QCD vector channel spectral function.
\end{abstract}

\maketitle

\section{Introduction}

In a relativistic plasma,
the rates of processes like particle production and damping 
are derivable from the imaginary part of a particle's self-energy 
\cite{Weldon1983jn,Bodeker2015}.
That quantity, also called the spectral function, 
depends on the energy $k_0$ and momentum $\bm k$ which
can occur only in the combination $K^2 \equiv k_0^2 - \bm k^2$ at
zero-temperature.
This is {\em not} so for thermal systems, 
where the medium's rest frame is distinguished
and the temperature $T\neq0$ joins $k_0$ and $k = |\bm k|$ 
as an important scale in the problem.
Introducing another scale can dramatically alter
 the naive weak coupling expansion:
New infrared singularities
 foreshadow that next-to-leading order (NLO) corrections are large,
 or even that resummation is obligatory.

One such instance is the photon spectral function in hot QCD
\cite{Baier1988,Altherr1989,Gabellini1989}.
Truncating the perturbative result for the self-energy to order $e^2$ in 
the electromagnetic interactions,
we denote by $\Pi_{(l)}$ the ensuing contribution from $g^{2l}\,$
to the strong coupling expansion.
The series then takes the form
\bea
\Pi\omn (K) &=&
e^2 \, 
\big[\, \textstyle\sum_{l=0}^\infty \, g^{2l} \, \Pi_{(l)}\omn \, \big] 
\ + \ {\cal O}(e^4) \, ,
\label{expansion def}
\eea
with a supposed ordering by powers of $g^2$.
For a strict loop expansion, the `coefficients' of $g^{2l}$
are themselves functions of $k_0$ and $k$ but independent of $g$.
However their dependence on the external momentum $K$
can (and does) spoil this power counting, \eg when $|K^2| \lsim g^2 T^2$.
In particular, for high-energy real photons (i.e. $k_0=k\sim T$) 
 resummation of thermal loops is a minimal requirement to prevent an
 unphysical log-singularity 
 \cite{Kapusta1991,Baier1991,Arnold2001,Aurenche2002,Ghiglieri2014}.

For many observables only leading-order (LO) or partial NLO results are known,
making it unclear where \eq{expansion def} actually breaks down.
To obtain an approximation that is justified for all $k_0\,$, 
the fixed order expansion can be `matched' with the resummed approach
(which works near the light cone).
That was the idea put forward in Ref.~\cite{dileptons},
where it was tested for $\Im\,\Pi_{\ \mu}^\mu$ with $k_0>k\,$.
Here we also consider energies below the light cone and 
separately the polarisation state
 $\Im\,\Pi_{(1)}^{00}\,$, as inspired by Ref.~\cite{Brandt2017}.

Our goal is to assist in the effort of quantifying another order in perturbation
theory by cataloging a general class of two-loop spectral functions.
(One of the earliest attempts in this spirit provided the first
correction to the gluon plasma frequency \cite{Schulz1993}.)
What follows is rather technical, but lays out a generic approach to 
evaluate those integrals frequently needed in NLO computations.
All code used for determining the finite thermal parts (defined as 
specified below) is supplied in Ref.~\cite{code}.
The primary task of that code is a phase space integration of 
amplitudes squared, with thermal weightings appropriate to each process.

\bigskip

To compute loop integrals at finite temperature, we apply
the imaginary time formalism for massless particles.
Free scalar propagators,
 carrying either bosonic ($s=+1$) or fermionic ($s=-1$) momentum,
are denoted by
\bea
\D_{s}(P) &=& \frac1{p_0^2-p^2} 
\ ; \quad
p_0 \ =\
  i \, [2n+\Theta(-s)]\pi T  \, . 
\label{props}
\eea
The integer $n$ specifies the Matsubara frequencies
and $\Theta$ is the Heaviside step function.

We regularise the spatial momentum in $d=3-2\epsilon$ 
dimensions with the modified minimal subtraction ($\msbar$)
scheme and renormalisation scale $\bar \mu$. 
The trace over momentum $P=(p_0, \bm p)$ at finite temperature is defined  by
$$
\sumint{\, P} = \int_{\bm p} \, T \sum_{p_0} \ ; \qquad
\int_{\bm p} = \l( \frac{e^\gamma \bar \mu^2}{4\pi} \r)^\epsilon 
\int\!\! \frac{d^d p}{(2\pi)^d} \, ,
$$
where $\gamma$ is Euler's constant.
We follow \cite{BP} to carry out the sums over $p_0\,$, defined in \eq{props}.

This paper is organised as follows.
In section \ref{list} a general class of master sum-integrals is introduced
and those considered here are specified.
They are then evaluated, one by one, 
in sections \ref{sec: 1}, \ref{sec: 2}, \ref{sec: 3} and \ref{sec: 4}.
(For completeness, and as an important cross-check on our results,
the $K^2 \gg T^2$ behaviour of each
sum-integral is derived analytically in Appendix~\ref{app: D}.)
Finally, 
we `sum up' in section \ref{end} and mention some potential applications.

\section{List of integrals\label{list}}

\begin{figure}[h]
  \includegraphics[scale=\figscale]{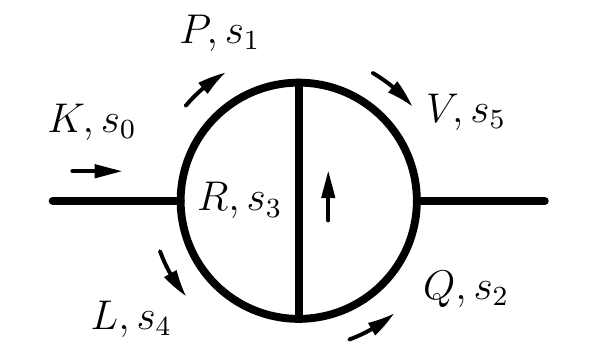}
  \caption{
    \label{fig: loop}
    Labelling of momenta and statistics for \eq{I def}.
  }
\end{figure}

Let us define, for generic sum-integrals ${\cal I}$
as functions of the external four-momentum $K=(k_0,\bm k)\,$,
a uniform notation
(for $m=n=0$, cf.~\cite{laine1})
\bea
{\cal I}_{abcde}^{(m,n)}
(K)
&=&
\sumint{\, P,Q} \, p_0^m \, q_0^n \,
\D_1^a
\D_2^b
\D_3^c
\D_4^d
\D_5^e 
\, ,
\label{I def}
\eea
where $\D_i \equiv \D_{s_i}(P_i)\,$.
The case $n=0$ is abbreviated by ${\cal I}_{abcde}^{(m)}\,$.
Together with $P$ and $Q$, the integration variables,
$K$ determines all the propagating momenta 
(as depicted in Fig.~\ref{fig: loop}),
\be
& P_1 \equiv P \ ,\quad
P_2 \equiv Q\ ,\quad
P_3 \equiv R = K-P-Q \ , \nonu \\
& P_4 \equiv L = K-P\ ,\quad
P_5 \equiv V = K-Q\ . \nonu 
\ee
(We introduced $P,Q$ etc.\! to avoid the proliferation of subscripts.)
The statistical signatures $s_0$ (for $K$), 
$s_1$ and $s_2$ fully determine the
others by their connections at each vertex: 
$$
s_3 = s_0s_1s_2\ , \quad s_4=s_0s_1\quad {\rm and} \quad s_5=s_0s_2 \, .
$$
Thus we summarise the statistical content of \eq{I def}
by $(s_0,s_1,s_2) = (\pm,\pm,\pm)\,$,
not including it explicitly on the notation.

At $T=0\,$, the statistics play no role and these integrals 
can be evaluated using well-known methods \cite{ItzyksonZuber}.  
The vacuum contributions will dominate
over the thermal ones for large $K^2$.
Relative corrections are suppressed by powers of $K^2$ that can be formally organised 
with an operator product expansion (OPE) \cite{CaronHuot2009ns}.
Thermal effects are important for $K^2$ being of similar order to $T^2$, 
the regime we consider here to calculate 
the imaginary part of \eq{I def}\footnote{%
  By evaluating it at an energy $k_0 + i 0^+\,$.
}.
The remaining limit, $K^2 \ll T^2\,$, is frequently associated with the 
need to resum all orders in perturbation theory to cure a diverging spectral
function.
With that in mind, 
we shall often also discuss the master integrals for $K^2 \approx 0\,$.

\subsection{Strategy: $m,n\geq1$\label{strategy}}

To take care of the powers of the energy in the numerator of \eq{I def}, 
namely $p_0^m$ and $q_0^n$ with positive integers
$m$ and $n\,$, we employ the following strategy.
In special cases, the corresponding graph has a symmetry in momenta $P$ and $Q$
(\ie if $a=b$ and $d=e$),
and one can take advantage of the same symmetry 
for the powers of $p_0$ and $q_0\,$.
But in general, one should make use of 
the Fourier representation of the (massive) scalar propagator
\bea
\Delta_s (\tau, \bm p) 
&=& -\, \frac{1}{ 2E_p } \sum_{v=\pm} f^v_s (E_p) e^{-vE_p \tau} \, ,
\label{IT prop}
\eea
where $E_p = \rt{\lam^2 + p^2}$ and $\lam$ is the particle mass.
Here we also introduced $f_s^- = s n_s$, 
$f_s^+ = 1+s n_s$ and the distribution function 
$n_s(E) = [ \exp(E/T)-s ]^{-1}$.

Beginning with the case $m=1$ and differentiating with respect to $\tau$ under 
the Fourier transformation,
one effectively multiplies\footnote{%
  Here we generalise \eq{props} to have a mass:
  $\D_s (P) = \bm( p_0^2 - p^2 - \lam^2 \bm)^{-1}\,$.
  This will help later on, as an infrared regulator.
  } by the conjugate variable, 
\bea
p_0\, \Delta_s (P) 
&=& -\int_0^{T^{-1}} \! d\tau \, e^{p_0 \tau}
\partial_\tau \big( \, \Delta_s (\tau, \bm p) \, \big) \, .
\label{diff trick}
\eea
This is inserted into \eq{I def} before carrying out the frequency sum.
It is straightforward to differentiate \eq{IT prop} with respect to $\tau$.
Hence \eq{diff trick} provides an extra factor of $(vE_p)$, counting
the minus sign from integrating by parts.

The case $m=2$ is also elementary from the relation
\bea
p_0^2 \, \D_s(P) 
&=&  1 + E_p^2 \, \D_s(P) \, .
\label{p0^2}
\eea
By applying \eq{p0^2} and \eq{diff trick} in sequence one can reduce $p_0^m$ 
for $m\geq 2$ in \eq{I def} to 
a sum of powers of $(vE_p)$ with simpler masters.
(And the same strategy works for $q_0^n\,$.)
The benefit of all this, is that the frequency sums are relatable
to cases with $m=n=0\,$; any complications will move to the integration
over the three momenta $\bm p$ and $\bm q$ that follows.

\subsection{Example: A QCD spectral function}

Integrals of the form \eq{I def} [with statistics $(+,-,-)$] can be used to
express the NLO photon self-energy in
an equilibrated QCD plasma at zero chemical potential.
The emission rate is derived from the (contracted) spectral
function $\Im\,\Pi^{\ \mu}_{\mu}$ \cite{McLerran1985,Weldon1990iw},
but here we also study $\Pi^{00}\,$. 
Due to the Ward identity at non-zero temperature, the polarisation tensor 
has two independent components.
identified with the longitudinal and transverse polarisations:
\bea
\Pi{_{\rm L}} = \frac{K^2}{\bm k^2} \, \Pi^{00} \ , & &
\Pi{_{\rm T}} = - \,\frac1{2}
\Big(\, \Pi_\mu^{\ \mu} + \frac{K^2}{\bm k^2} \Pi^{00} \, \Big) \, .
\label{TL}
\eea
The difference between $\Pi_{_{\rm L}}$ and $\Pi_{_{\rm T}}$ 
is purely thermal, at zero temperature there is none \cite{Brandt2017}.
Accordingly, $\Pi_{\ \mu}^{\mu}$ and $\Pi^{00}$ are enough to
completely specify $\Pi\omn$ at finite temperature.

Denoting the number of colours by $N$ and the group factor by
$\cf \equiv (N^2-1)/(2N)\,$,
they read (with $\epsilon \to 0$)
\bea
g\mn \Pi_{(1)}\omn &=&   \label{masters, mn}
- 8(1-\epsilon) N \cf  \Big\{ \ 
  2(1-\epsilon) \\ &\times&
 \big[\, 
 {\cal I}_{00120}^{(0)} - {\cal I}_{01020}^{(0)}
 + K^2 \big({\cal I}_{11020}^{(0)} - {\cal I}_{10120}^{(0)}\big) \,\big]
+ 2\,{\cal I}_{11010}^{(0)}  
+ 2\epsilon \, \big( 
{\cal I}_{11100}^{(0)}  
-{\cal I}_{01110}^{(0)} 
-{\cal I}_{10110}^{(0)} 
\big) \nonu\\
&-&
 \tfrac12(3+2\epsilon) K^2 {\cal I}_{11011}^{(0)} 
- 2(1-\epsilon) {\cal I}_{1111(-1)}^{(0)}  
+ 4K^2 {\cal I}_{11110}^{(0)} 
- K^4 {\cal I}_{11111}^{(0)} 
 \ \Big\}  \ ,\nonu \\
\Pi_{(1)}^{00} &=& \label{masters, 00}
- 4 N \cf  \Big\{ \ \\
  & & 2(1-\epsilon) \Big[\,
 {\cal I}_{00120}^{(0)} - {\cal I}_{01020}^{(0)}
 + K^2 \big( {\cal I}_{11020}^{(0)} - {\cal I}_{10120}^{(0)} \big) 
- 4k_0 \big( {\cal I}_{11020}^{(1)} - {\cal I}_{10120}^{(1)} \big)  \nonu\\
&+& 4 \big( {\cal I}_{11020}^{(2)} - {\cal I}_{10120}^{(2)} \big) \, \Big]
+ 
 2(1-\epsilon){\cal I}_{10110}^{(0)}  
+  2\epsilon \, \big( 
{\cal I}_{11100}^{(0)}
- {\cal I}_{01110}^{(0)} 
\big)
+ (1+\epsilon) k^2 \, {\cal I}_{11011}^{(0)}
\nonu \\ 
&-& 2(1-\epsilon) {\cal I}_{1111(-1)}^{(0)} 
+
 4\big[\,(1-2\epsilon) k_0^2  -  k^2\,\big]\,{\cal I}_{11110}^{(0)}  
+ 8\epsilon \, k_0 \, {\cal I}_{11110}^{(1)} 
- 8(1-\epsilon) k_0 \, {\cal I}_{11110}^{(0,1)} 
\nonu \\
&+& 
  \big[\,(1-2\epsilon) k_0^2 + k^2\,\big] \,K^2 {\cal I}_{11111}^{(0)}
+ 4\epsilon \, K^2 \, {\cal I}_{11111}^{(1,1)} 
- 4(1-\epsilon) K^2\, {\cal I}_{11111}^{(2)}
\ \Big\} \ . \nonu
\eea
As part of the procedure to reduce $\Pi_{00}$ and $\Pi_{\mu}^{\ \mu}$
to a minimal set of integrals, we removed angular variables in the numerator
thanks to relations like
$\bm p \cdot \bm k = p_0 k_0 + \frac12 ((K-P)^2 - P^2 - K^2)\, .$
These replacements put frequencies in the numerator and bring about 
other (usually) simpler master integrals.

This motivates our study of the following set of master functions.
[Values for $m,n$ are from \eq{masters, mn} and \eq{masters, 00}.]

\bean
\renewcommand{\arraystretch}{1.5}
\renewcommand{\tabcolsep}{4mm}
\begin{tabular}{cl|rl}
$\vcenter{\hbox{\includegraphics[scale=1.3]{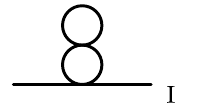}}} :$ & 
  $ {\cal I}_{01020} \, , \ {\cal I}_{00120} $ 
  & $(m,n)\ =$ \!\!\!\!\!\!\!\!\!\! & $(0)$ 
  \\

  $\vcenter{\hbox{\includegraphics[scale=1.3]{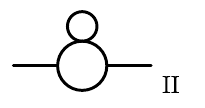}}} :$ & 
  $ {\cal I}_{11020} \, , \ {\cal I}_{10120}$ 
  & & $(0),\ (1),\ (2)$ 
  \\

$\vcenter{\hbox{\includegraphics[scale=1.3]{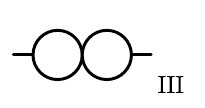}}} :$ & 
  $ {\cal I}_{11011} $ 
  & & $(0),\ (1),\ (0,1),\ (1,1)$ 
  \\
$\vcenter{\hbox{\includegraphics[scale=1.3]{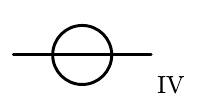}}} :$ & 
  $ {\cal I}_{01110} \, , \ {\cal I}_{11100} $ 
  & &  $(0)$ 
  \\

$\vcenter{\hbox{\includegraphics[scale=1.3]{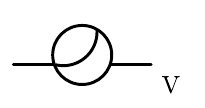}}} :$ & 
  $ {\cal I}_{11110} \, , \ {\cal I}_{1111(-1)} $ 
  & & $(0),\ (1),\ (0,1)$ 
  \\

$\vcenter{\hbox{\includegraphics[scale=1.2]{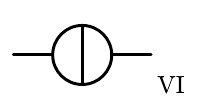}}} :$ & 
  $ {\cal I}_{11111} $ 
  & & $(0),\ (1),\ (0,1),\ (1,1),\ (2),\ (0,2)$ 
  \\
\end{tabular}
\eean
We have grouped the master integrals into classes 
(designated by the numeral on the graph), 
 according to the associated topology.
The topology of the class does not necessarily follow directly from the
assignment of loop momenta in Fig.~1.
A change of integration variables is sometimes required to relate them.
The first three classes are all presented together in Sec.~\ref{sec: 1} 
because they consist of simpler one-loop subgraphs that factorise.
Classes IV, V and VI can be considered {\em genuinely} two-loop 
and will receive the most attention, being discussed in 
Secs.~\ref{sec: 2}, \ref{sec: 3} and \ref{sec: 4} respectively.

Before moving on, a brief comment on one-loop diagrams is in order.
They have been studied extensively in the literature and are
usually considered in the hard thermal loop (HTL) approximation.
(Higher order HTL results have also been investigated, cf. Ref.~\cite{Mirza2013}.)
This is common in the high temperature limit \cite{BP} because it
affords analytic expressions for the self energy, 
and supplies results that are automatically gauge invariant.
If one relaxes the HTL assumption that the external momentum $K^2$ 
is much smaller than $T^2\,,$ the complete
self-energies must be evaluated numerically \cite{Peshier1998}.
Effective field theory methods have also been developed recently
to compute associated power corrections \cite{Manuel2016}.
The imaginary parts involve phase space integrals for
$1\leftrightarrow 2$ `decays' 
which are needed for our master diagrams II and III as well as certain
terms arising in V and VI.
Appendix~\ref{app: A} gives details on the integration measure,
where we also discuss the $2\leftrightarrow 2$ and $1\leftrightarrow 3$ 
processes to be utilised when more intermediate states can go on-shell.

\section{Diagrams I-III (factorisable topologies) \label{sec: 1}}

Those diagrams we assigned to classes I, II and III are reducible to products
of simpler one-loop integrals.
They can all be recast as those having $c=0$ in \eq{I def}\footnote{%
For example, because ${\cal I}_{10120}$ is equal to ${\cal I}_{11020}$ with $s_2 \to s_3\,$.
}, 
which implies that $P$ and $Q$ 
dependence of the integrand does not mix.
It is thus useful to recap a general one-loop function, defined by
\bea
{\cal J}_{\ ab}^{(m)}(K) &=& 
\sumint{\, P} p_0^m \, \D^a_{s_1}(P) \D^b_{s_2} (K-P) \, .
\label{def J}
\eea
The frequency sum over $p_0$ is well known \cite{BP}, and
we organise the subsequent integration over spatial momentum $\bm p$
 according to Appendix~\ref{app: A}\,.

The cases where $b=0$ are local contributions.
For the one with $a=1$ we abbreviate the integral by
$$
I_m(s_1) \equiv {\cal J}_{\ 10}^{(m)} =
T^2 \Big[\, \frac{\Theta (-s_1)}{2^{m+1}} -1 \, \Big] \,
(2\pi T)^m \z (-m-1) \, ,
$$
where $\z$ is the Riemann zeta function \cite{Schroder}.
Type I self-energies are then constant and we need not discuss them
because they have no imaginary part.
Moreover, since $I_n$ is zero in vacuum (for $m \geq 0$),
the type II integrals are entirely thermal corrections.

For $a=b=1\,$, integrals of the form \eq{def J} are usually considered 
in the limit $k_0 \sim k\,$ for which the HTL functions can be used.
But in general, the emerging integral
expressions must be evaluated numerically \cite{Peshier1998}.
Only when taking the imaginary part, thus putting internal momenta on-shell,
is the integral doable analytically.

Let us introduce three useful functions $F,\ G$ and $H$ that 
make the statistics explicit,
\be
& 
F_m(K;\, s_1,s_2) 
\ =\ 
\Im \ {\cal J}_{\ 12}^{(m)} 
\, ; & \quad m=\{0,1,2\} \, , \nonu \\
& 
G_m(K;\, s_1,s_2) 
\ =\ 
\Re \ {\cal J}_{\ 11}^{(m)} 
\, ,  \nonu \\
& H_m(K;\, s_1,s_2) 
\ =\ 
\Im \ {\cal J}_{\ 11}^{(m)} 
\, ; & \quad m =\{0,1\} \, .
\label{def FGH}
\ee
With help from these intermediate functions, 
the imaginary part of our relevant two-loop master integrals can be written
\bea
\Im \ {\cal I}_{11020}^{(m,n)} &=& I_n(s_2) F_m (K;\, s_1,s_4) \, , 
\label{V and VI}\\
\Im \ {\cal I}_{11011}^{(m,n)} &=& 
\big[ G_m (K;\, s_1,s_4) H_n (K;\, s_2,s_5) \big] \  
    + \big[ \, m \leftrightarrow n \, , \
            s_1\leftrightarrow s_2\, ,\ 
            s_4\leftrightarrow s_5 \, \big] \, .\nonu
\eea
(The same spectral functions in Ref.~\cite{laine2} 
were labelled by a `d' and `g' respectively.)

Since $I_n$ was given above, we now 
turn to the $K$-dependence of $F_m\,$, for the particular cases needed.
As derived in Appendix \ref{app: B}
(and applicable for both $k_0>k$ and $k_0<k$)
\bea
& & F_m (K;\, s_1,s_2) \ = \
-\,s_1 s_2 \, \frac{n_{0}^{-1}}{64 \pi\, k} 
\label{F0}  \\
& & \qquad \times \, \Big(\,
  \frac{k_+^m}{k_-} n_{s_1}(k_+)   n_{s_2}(k_-) -
  \frac{k_-^m}{k_+} n_{s_1}(k_-)   n_{s_2}(k_+)  
\,\Big) \, , \nonu
\eea
where the distribution function $n_s$ was defined below \eq{IT prop}
and is evaluated at light cone momenta $k_\pm = (k_0\pm k)/2\,$.
We abbreviated the quantity $s_0 n_{s_0}(k_0)$ by $n_0\,$.

\begin{figure}[t]
  \includegraphics[scale=\figscale]{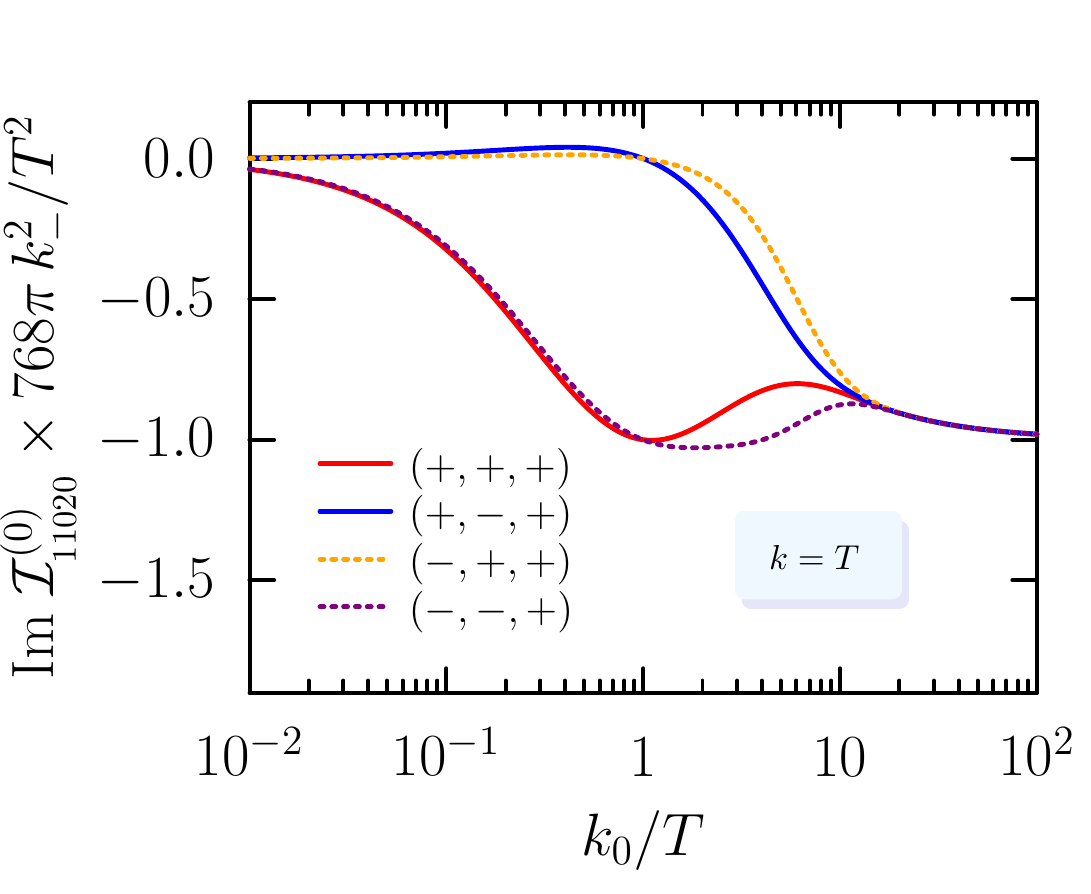}
  \caption{
    Energy dependence of the type II master integral with $m=n=0\,$, 
    for various statistics  at $k=T\,$.
  }
  \label{fig: 5}
\end{figure}

Figure \ref{fig: 5} shows the associated master integral with $m=n=0\,$.
We display all permutations of $s_0$ and $s_1\,$; the value of $s_2$
plays no role other than to change the vertical scale via $I_n (s_2)$ in Eq.~\eq{V and VI}.
($F_m$ is evaluated at $s_1$ and $s_4=s_0 s_1\,$.)
Note that the entire master has been multiplied by $k_-^2\,$,
which clarifies the nature of the pole at $k_0=k\,$: It is
simple if $s_1=-1$ and repeated if $s_1=+1\,$.

Moving along to the functions $H_m\,$, for $m=\{0,1,2\}$, 
after the frequency sum we have
\bea
H_m &=&
- \sum_{v} \int_{\bm p,\bm q} \ 
\frac{(2\pi)^{d+1}}{4E_1E_2} \d^{(d+1)} (K - v_1 P - v_2 Q ) 
 \nonu \\
&\times&   \big( \, 
 f_{s_1}^{ v_1} f_{s_2}^{ v_2} -
 f_{s_1}^{-v_1} f_{s_2}^{-v_2} \,  \big)  \,(v_1 E_1)^m \, , 
 \label{Hm int} 
\eea
where the summation extends over $v_{1,2}=\pm1$
[a definition of $f_s^v$ is given below Eq.~\eq{IT prop}].
The strategy discussed in Sec.~\ref{list} was applied to cover the cases 
$m\geq 0\,$.
Dimensional regularisation is adopted because the related function
$G_m$ will include a customary ultraviolet divergence:
Take $G_0$ and $H_0$ for instance, which are real and imaginary parts 
(respectively)
of the same function.
Their zero temperature limits can be read from
\bea
\lim_{T\to 0} {\cal J}_{\ 11}^{(0)} &=&
\frac1{(4\pi)^2} 
\Big[\, \frac1{\epsilon} + \kappa +2 
+ {\cal O}(\epsilon) \, \Big] \, ; 
\label{J11} \\
\kappa &\equiv& \log \frac{\bar \mu^2}{K^2} 
- i \, \pi \Theta(K^2) \, . \nonu
\eea
Branches of the logarithm are made explicit;
we write $\log X$ to mean $\log |X|$\,.
Thus $G_0$ bears an ultraviolet divergence and so ${\cal O}(\epsilon)$ 
terms must be kept in $H_0$ when multiplying them together.
For that reason we write
\bea
G_m &=& \,
\frac1{(4\pi)^2} \Big[ \
G_m^{[-1]} 
\, \Big( \, \frac1{\epsilon} + \log \frac{\bar \mu^2}{K^2} + 2 \, \Big)
+ G_m^{[0]}  + \ldots \ \Big] \, ,
\nonu \\
H_m &=&
-\,\frac{1}{16\pi} \Big[ \
H_m^{[0]} \, \Big( \, 1 + \epsilon \log \frac{\bar \mu^2}{K^2} \, \Big)
+ \epsilon \, H_m^{[1]}  + \ldots \ \Big] \, , 
\label{H order eps}
\eea
to make the dependence on the scale $\bar \mu$ explicit.
Setting $d=3$ in \eq{Hm int} allows us to find $H_m^{[0]}$: With help
from the moments $\psi^{(m)}$, provided in Eq.~\eq{psi} of
Appendix~\ref{app: A}, 
one can express
$$ H_m(K; s_1,s_2) \big|_{\epsilon\to 0} =
- k_0^m \,\psi^{(m)}_{s_1,s_2} \, \big/\, (16\pi) \ . $$
Thus $H_m^{[0]} = k_0^m \psi_{s_1,s_2}^{(m)}$ is easily read off.
Also needed are $G_0^{[-1]} = 1$ and $G_1^{[-1]}=\tfrac12 k_0\,$.
The order $\epsilon$ terms $H_m^{[1]}$
are presented in Appendix~\ref{app: B},
as is the function $G_m^{[0]}$ for $m=\{0,1\}\,$.

\begin{figure}[t]
  \includegraphics[scale=\figscale]{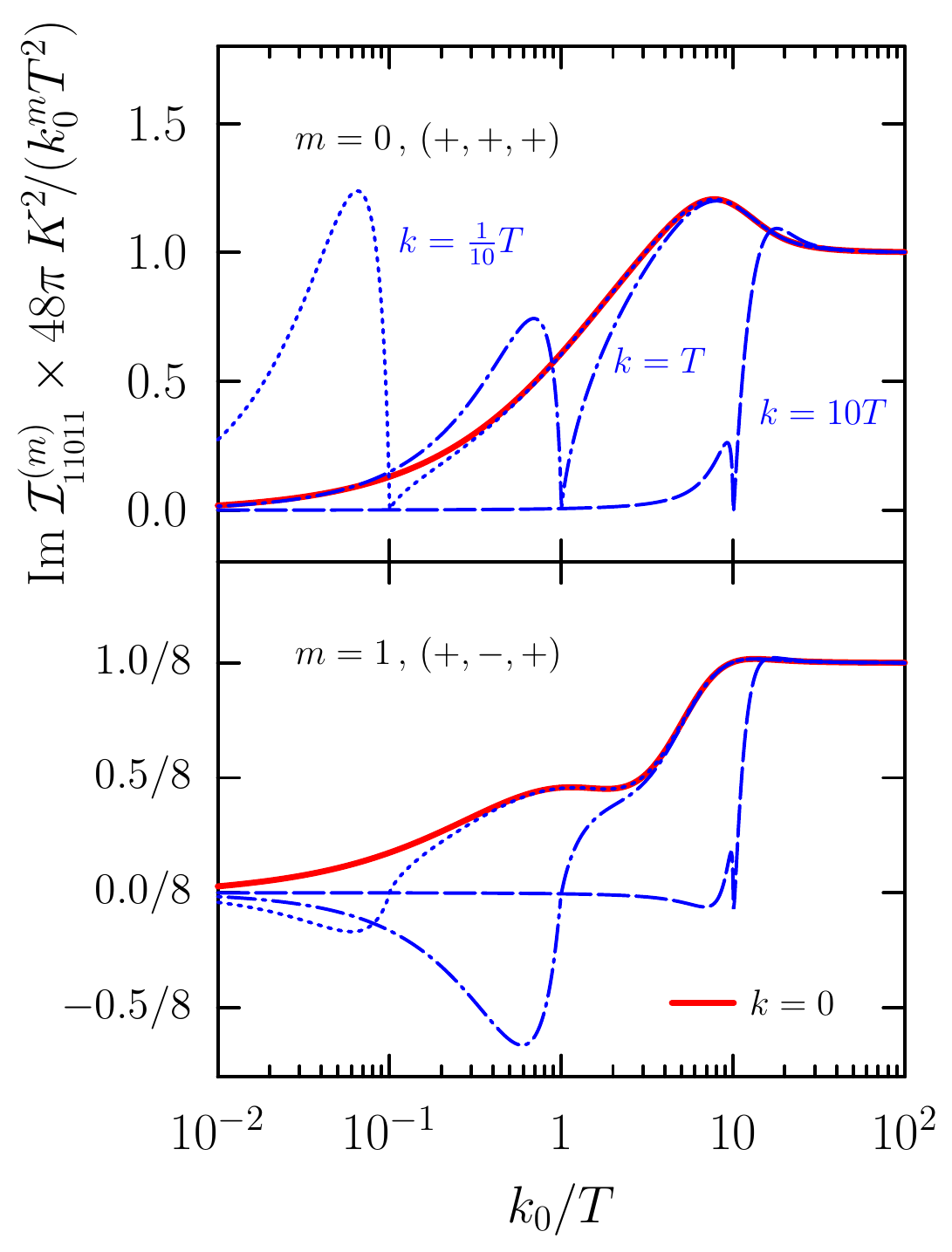}
  \caption{
    The finite part of the type III master integral, as a function of energy.
    We show here, for $k=\{0,\tfrac1{10}, 1, 10\}T\,$,
    the two cases $m=0$ (upper) and $m=1$ (lower).
  }
  \label{fig: 6}
\end{figure}

Turning to the class III integral, according to Eq.~\eq{V and VI} it can be written
\bea
\Im \ {\cal I}_{11011}^{(m,n)} \ =\
 \frac{-1}{4(4\pi)^3} &\Big[ &
  G_m^{[-1]}  H_n^{[0]} 
\Big( \, \frac1{\epsilon} + 2 \log \frac{\bar \mu^2}{K^2} +4 \, \Big)  \nonu \\
&-&\big( 2\, G_m^{[-1]} - G_m^{[0]} \big)  H_n^{[0]} 
+ G_m^{[-1]} H_n^{[1]} 
+ \,{\rm sym.} \,\Big] \, , 
\nonu
\eea
where the last term is from a symmetry as specified in \eq{V and VI}.
The first line above includes all divergences and yields the entire result
for $T=0\,$.
The second line is purely a finite thermal function which is not present
in vacuum.
Note that the divergent first line is also a function of the temperature
and we omit it in Fig.~\ref{fig: 6},
where the master integral is displayed.

On the light cone, a logarithmic 
singularity in this thermal part may arise from $\psi\,$.
This divergence is softened by the extra weight $K^2/T^2$ that
was used in Fig.~\ref{fig: 6}, and the plotted function is
zero at $k_0=k\,$.
We note that if $s_0=1$ (implying $s_1=s_4$ and $s_2=s_5$),
some simplifying relations hold
$$
{\cal I}_{11011}^{(1)} \ = \ 
{\cal I}_{11011}^{(0,1)} \ = \
\frac{k_0}2 \, {\cal I}_{11011}^{(0)}  \, .
$$

\section{Diagram IV (setting sun) \label{sec: 2}}

We now consider the first {\em genuine} two-loop structure, specifically
 the integral ${\cal I}^{(0)}_{11100}\,$, which gives 
 \eg the first non-zero contribution to the imaginary part of the 
self energy in a scalar $\varphi^4$-theory \cite{Ramond}. 
Another master of the same class, ${\cal I}^{(0)}_{01110}=0\,$,
is identically zero due to 
 integration by parts identities \cite{Schroder}.
The `setting sun' graph is given in vacuum by
\bea
\lim_{T\to 0}\, {\cal I}^{(0)}_{11100}
\!\!&=&\!
\frac{-K^2}{4(4\pi)^4} 
\Big[\, \frac1{\epsilon} + 2\, \kappa
+ \frac{13}2
+ {\cal O}(\epsilon) \, \Big] \, , \quad 
\label{sunset vac}
\eea
where $\kappa$ was introduced in Eq.~\eq{J11}.
This vacuum result has an imaginary part 
for $K^2>0$, associated with the threshold for massless particle production.
In a thermal medium, the Landau-damping mechanism explains why this imaginary
part also builds up below the light cone \cite{Wang1995qf}.
Explicitly, 
\bea
 \Im\
  {\cal I}^{(0)}_{11100} &=&
\label{im phi} 
 \sum_{v} \int_{\bm p,\bm q,\bm r} 
\frac{(2\pi)^6}{8E_1E_2E_3}  \\
 &\times& \d^{(4)} (K - v_1 P - v_2 Q - v_3 R)  
 \, \Big( \  
 f_{s_1}^{ v_1} f_{s_2}^{ v_2} f_{s_3}^{ v_3} -
 f_{s_1}^{-v_1} f_{s_2}^{-v_2} f_{s_3}^{-v_3} \  \Big) \, ,
 \nonu
\eea
where $f_s^v$ was defined just below \eq{IT prop} and takes
the arguments at
energies $E_1 = |\bm p|$, $E_2=|\bm q|$ and $E_3 = |\bm r|$ which are on-shell.
This integral is labelled `f' in Ref.~\cite{laine2}.

Let us clarify the physical content of Eq.~\eq{im phi}.
The sum over the signs $v_i=\pm1$ 
enumerates eight distinct physical interactions,
with external momentum $K=(k_0,\bm k)$\,.
We denote the corresponding fields $\phi_i$ for argument's sake with $s_i=+1$.
As an example, the term with $v_1=v_2=v_3=+1$ represents the probability 
for decay $\phi_0 \to \phi_1 \phi_2 \phi_3$, with a statistical weight of
$(1+n_{_B})(1+n_{_B})(1+n_{_B})$ for spontaneous emission, 
minus the probability for creation $\phi_1 \phi_2 \phi_3 \to \phi_0$, with a 
weight $n_{_B} n_{_B} n_{_B}$ for absorption.
There are many other processes, such as $\phi_0 \phi_2 \phi_3 \to \phi_1$
minus $\phi_1 \to \phi_0 \phi_2 \phi_3$ and so on \cite{Weldon1983jn}.

Equation \eq{im phi} may be simplified into a two-dimensional integral
(now for general $s_i$)
\bea
\Im\ 
{\cal I}^{(0)}_{11100} &=&
\frac{n_0^{-1}}{(4\pi)^3} 
\int \! d p \, dq \ W_{\rm I\!V}(p,q) \
 n_1 \, n_2 \, n_3 \ , 
\eea
where we abbreviated
$n_i = s_i n_{s_i}\,$ and
agree that the arguments\footnote{%
  To avoid possible ambiguity, but referring ahead, \eq{ni} summarises
our shorthand notation for the distribution functions explicitly.}
of the distribution functions may be negative.
The `kernel' $W_{\rm I\!V}$ (defined below) also depends on $k_0$ and $k\,$, 
but not on the temperature.
The momentum moduli $p$ and $q$ have been generalised 
to negative values, which implicitly incorporates the 
sum over the signs $\{v_i\}$.
And the statistical weight has accordingly been re-expressed using
\bea
 f_{s_1}^{+} f_{s_2}^{+} f_{s_3}^{+} -
 f_{s_1}^{-} f_{s_2}^{-} f_{s_3}^{-} 
 &=&
  \ n_0^{-1}\,  n_1 \, n_2 \, n_3 \ ,
 \label{gain minus loss}
\eea

An explanation that starts with Eq.~\eq{im phi}
is given in appendix \ref{app: C},
where we also show how to calculate $W_{\rm I\!V}$
from kinematic constraints.
Here we simply state the result:
\bea
W_{\rm I\!V}(p,q) \ =\  \frac{1}{2k} 
&\Big\{& 
|p-k_+| + |q-k_+| - |p+q-k_+| 
\label{W formula}\\
&-&      
|p-k_-| - |q-k_-| + |p+q-k_-| 
- \min [ k_0, k ] \ \, \Big\} \, ,
\nonu
\eea
where $k_\pm = (k_0 \pm k)/2\,$ are the light cone momenta.

Of note is that $W_{\rm I\!V}(p,q) \simeq p/k$ for $p\to 0$, 
which suppresses the log divergence from $n_{_{B}}(p)$\,.
Furthermore, $W_{\rm I\!V}=0$ in regions that are kinematically forbidden,
providing limits on the $p$ and $q$ integrals.
Continuity of $\Im\ {\cal I}^{(0)}_{11100}$ at $k_0 = k$ follows 
from the very same property in \eq{W formula}.
We note that the limit $k\to 0$ is also well-defined and leads to
${W_{\rm I\!V} = \sgn \bm(\,pq(k_0-p-q)\,\bm)}$ where it has non-zero support.

The statistical factor \eq{gain minus loss} includes  the vacuum
contribution for $v_1=v_2=v_3=+1\,$, 
i.e.~where $p$ and $q$ are positive and $p+q<k_0\,$.
It is the leading term in Eq.~\eq{gain minus loss}, 
after expanding in combinations of the distribution functions
$$
 f_{s_1}^{+} f_{s_2}^{+} f_{s_3}^{+} -
 f_{s_1}^{-} f_{s_2}^{-} f_{s_3}^{-} 
 =
 1 + \textstyle\sum_i n_i + \sum_{i<j} n_i n_j \ .
$$

\begin{figure}[h]
  \includegraphics[scale=\figscale]{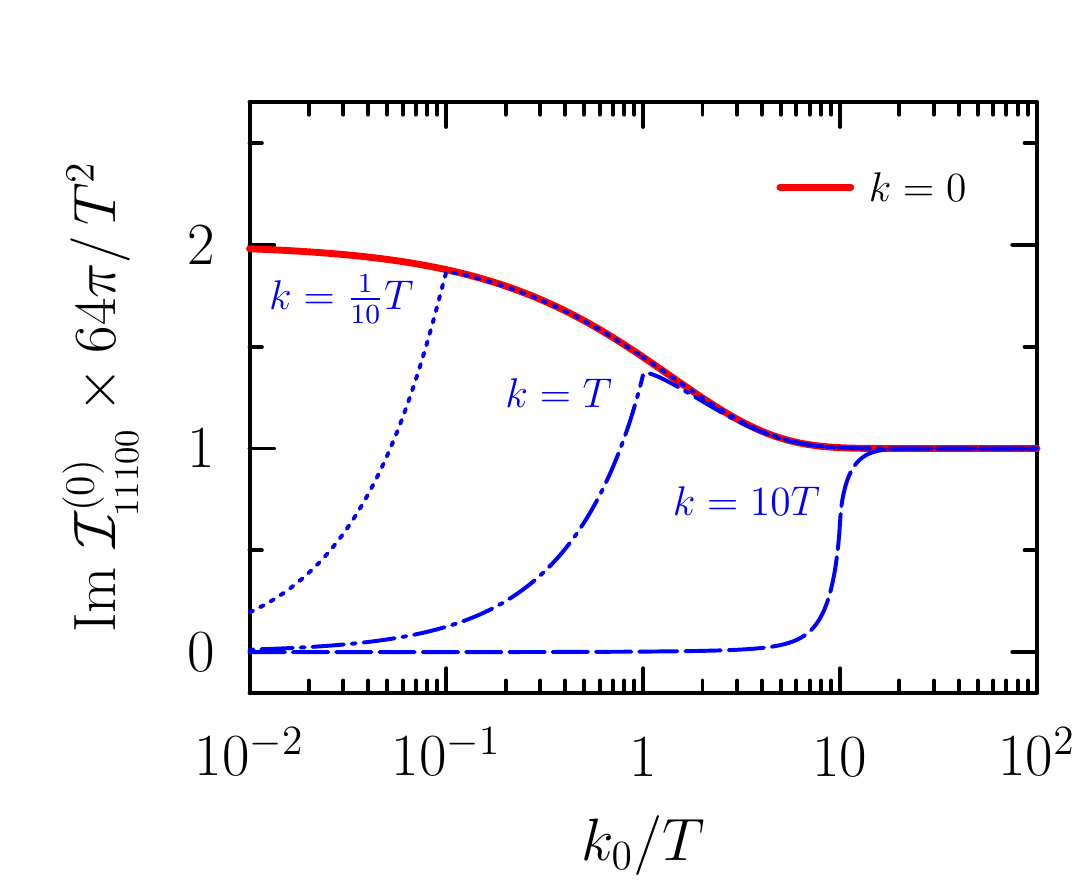}
  \caption{
    The imaginary part of the scalar self energy (with $s_i=+1$), 
    according to Eq.~\ref{sunset plot}.
    Shown here is the dependence on $k_0$ for various values 
    of the external momentum.
  }
  \label{fig: im pi}
\end{figure}

In Fig.~\ref{fig: im pi}, the energy dependence of
$\Im\ {\cal I}_{11100}^{(0)}$ is shown for $k=\{ 0,\tfrac1{10},1,10 \}\times T$.
Here the vacuum result \eq{sunset vac} was subtracted, \ie 
we actually plot 
\be
\Im\ {\cal I}_{11100}^{(0)} - \Theta (K^2) \, \frac{K^2}{8(4\pi)^3}\, .
\label{sunset plot}
\ee

Because the master integrals are holomorphic in the 
upper half of the complex $k_0$-plane, \eq{im phi}
is an odd function of real energies.
(Meaning, in particular, that it is zero for $k_0=0\,$.)
The exception, for massless particles, 
occurs when $\bm k =0$ so that there is an essential
singularity at $k_0 = 0$ \cite{Weldon2001vt}.
Hence the zero momentum curve in Fig.~\ref{fig: im pi} is finite
for $k_0 \to 0$ and not equal to the same limit at fixed  $|\bm k|>0\,$.

\subsection{Kinematics}

The region(s) where Eq.~\eq{W formula} provides non-zero support 
for $W_{\rm I\!V}$ can be understood by elementary kinematic reasoning.
It is necessary to belabour this point because the argument will reveal 
why it holds {\em in general} for the real corrections.
To illustrate, we first consider the simpler case $k\to 0$ 
and then explain what happens when $k<k_0$ and $k>k_0$ separately.

\begin{figure}[t]
  \includegraphics[scale=\figscale]{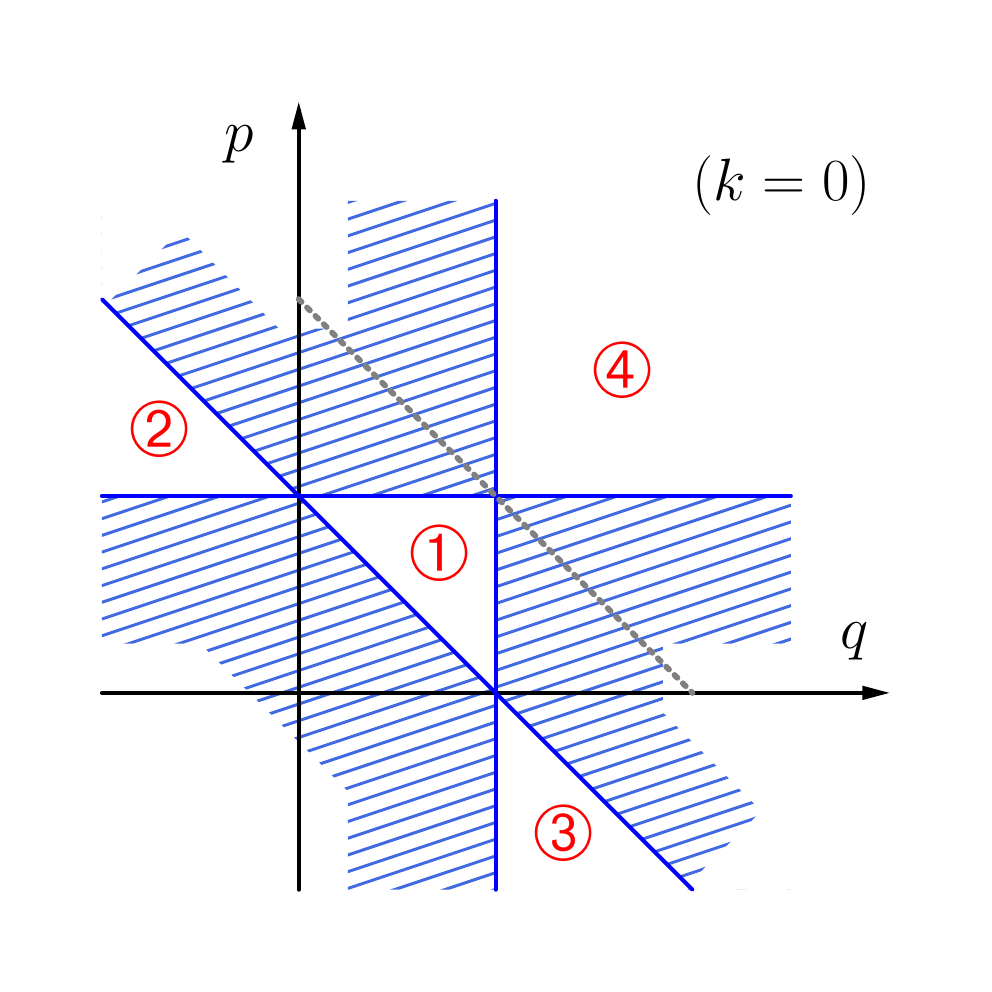}
  \caption{
    The boundaries to the integration regions for the four
    (real) corrections. 
    Each region corresponds to a physical process,
    \eg `1' is described in the paragraph below \eq{im phi}.
    The dashed line represents momenta that satisfy $p+q=k_0\,$.
  }
  \label{k=0}
\end{figure}

The function $W_{\rm I\!V}$ is not  Lorentz invariant --
if it were, we could perform the whole calculation in the rest frame.
Nevertheless, specialising to $k=0$ will be useful as a starting point.
For example, in the channel with $v_1=v_2=v_3=+1$ we require vectors
$\bm p$ and $\bm q$ that satisfy 
\bea
 k_0 &=& |\bm p| + |\bm q| + |\bm p+ \bm q|\, .
 \label{+++}
\eea
Clearly $p+q \leq k_0$ in general, 
with equality if and only if $\bm p = -\bm q$ and $|\bm p| = \frac12 k_0$.
Moreover, by the triangle inequality we have 
$|p-q| < |\bm p + \bm q| < p+q$ (here $p,q>0$).
The upper limit thus gives $\frac 12 k_0 < p+q$, while the lower limit leads to
$p < \frac12 k_0$ if $p>q$ and $q< \frac12 k_0$ if $p<q$.
That defines the relevant domain in the $(p,q)$-plane; 
see `1' in Fig.~\ref{k=0}, where $W_{\rm I\!V}=+1$.

If $v_1=v_2 = +1 = -v_3$, instead of \eq{+++} we need 
\bea
k_0 = |\bm p| + |\bm q| - |\bm p + \bm q| \, .
\eea
This time $p+q \geq k_0$ and 
the very same triangle inequalities give $p>\frac12k_0$ 
if $q>p$ and $q>\frac12k_0$ if $q<p$.
Similar results hold if $v_3 = +1$ and if 
exactly one of $v_1$, $v_2$ is equal to $-1$.
However if two or more of $\{v_1,v_2,v_3\}$ are negative, 
the equivalent of \eq{+++} 
cannot be satisfied (for $k_0>0$).
Hence in the sum over $\{ v_i \}$, only half of the summands contribute.

This is summarised by the wedge-shaped regions `2', 
`3' and `4' in Fig.~\ref{k=0}.
In each of these three regions, $W_{\rm I\!V}=-1$.
Although they include arbitrary large momenta (in absolute value), 
those much larger than the temperature are cut off by the thermal distribution
functions.

\bigskip

\begin{figure}[t]
  \includegraphics[scale=\figscale]{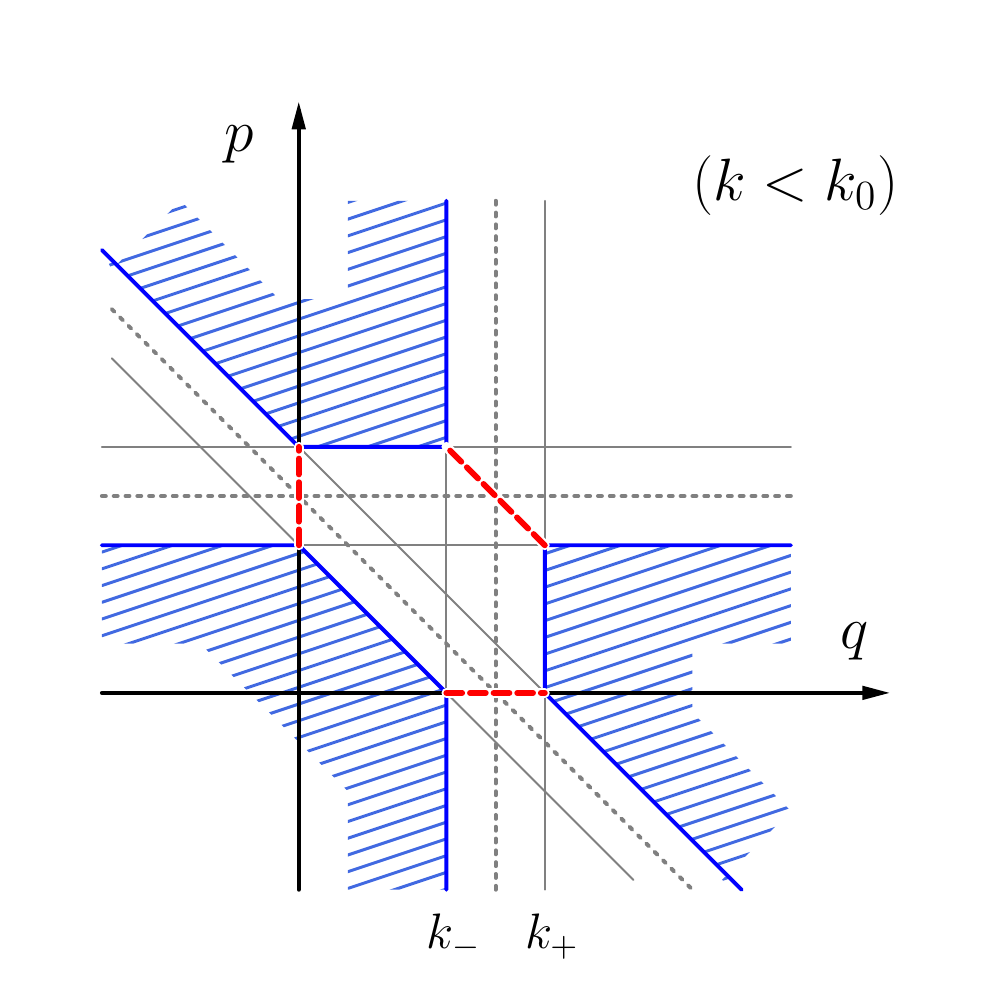}
  \caption{
    As before, now with $0<k<k_0$. 
    The red dashed lines give the boundaries between different channels;
    \ie where one of $p$, $q$ or $r=(k_0 -p-q)$ changes sign.
  }
  \label{k<k0}
\end{figure}

We now consider $k>0$, but still less than $k_0$ so that 
$k_-=\frac12 (k_0-k)$ is positive.
The external vector $\bm k$ now plays a role, \ie \eq{+++} 
is supplanted by
\bea
k_0 &=& |\bm p| + |\bm q| + |\bm k - \bm p - \bm q| \, .
\label{energy conservation}
\eea
Of course $p+q \leq k_0$ still holds, but now equality can occur for all $q\in[k_-,k_+]$.
The triangle inequality gives $k+|\bm p + \bm q| > |\bm k - \bm p -\bm q|$, and 
therefore
\bean
2k_- = (k_0 - k) &<& |\bm p| + |\bm q| + |\bm p + \bm q|  \ <\  2(p+q) \, .
\eean
Hence the lower bound on $p+q$ is diminished to $k_-$.
Similarly, 
the other side of the triangle inequality gives 
$|\bm k - \bm p - \bm q| > |\bm p + \bm q| - k$.
That produces $k_+ \geq \max (p,q)$, which is higher than the upper bound for $k=0$.
Generalising to other channels is trivial; see Fig.~\ref{k<k0}.
These restrictions are reflected in the function $W_{\rm I\!V}$.
In the `new' bands that open up for $p,q,r \in [k_-, k_+]$,
tilted facets make $W_{\rm I\!V}$ a continuous 
function compared to the case where $k=0$.
Moreover, the exact form \eq{W formula} renders the 
product of distribution functions integrable.

\bigskip

As $k$ is increased beyond $k_0$, the virtuality $K^2 = 4 k_+k_-$ becomes negative.
The channel with $v_1=v_2=v_3=+1$ ceases to be accessible; conservation of energy
\eq{energy conservation} cannot be satisfied if $k_0 < k$.
This simply means that there is no vacuum contribution below the light cone,
as expected.

\begin{figure}[t]
  \includegraphics[scale=\figscale]{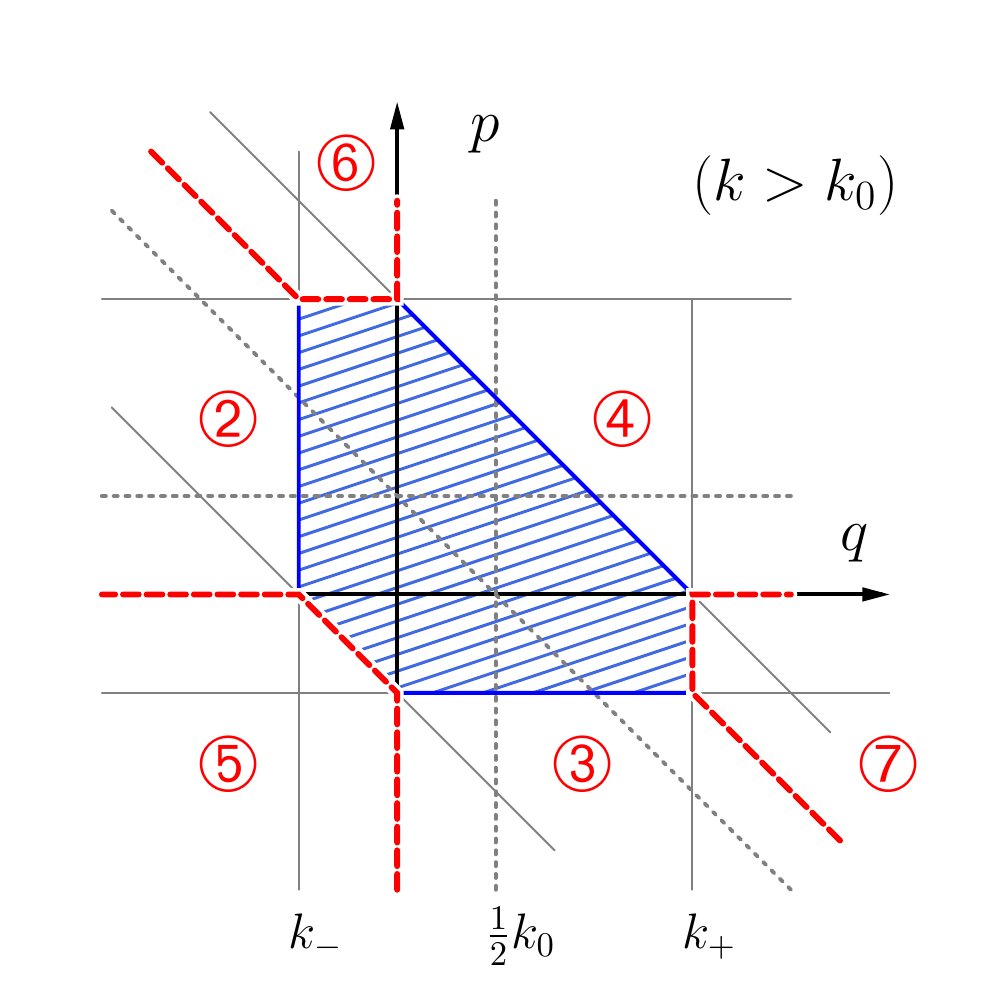}
  \caption{
    As before, now with $k>k_0$ so that $k_- < 0$.
    (The blue hatched region is forbidden by kinematics. 
    It is, incidently, also where the vacuum contribution comes from.)
  }
  \label{k>k0}
\end{figure}

New wedges open up in the $(p,q)$-plane -- they
correspond to having {\em exactly} two of $\{v_1, v_2, v_3\}$
equal to $-1$.
In Fig.~\ref{k>k0} they are labelled `5`, `6' and `7'.
Regions `2', `3' and `4' are carried over from the case $k<k_0$, 
with modified boundaries:
For example in `2', 
the same triangle inequality as before gives $|\bm q|<-k_-$\,.
Region `5' has $v_1=v_2=-1$ and $v_3=+1$\,, 
so in some sense it is the {\em intersection} of `2' and `3'. 
In `5' we have $p$ and $q$ negative such that $p+q<k_-$\,.
(And there are similar constraints for `6' and `7'.)
We note that adjacent regions are separated by lines
where $p$, $q$ or $r$ is zero.
These boundaries are important because they mark the location
of potential singularities coming from bosonic distribution functions.

\bigskip

The expression in \eq{W formula} can be simplified in each of 
the demarcated regions discussed for $k_0>k$ and $k_0<k\,$.
In the former case, we recover Eqs.~(33-43) of Appendix B in 
Ref.~\cite{laine2} (with an adjustment in variables, namely $p\to k_0-p$).
For the more complicated master integrals still to be studied, the
same regions must be considered, though the kernel functions
will be different.
Hence, the foregoing analysis serves to spell out what must
be included for the task of numerical integration.

\section{Diagram V (the squint) \label{sec: 3}}

\begin{figure}[h]
  \includegraphics[scale=2.1]{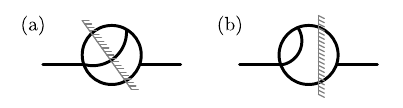}
  \caption{
    Cut type II diagrams. Real (a) and virtual (b).
  }
  \label{fig: 11110}
\end{figure}

In the previous section there is only one
`contribution' to the imaginary part: all of the three internal energies 
are set to their on-shell value giving \eq{im phi}.
But more generally, the discontinuity may always be written in terms of products of 
two amplitudes that are separated by on-shell `cut' propagators 
\cite{Weldon1983jn}.
For classes V and VI, there is more than one way to do this; 
see Fig.~\ref{fig: 11110}.
These contributions, which may be readily identified after carrying out the 
Matsubara sums, are separately infrared divergent.
They differ by one extra loop momentum being on-shell in the {\em real} case,
yielding a tree-level decay that may be treated in
a similar manner to that of the previous section.
The {\em virtual} correction has an internal loop due to one fewer final 
state particle than before and includes a two-body phase space integration.
The latter is also ultraviolet divergent, seen in the vacuum result
[$\kappa$ was introduced in Eq.~\eq{J11}]
\bea
\lim_{T\to 0}\, {\cal I}^{(0)}_{11110}
&=&
\frac1{(4\pi)^4} 
\Big[\  \frac1{2\epsilon^2} \,+\,
\frac{2\kappa+5}{2\epsilon} \,+\, \kappa(\kappa+5) 
+ \, \frac{19}2 \,-\, \frac{\pi^2}{12}
\,+\, {\cal O}(\epsilon) \, \Big] \, ,
\label{11110 vac}
\eea
which, for $K^2$ positive, contains an imaginary part $\propto 1/\epsilon\,$.
This divergence will then acquire a temperature dependence in the medium.
When assembled together, these infinite parts cancel in actual observables.

The two terms shown in Fig.~\ref{fig: 11110} also have separate 
collinear divergences which exactly compensate {\em only} in their sum.
This cancellation is somewhat intricate at finite temperature
and affirms the Kinoshita-Lee-Nauenberg (KLN) theorem, in this case
applied to individual graphs \cite{kln1,kln2}.
The amplitude (a) in Fig.~\ref{fig: 11110} may
contain a large `eikonal factor' if the denominator
of an internal line is zero,
\be
L^2 
\ =\  
K^2 - 2(v_1p)\, \big(\, k_0 - k \cos \theta_{kp}\,\big) 
\ \approx \ 0 \, ,
\label{collinear}
\ee
where $\theta_{kp}$ is the angle between $\bm p$ and $\bm k$.
Such configurations exhibit two collinear outgoing
(massless) particles that should be removed from the definition of
a {\em physical} production rate, \ie by including (b) from Fig.~\ref{fig: 11110}.
The conditions for some $\theta_{kp}$ to satisfy
Eq.~\eq{collinear} are inferred according to 
the sign of $K^2\,$:\
If $K^2>0$ then $(v_1 p)$ must be in the interval $\Omega=[k_-,k_+]$\ 
(spanned by the light cone momenta).
While if $K^2<0$, \eq{collinear} can be satisfied if and only if $(v_1p)$ 
is in the complementary
region $\Omega^c = \mathbb{R}\backslash \Omega\,$.

A fictitious mass $\lam\,$ (for the momentum $R$) is convenient to 
use as a calculational tool \cite{ItzyksonZuber}\footnote{%
  The parameter $\lam$ is not related to renormalization.
  }.
It prevents $L^2 \to 0$ in \eq{collinear} and proves that the whole expression
\bea
\Im\ 
{\cal I}^{(m,n)}_{11110} & & \!\!\!\!\! =
\frac{n_0^{-1}}{(4\pi)^3} 
\int \! dp \, p^m \,
\Big[ \label{11110} \\
  \int \!  dq &&\!\!\!\!\!\! q^n\, W_{\rm V}(p,q\,;\lam) \
n_1 n_2 n_3  
+
U^{(n)}_{\rm V} ( p\, ;\lam) \ n_1 n_4 \ \Big] \, ,
\nonu
\eea
is rendered finite as $\lam \to 0\,$. 
[See ahead \eq{ni} for the abbreviations $n_i\,$.]
The first term above represents the real correction, and involves
the thermal weight \eq{gain minus loss} from before.
Virtual corrections lead to the second term and depend on the mass $\lam$
so as to compensate for the contingent singularity in the first.
Equation \eq{11110} is valid for $m,n\leq1$, 
otherwise there could be more terms.

We may obtain $W_{\rm V}(p,q\,;\lam)$ by the techniques laid out in
Appendix~\ref{app: C}.
A compact formula for it can be given with the help
of some funny notation: 
$$\hat p = \max\big[\,k_-,\, \min[k_+,p]\,\big] \, .$$
To also indicate whether $p$ is in the interval
$\Omega = [k_-,k_+]$, we define
$$ o_p = \Theta\big((k_+-p)(p-k_-)\big) = 
\l\{
\begin{array}{ll}
  1  &  {\rm if} \ p \in \Omega \\
  0  &  {\rm otherwise.}
\end{array} \r. $$
Thus the complementary set $\Omega^c = \mathbb{R} \backslash \Omega$ contains $p$ 
if the value of $\bar o_p = 1-o_p \,$ is equal to unity.
Using $o_p$ and $\bar o_p$ will signal terms that are included or not,
depending on the relative ordering of $k\,$, $k_0$ and $p\,$.
With these definitions, and regulating through $\lam \to 0$, the weight function can be expressed 
as\footnote{%
  Compare with Eqs.~(57)-(67) of Appendix B in Ref.~\cite{laine2}.
  }
\bea
W_{\rm V}(p,q; \lam) \ =\ \frac1{4k\,\ell} \!\!
&\Big\{&
 \log W_{\rm V}^\prime 
\label{W_V} 
+ \Theta(k_-) \log W_{\rm V}^{\prime\prime} \\
&+& \bar o_p\log W_{\rm V}^{\prime\prime\prime} 
+  \big( \Theta(k_-)-\bar o_p \big) \log \frac{4\ell r}{\lam^2}
\, \Big\} \, , 
\nonu
\eea
where we have introduced the ratios
\bean
&\displaystyle W^\prime_{\rm V}
= \frac{(\ell - \hat q)(\ell - \hat r)}{\hat q\, \hat r} \, ,
\qquad
W^{\prime\prime}_{\rm V}
= \frac{k_-}{p-k_+} \, ,
\qquad
\displaystyle W^{\prime\prime\prime}_{\rm V}
= \frac{k_0-\hat p}{p-\hat p} \, .
\eean
An explicit log-divergence in \eq{W_V} lingers 
if $p\in\Omega$ (for $k_0>k$) or if $p \in \Omega^c$ (for $k_0<k$).
This implies that it is too soon to set $\lam = 0$ in those domains
of the $(p,q)$-plane and brings us to incorporate the
missing virtual pieces.

Again deferring details to Appendix~\ref{app: C}, we write, for $\lam \to 0$,
\bea
U^{(0)}_{\rm V} (p\,;\lam)  =  \frac{\bar o_p - \Theta(k_-)}{4k} \!\!
&\Big[&
  \label{U_V} \\
 &-& \frac{1}{\ell} \int_{-\infty}^{+\infty} \!\! dq
 \Big\{\,
  \frac{n_2 n_3}{n_4} \log \frac{\lam^2}{4\ell q}
   +
  s_3  \sgn (r) n_{s_3} \big(|r|\big) 
  \log \frac{q^2}{r^2} \,\Big\} \nonu \\
 &+&
  \frac1{\epsilon}  + 
  2 \log \frac{\bar \mu^2}{K^2} 
+   \log \frac{K^2k^2}{4\ell^2(p-k_-)(p-k_+)} + 2 \nonu
\, \Big] \, .
\eea
This reveals how the anticipated ultraviolet divergence in 
\eq{11110 vac} emerges from the loop in (b) of Fig.~\ref{fig: 11110}.
At the same time, it can be seen that 
the $\lam$-dependence in \eq{11110} cancels for $\lam=0$, 
with the logarithmic mass singularities compensating perfectly,
$$
\log \frac{4\ell r}{\lam^2} + \log \frac{\lam^2}{4\ell q} 
= \log \frac{r}{q} \, ,
$$
and in exactly the domains where \eq{collinear} is satisfied.
They coalesce in this way both above and below the light cone.

\begin{figure}[t]
  \includegraphics[scale=\figscale]{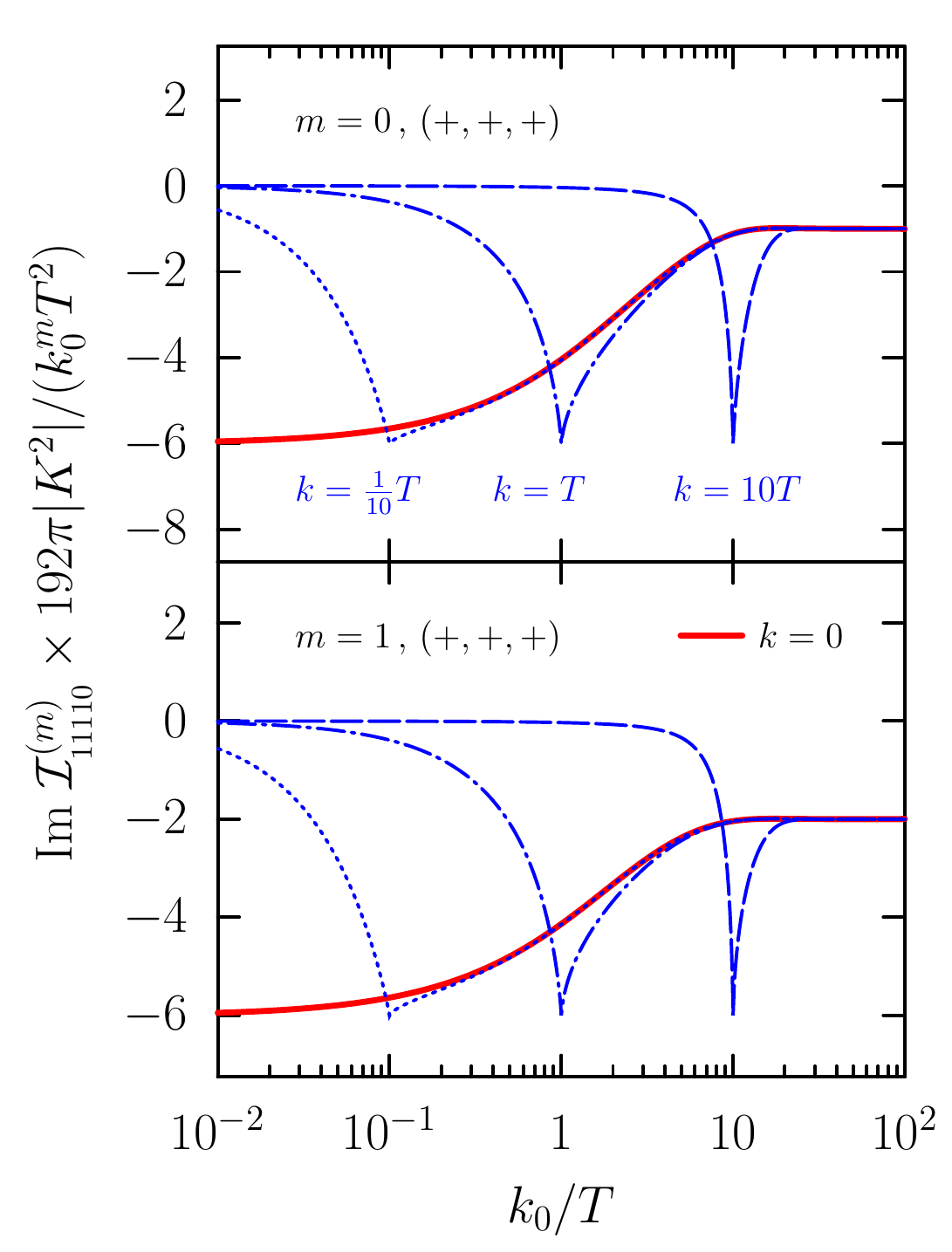}
  \caption{
    The imaginary part of the type V master (with all $s_i=+1$), 
    in units of the asymptotic result.
    Shown here is the dependence on $k_0$ for various values of 
    the external momentum, including $k=0$.
  }
  \label{fig: im V}
\end{figure}

For the purpose of plotting, we subtract a piece that is 
ultraviolet divergent from \eq{11110} 
and coincides with its vacuum result for $T=0$.
What remains is thus finite and proportional to $T^2$ for 
large photon virtualities, see Fig.~\ref{fig: im V}.
The function is continuous across the light cone unless it diverges there,
in which instance the singularity is the same if  $k_0 \to k$ is approached
from either above or below.
That is why, 
in Fig.~\ref{fig: im V}, we multiply the whole function by $|K^2|$ which
is enough to render the blow-up finite on the light cone. 
(It also gives the whole master integral a dimension of $T^2$.)
To be clear, and following \cite{laine2} 
(in which this master is labelled `h'), 
the part subtracted is
\be
\Im\ 
{\cal I}^{(0)}_{11110} \Big|_{\rm div.} &=&
- \frac{\psi^{(0)}_{s_1,s_4}}{4 (4\pi)^3}
\Big( \, \frac1{\epsilon} + 2 \log \frac{\bar \mu^2}{K^2} + 5 \, \Big) \, ,
\ee
where $\psi$ is defined in Eq.~\ref{psi} of the appendices.
The large $k_0$-expansion of the whole master integral
is given in Eq.~\eq{ope: V 00}.

\bigskip

Next, let us consider the case $m=1$ and $n=0\,$. 
The modification in \eq{11110} is trivial; 
only an extra factor of $p$ needs to be included in the integrand.
The kernels $W_{\rm V}$ and $U_{\rm V}^{(0)}$ are unaffected and hence the
$\lam$-dependence (and ultimate lack thereof) is the same as before.
For plotting in Fig.~\ref{fig: im V}, we subtract
\be
\Im\ 
{\cal I}^{(1)}_{11110} \Big|_{\rm div.} &=&
-\frac{k_0 \psi^{(1)}_{s_1,s_4}}{4 (4\pi)^3}
\Big( \, \frac1{\epsilon} 
      + 2 \log \frac{\bar \mu^2}{K^2} + \frac{11}2 \, \Big) 
\ee
from the result. 
The shape is similar to 
$\Im\ {\cal I}^{(0)}_{11110}\,$,
but ratios of the large-$k_0$ limit to its value on the light cone
are different.

For $m=0$ and $n=1\,$, the virtual parts need to be reconsidered. 
The result, with details available in 
Appendix~\ref{app: C}, can be written, for $\lam \to 0\,$,
\bea
 U^{(1)}_{\rm V} (p\,;\lam)  = \frac{\bar o_p - \Theta(k_-)}{4k} \!\!
&\Big[&
  \label{U1_V} \\
 &-& \int_{-\infty}^{+\infty}  \! dq \, \frac{q}{\ell}
 \Big\{\,
  \frac{n_2 n_3}{n_4} \log \frac{\lam^2}{4\ell q}
  + 
  s_3  \sgn (r) n_{s_3} \big(|r|\big) 
  \log \frac{q^2}{r^2} \, \Big\} \nonu \\
 &+&
  \frac{\ell}2 \, \Big\{ \, \frac1{\epsilon} + 
  2 \log \frac{\bar \mu^2}{K^2} + 
  \log \frac{K^2k^2}{4\ell^2(p-k_-)(p-k_+)} + 1 
  \, \Big\}
  \nonu
\  \Big] \, . 
\eea
\begin{figure}[t]
  \includegraphics[scale=\figscale]{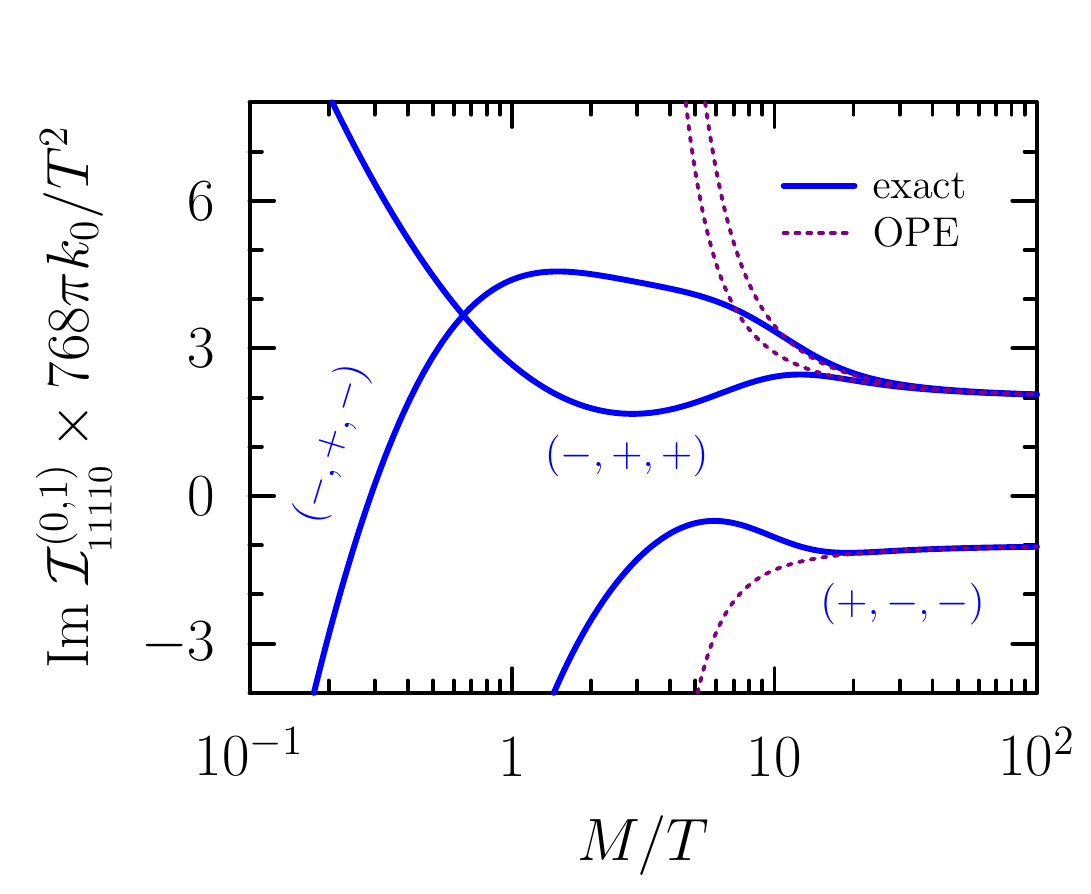}
  \caption{
    Here we display the type V master with $m=0$ and $n=1\,$,
    minus the divergent part.
    Various statistical configurations are shown with $s_0 \neq s_1$
    (so that these cannot be reduced to a linear combination
    of master integrals with $n=0$).
    The first two orders from the OPE asymptotics, Eq.~\eq{11110_01},
    start agreeing from about $M \gsim 30T\,$.
  }
  \label{fig: im V 01}
\end{figure}

Figure~\ref{fig: im V 01} displays the result
as a function of positive invariant mass $M = \rt{k_0^2 - k^2}\,$.
We have followed Refs.~\cite{laine1,laine2} by using $M$ to
define
\be
k_{\rm ave}^2 (M) &\equiv&
\frac{3 MT K_3\big(M/T\big) }{K_2\big(M/T\big)} \, ,
\label{k2ave}
\ee
as a proxy for the average three momentum squared. 
($K_\nu$ are modified Bessel functions.)
This quantity assumes Boltzmann distribution functions
to average $k^2$ for a fixed mass.
The curves shown in Fig.~\ref{fig: im V 01} use
$k \to \rt{k_{\rm ave}^2(M)}$ as a rather crude substitute
for `typical' momenta.
As $M\to 0\,$, we note the clear log-divergent behaviour of the master integral
for the statistics shown in the figure.

The divergent piece that was subtracted is given by [note the 
ordering of $s_4$ and $s_1$ on $\psi^{(1)}$ ]
\be
\Im\ 
{\cal I}^{(0,1)}_{11110} \Big|_{\rm div.} &=&
-\frac{k_0 \psi^{(1)}_{s_4,s_1}}{8 (4\pi)^3}
\Big( \, \frac1{\epsilon} 
+ 2 \log \frac{\bar \mu^2}{K^2} + \frac92 \, \Big) \, . 
\nonu 
\ee
We have also checked numerically that 
$\Im\ {\cal I}_{11110}^{(0,1)} = \frac12 \big(
k_0\, \Im\ {\cal I}_{11110}^{(0)} - \Im\ {\cal I}_{11110}^{(1)}\big)$
provided $s_2=s_3\,$. 
This relation follows by a shift of integration variables,
but it seems difficult to discern this rule by simply
looking at the explicit form of the integrand.

In Sec.~\ref{list} we listed among the type V integrals, one
with a propagator of negative power $e=-1\,$. 
It is significant because of a logarithmic divergence for $K^2 \ll T^2$
that makes it dominant over those that are merely finite on the light cone
(when scaled by $K^2$ to have the appropriate dimension).
This brings about its appearance in many applications.
Moreover, the continuity across the light cone (that some previous masters
seemed to enjoy) is no longer guaranteed.
This issue reveals itself explicitly in the QCD corrections to the photon 
spectral function \eq{masters, mn} [used in \eq{expansion def}], 
\bea
\Im\, \Pi_\mu^{\ \mu} &\simeq&
 e^2 g^2 \,  T^2\ \frac{N\cf}{16\pi} 
\big[\, 1 + 2\, n_{_-}(k_0) \,\big]
\log \frac{T^2}{K^2} 
\label{log div}
\eea
for $K^2 \ll T^2$ \cite{Baier1988,Altherr1989,Gabellini1989}.
This is the singularity alluded to in the Introduction, which mandates
screening effects to be incorporated through resummation.
The log-singularity on the light cone can be traced back to 
the integral we are about to discuss.

Let us consider the particular combination of master integrals defined by
\bea
{\cal I}_{11110}^\text{\,\large $\star$} &\equiv&
{\cal I}_{10110}^{(0)}
+ K^2 \, {\cal I}_{11110}^{(0)}
- {\cal I}_{1111(-1)}^{(0)} 
\label{star} \\ &=&
\sumint{\,P,Q} \frac{2\, K\cdot Q}{P^2 \, Q^2 \, (K-P-Q)^2 \, (K-P)^2} \, .
\nonu
\eea
(This one is labelled by `h$^\prime$\,' in Ref.~\cite{laine2}.)
After carrying out the Matsubara sum, one arrives at Eq.~\eq{C2 - 1b}
in the Appendix.
Along the same lines as \eq{11110}, this can be expressed by
\bea
\Im\ {\cal I}_{11110}^\text{\,\large $\star$} \,=\,
\frac{n_0^{-1}}{(4\pi)^3} 
\int \! dp \!\! &\Big[& \int \! dq \,
  W_\text{\large $\star$}(p,q\,;\lam) \, n_1 n_2 n_3  
+ U_\text{\large $\star$} ( p\, ;\lam) \, n_1 n_4 \  \Big] \ ,
\label{star int}
\eea
where the two terms are the real and virtual parts respectively.
The first weight includes the same manner of $\lam$-dependence as that
previously defined in \eq{W_V}, which allows the previous argument
to be partially recycled here.
To explicitly define $W_\text{\large $\star$}\,$, let
$$g(x) \equiv (k_0^2 + k^2 -2k_0p -2k \ell x)^2\, .$$
Accordingly, in the first summand of \eq{star int}, we have
\bea
& & \hspace{-.4cm} W_\text{\large$\star$}(p,q; \lam) \,=\, 
\frac{q}{\ell} \, \Big\{ \label{C2 - 2} 
 K^2 \, W_{\rm V} 
-  \frac{\sgn(p)}{4k\, \ell} 
\big( \, \rt{g(x_2^{\rm max})} - \rt{g(x_2^{\rm m\i n})} \, \big)
\  \Big\} \, . 
\eea
The arguments of $g$ are defined in Appendix~\ref{app: C}; 
see Eq.~\eq{C1 - 5}.
Virtual corrections in \eq{star int} 
require the function
\bea
U_\text{\large $\star$} (p\,;\lam) &=&
\frac1{\ell}\, K^2 U_{\rm V}^{(1)} +
\frac{\bar o_p - \Theta(k_-)}{4k} 
\big(\, K^2 - 2 \ell\, k_0 \,\big) 
\nonu\\
&\times&
\Big[\, 1 - \frac2{\ell^2} \int_{-\infty}^{+\infty} \!\! dq \, q\ 
\frac{n_2 n_3}{n_4} \, \Big]\, .
 \label{Ustar} 
\eea
We see that the auxiliary mass $\lam$ enters only in the previously 
defined functions, $W_{\rm V}$ and $U_{\rm V}^{(1)}$, 
so that the pattern of cancellation is unchanged and allows us to set $\lam=0\,$.
The divergent piece that we choose to subtract is
\be
\Im\ 
{\cal I}^\text{\large $\star$}_{11110} \Big|_{\rm div.} &=&
- \frac{\psi^{(0)}_{s_1,s_4}}{8 (4\pi)^3}
\Big( \, \frac1{\epsilon} 
+ 2 \log \frac{\bar \mu^2}{K^2} + \frac92 \, \Big) \, . \nonu
\ee
We have checked numerically that 
$\Im\ {\cal I}_{11110}^\text{\large $\star$} = \frac12 \big(
K^2 \, \Im\ {\cal I}_{11110}^{(0)} + \Im\ {\cal I}_{11100}^{(0)}\big)$
if $s_2=s_3\,$. 
As far as the nature of the function across the light cone is concerned 
in this case,
it may be discontinuous if $\Im\ {\cal I}_{11110}^{(0)}$ is singular; see
the earlier discussion about that master integral.
The master $\Im\ {\cal I}_{11100}^{(0)}$ is continuous.

\begin{figure}[t]
  \includegraphics[scale=\figscale]{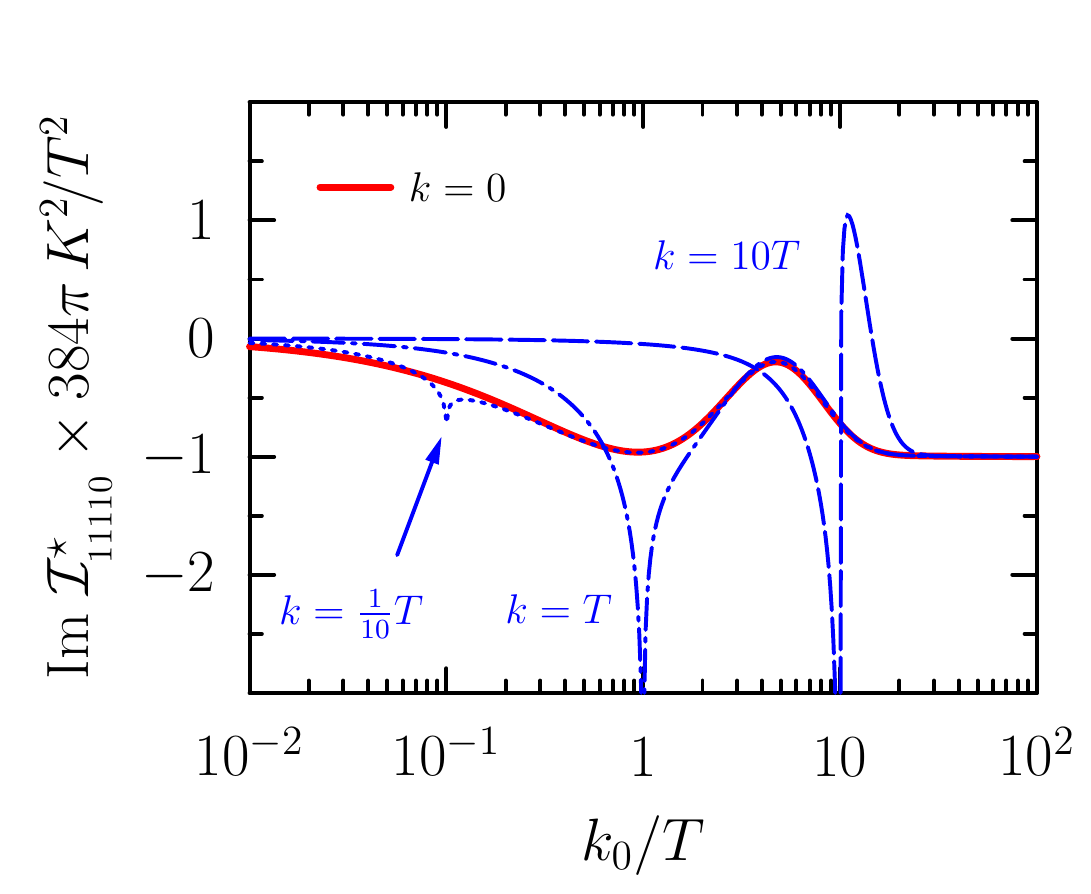}
  \caption{
    The special master diagram defined in \eq{star}, for $s_0=+1$
    and $s_1=s_2=-1\,$.
    With these statistical factors, the function
    cannot be reduced to any masters already discussed. 
    It also bears a logarithmic
    divergence on the light cone.
  }
  \label{fig: I star}
\end{figure}

That brings us to the case $s_2 \neq s_3\,$, 
shown for $s_0=+1$ and $s_{1,2}=-1$ in Fig.~\ref{fig: I star},
for which this master integral is a new object.
Indeed the outstanding $q$-integral in \eq{Ustar} 
can be found by recalling $s_4 = s_2 s_3$ and mapping the arguments
of the distribution functions to positive values, {\em viz.}
\bean
\int_{-\infty}^{+\infty} \!\! dq \, q\ 
\frac{n_{s_2}n_{s_3}}{n_{s_4}} 
&=&
 \int_0^{\ell} \!\! dq \, q 
 \,+\, 2 \int_0^{\infty} \!\! d q \, q \, 
\big(\, s_2 n_{s_2}(q) - s_3 n_{s_3}(q)  \, \big)  \\
&=& \frac{\ell^2}2 + T^2(s_2-s_3) \frac{\pi^2}4 
\, .\nonu
\eean
This makes the square brackets in \eq{Ustar} a 
difference between two thermal moments; 
see \eq{p integrals}, for the particular statistics $s_2$ and $s_3\,$.
If they are the same, it gives zero.
A noteworthy feature is the behaviour for $K^2 \to 0\,$,
where the functions $W_\text{\large $\star$}$
 and $U_\text{\large $\star$}$ simplify.
There is a discontinuity due to the latter, cf. Ref.~\cite{JL},
defined by the function's limit $k_0 \to k + 0^+$ minus  $k_0 \to k - 0^+\,$.
For a general statistical configuration, it is given by
\bea
\Im\ 
{\cal I}^\text{\large $\star$}_{11110} \,
\Big|_{\rm disc.}
&=&
T^2 (s_3-s_2) \, \frac{n_0^{-1}}{256\, \pi } \,
\int_{-\infty}^{+\infty} \!\! dp  
\  \frac{n_1 n_4}{k_0-p} \, , 
\label{star disc}
\eea
where the integration is meant in the principal valued sense.
%

\section{Diagram VI (cat's eye) \label{sec: 4}}

The most complicated form of \eq{I def} that we consider here 
is $a=b=c=d=e=1\,$,
which requires a careful cancellation of real and virtual diagrams.
We can deploy the same strategy as in Sec.~\ref{sec: 3}, albeit
now with {\em two} real and {\em two} virtual amplitudes. (Each is related by the 
symmetry $s_1 \leftrightarrow s_4$ and $s_2 \leftrightarrow s_5\,$.)
An auxiliary mass $\lam$ is again attached to $E_3 = \rt{r^2+\lam^2}$ so that collinear
singularities can be regulated.

In Ref.~\cite{laine1}, the function $\Im\ {\cal I}_{11111}^{(0)}$ 
(it was labelled with `j' there) was computed above the light cone.
For that case, with $m=n=0\,$, we do not need to worry about 
ultraviolet divergences. The vacuum result, given by
\bea
\lim_{T\to 0}\, {\cal I}^{(0)}_{11111} &=&
\frac{6\zeta(3)}{(4\pi)^4 \, K^2}  + {\cal O}(\epsilon) \, ,
\label{11111, vacuum}
\eea
is finite and has no imaginary part.
The leading contribution to $\Im\ {\cal I}^{(0)}_{11111}$ is therefore
thermal.
Those masters with $m,n \neq 0$ are more complicated, and can
have non-zero vacuum parts with ultraviolet divergences.

For $m,n \leq 1\,$, one can directly use \eq{diff trick}
to express 
\bea
 \Im\ 
{\cal I}^{(m,n)}_{11111} =
\frac{n_0^{-1}}{(4\pi)^3} 
&\bigg[&   
\nonu
   \int \! dp dq \, p^m \, q^n \,
  \Big( \, 
  W_{\rm V\!I}(p,q\,;\lam) n_1 n_2 +
  W_{\rm V\!I}(\ell,v\,;\lam) n_4 n_5 \, \Big) \, n_3 
  \nonu \\
& & \hspace{-.4cm} +
\int \! dp \, p^m U_{\rm V\!I}^{(n)}(p\,;\lam)\, n_1 n_4 +
\int \! dq \, q^n U_{\rm V\!I}^{(m)}(q\,;\lam)\, n_2 n_5
\ \bigg] \, .
\label{11111}
\eea
The real (virtual) contributions are in the first (second) line above,
inheriting the notation from earlier sections.
As before, $p$ and $q$ may
take on negative values.
Here the arguments of the distribution functions were omitted, they are
\be
n_0 &=\ s_0 n_{s_0}(k_0)  &  , \label{ni} \\
n_1 &=\ s_1 n_{s_1}(p)    &  , \nonu \\
n_2 &=\ s_2 n_{s_2}(q)    &  , \nonu \\
n_3 &=\ s_3 n_{s_3}(r)    &  ; \quad r=k_0-p-q  \, ,\nonu \\
n_4 &=\ s_4 n_{s_4}(\ell) &  ; \quad \ell=k_0-p \, , \nonu \\
n_5 &=\ s_5 n_{s_5}(v)    &  ; \quad v=k_0-q \, . \nonu
\ee

The weight function $W_{\rm V\!I}$ that is needed in \eq{11111}, 
carrying over some notation from
\eq{C2 - 2}, reads
for $k_0>k$ and $\lam \to 0\,$,
\bea
W_{\rm V\!I} (p,q;\,\lam) \ =\ \frac1{4 k K^2 r} \!\!\! &\Big\{&
\label{C3 - 2b}
 \log W_{\rm V\!I}^\prime + \bar o_p \log W_{\rm V\!I}^{\prime\prime} 
+ \bar o_q \log W_{\rm V\!I}^{\prime\prime\prime} 
\\
&+& o_p \log \frac{K^2 r^2 (p-q)(p-k_-)}{\lam^2 p \ell (q-k_-)(q-k_+)}
+ o_q \log \frac{K^2 r^2 (p-q)(q-k_-)}{\lam^2 q v (p-k_-)(p-k_+)}
\,\Big\} \, .\nonu
\eea
Below the light cone ($k_0<k$), the function should instead be
\bea
& & \hspace{-.4cm} W_{\rm V\!I}(p,q;\,\lam) \,=\,  \frac1{4k K^2 r} \Big\{ 
 \log W_{\rm V\!I}^{\prime\prime\prime\prime} 
 - \bar o_p \log \frac{K^2 r^2 (p-\hat p)}{\lam^2 p \ell (v-\hat p)}
   - \bar o_q \log \frac{K^2 r^2 (q-\hat q)}{\lam^2 q v (\ell -\hat q)}
   \  \Big\} \, .\nonu
\eea
In Eq.~\eq{C3 - 2b} the following ratios were defined:
\bean
& \displaystyle W^\prime_{\rm V\!I}
= \frac{(\ell - \hat r)(v - \hat r)}{(q-\hat p)(p-\hat q)} \, ,
\qquad
W^{\prime\prime}_{\rm V\!I}
= \frac{p-k_-}{p-\hat p} \, ,\\
&\displaystyle W^{\prime\prime\prime}_{\rm V\!I}
= \frac{q-k_-}{q-\hat q} \, ,
\qquad W^{\prime\prime\prime\prime}_{\rm V\!I}
= \frac{(\ell-\hat r)(v-\hat r)}{pq} \, .
\eean
We note the appearance of a log-divergence, just where expected
and signaled by the coefficients $o_p$ and $o_q$\,.
The formulas for $W_{\rm V\!I}$ are symmetric in arguments $p$ and $q\,$,
as is (then) the other real correction, which comes from $p\to k_0-p$ and $q\to k_0-q\,$.

The case $m=2$ and $n=0$ can also be written in the form of \eq{11111}.
One way of seeing this, is to  follow \eq{p0^2} and rewrite
$$
{\cal I}_{11111}^{(2)} \ = \ 
{\cal I}_{01111}^{(0)} + \sumint{\,P,Q} E_1^2 \D_1 \D_2 \D_3 \D_4 \D_5\, ,
$$
where the explicit integrands for the two terms can be found in 
Appendix~\ref{app: C}.
(After $s_1$ is interchanged with $s_5\,$ in the first term above.)
Equation \eq{11111} can be recovered after some manipulations of 
integration variables.

The virtual corrections are triangle diagrams, one of which is given
 by the second line of \eq{C3 - 1} in the Appendices.
Their calculation is similar to those studied in the previous section and 
is given explicitly in Appendix~\ref{app: C}.
Taking up the first term in the second line of Eq.~\eq{11111} 
(the other term follows
analogously), for $\lam \to 0$ it can be expressed by
\bea
U_{\rm VI}^{(n)} (p; \lam) \ = \ 
\frac{\bar o_p - \Theta(k_-)}{4k K^2} 
\int \!\! \frac{dq\, q^n}{r} \!\! &\Big[&
 \big( n_2 - n_5 \big) \log \frac{(q-k_-)(q-k_+)}{qv} 
 \\
&+&
\frac{n_2n_3}{n_4} \log \frac{\lam^2 \ell v}{K^2 r^2}
\ -\ \frac{n_3n_5}{n_1} \log \frac{\lam^2 pq}{K^2 r^2}
\  \Big] \ . \nonu 
\eea
Together, the last two terms in \eq{11111} are seen to combine with
the real corrections [see \eq{C3 - 2b}] so that the complete expression is $\lam$-independent.

\begin{figure}[t]
  \includegraphics[scale=\figscale]{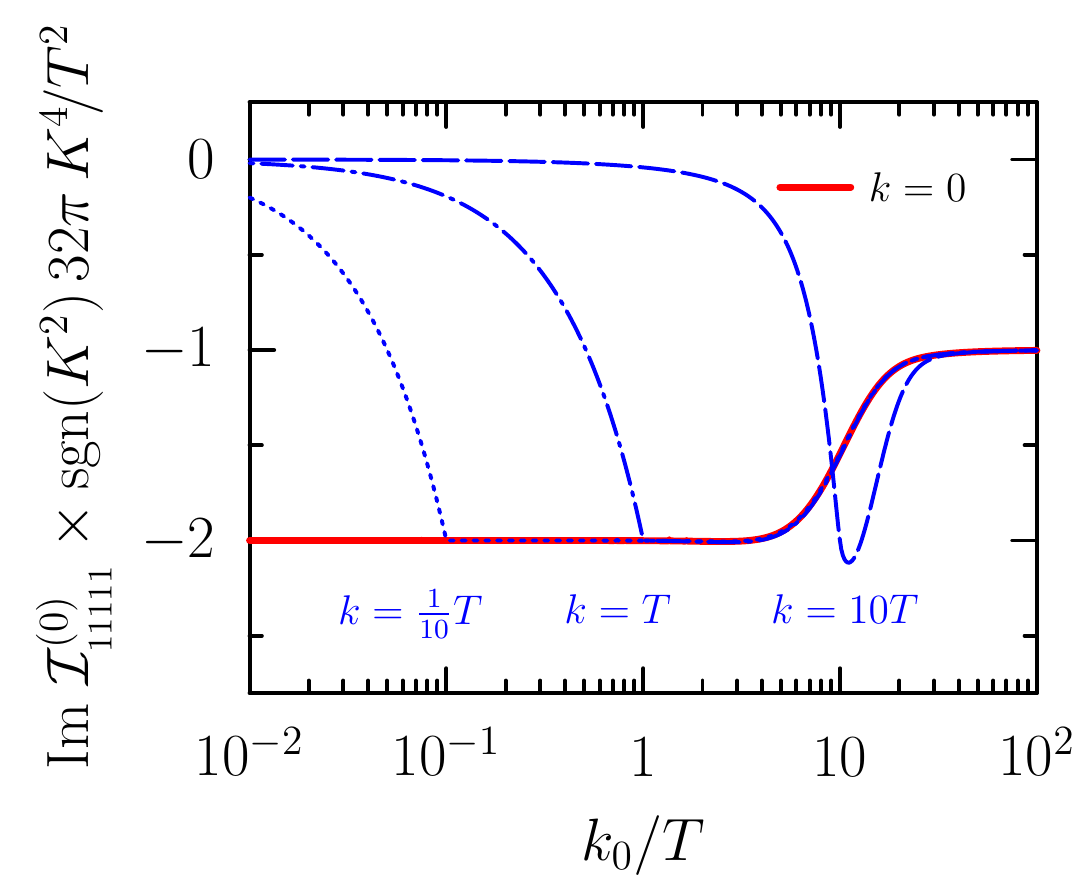}
  \caption{
    The imaginary part of the type VI master (with $s_i=+1$), in units of the
    asymptotic result.
    Shown here is the dependence on $k_0$ for various 
    values of the external momentum.
  }
  \label{fig: diagram 6}
\end{figure}

We show the case $m=n=0$ in Fig.~\ref{fig: diagram 6}, for all bosonic statistics.
(This figure confirms Ref.~\cite{laine1} above the light cone.)
The curves appear continuous across the light cone because we multiplied the whole
function by $\sgn(K^2) K^4 = |K^2| K^2\,$.
That supports the symmetrical nature of the discontinuity at $k_0 = k\,$.
No subtraction is necessary for large-$k_0\,$, since according to \eq{11111, vacuum} 
the whole master has no imaginary part in vacuum. 

If we consider $m=1$ and $n=0\,$, see \eq{11111_10}, no vacuum subtraction is necessary.
The symmetric case $m=0$ and $n=1$ is obtained by an appropriate exchange of statistics:
$s_1 \leftrightarrow s_2$ and $s_4 \leftrightarrow s_5\,$.
Moreover, 
the case $s_0=+1$ implies $s_1 = s_4$ and $s_2 = s_5$ which allows a change of integration
variables to show 
$\Im\,{\cal I}_{11111}^{(1)} = \frac12 k_0 \, \Im\,{\cal I}_{11111}^{(0)}$\,.
We have checked this numerically.

\begin{figure}[t]
  \includegraphics[scale=\figscale]{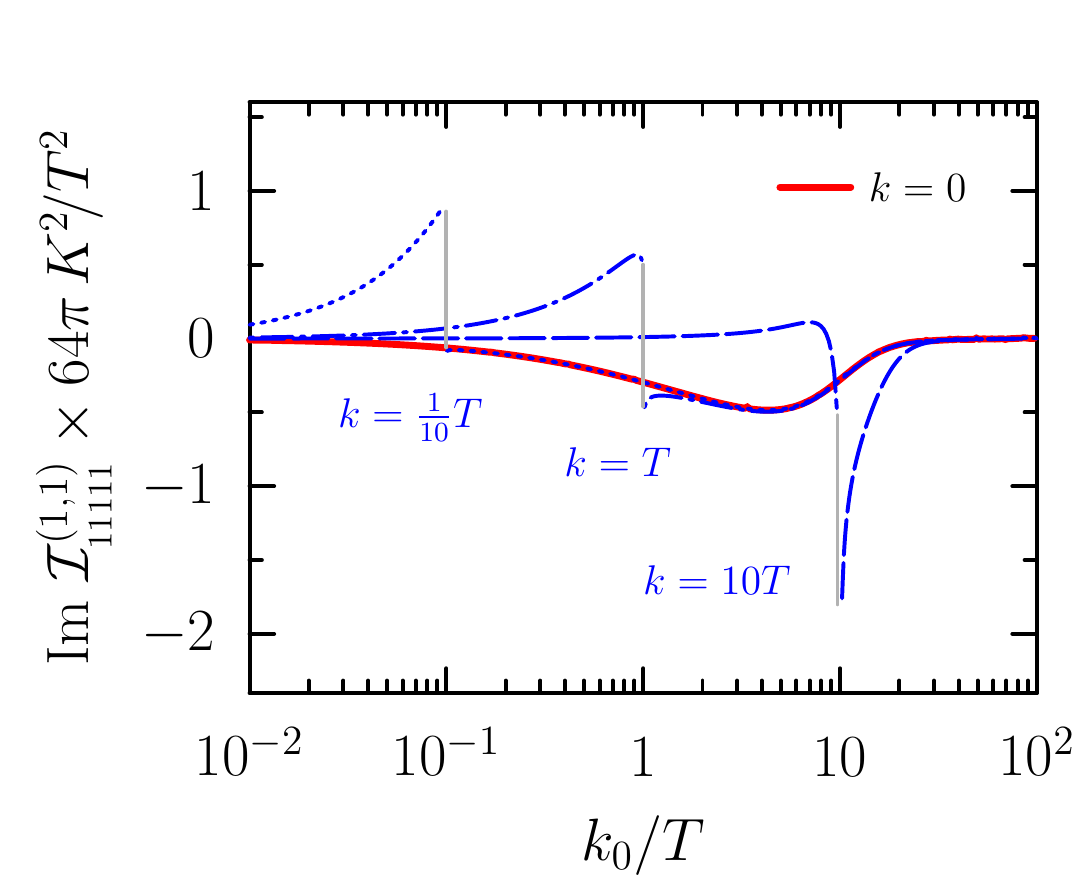}
  \caption{
    Like Fig.~\ref{fig: diagram 6}, but with $m=n=1$ and
    with all $s_i = +1\,$. 
    The  discontinuity is indicated by a vertical line.
    Large $k_0$ behaviour was subtracted according to \eq{11111_11 sub}.
  }
  \label{fig: diagram 6 11}
\end{figure}

For $m=n=1\,$, there is a leading vacuum term in \eq{11111_11}.
At finite temperature it originates from the two symmetrical virtual expressions.
They do not diverge, but we subtract them nonetheless:
\be
\Im\ 
{\cal I}^{(1,1)}_{11111} &-&
\frac{
  \psi^{(0)}_{s_1,s_4} + \psi^{(0)}_{s_2,s_5}
}{32 (4\pi)^3} \ .
\label{11111_11 sub}
\ee
Based on a numerical study,
I conjecture that the discontinuity across the light cone takes the simple form
$$
K^2\ \Im\ 
{\cal I}^{(1,1)}_{11111} \,
\Big|_{\rm disc.} = \frac{T^2}{512 \pi} (s_1+s_2+2)(s_3+1) \, .
$$
There thus seems to only be a discontinuity in the statistical configurations:
$(+,+,+)\,$, $(-,+,-)$ and $(-,-,+)\,$,
the first of which case is shown in Fig.~\ref{fig: diagram 6 11}, 
plotted after Eq.~\eq{11111_11 sub}'s subtraction was made.

\begin{figure}[t]
  \includegraphics[scale=\figscale]{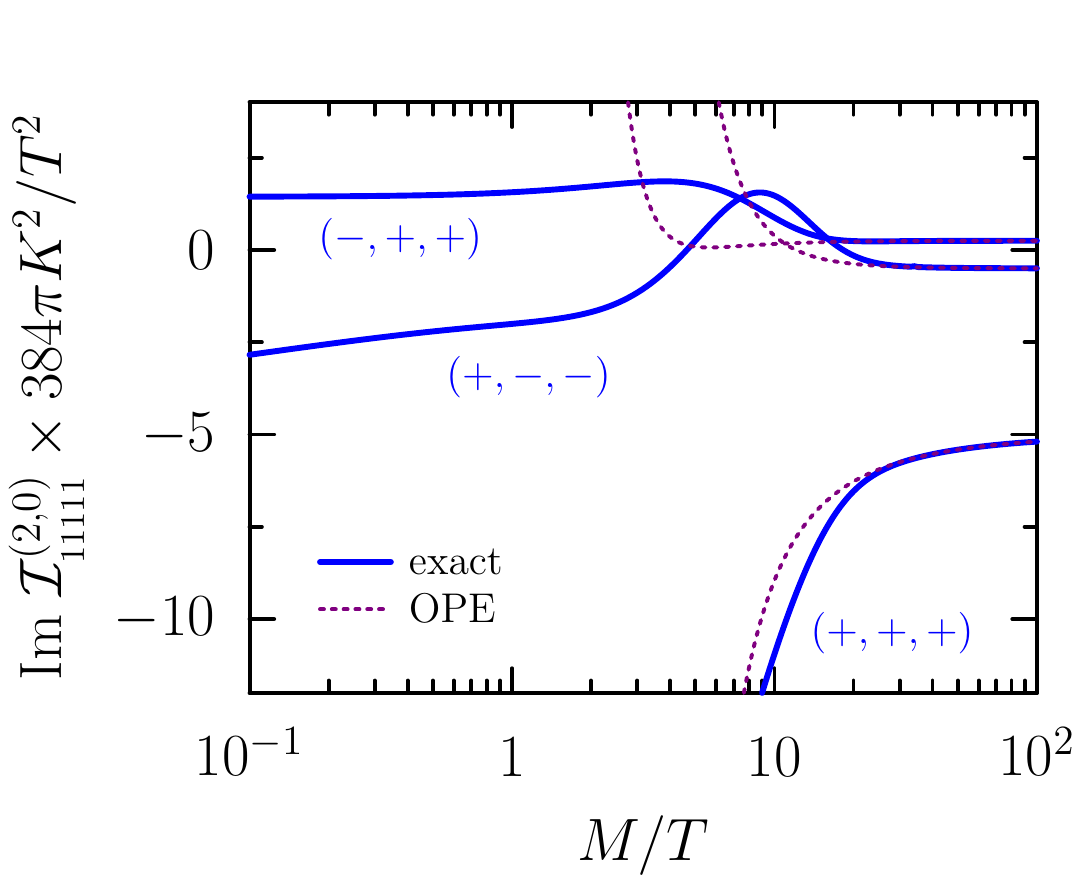}
  \caption{
    The dependence of the type VI master,
    with $m=2$ and $n=0\,$, on the invariant mass.
    An ultraviolet divergent part was subtracted, see Eq.~\eq{VI 20 div}.
  }
  \label{fig: diagram 6 20}
\end{figure}

One needs to be careful for $m=2$ because 
a genuine ultraviolet divergence must be subtracted.
That divergence is temperature dependent and originates from the virtual contribution
$U_{\rm VI}^{(2)}(q)\,$.
It is equal to 
\be
\Im\ 
{\cal I}^{(2)}_{11111} \Big|_{\rm div.} &=&
-\frac{\psi^{(0)}_{s_2,s_5}}{16 (4\pi)^3}
\Big( \, \frac1{\epsilon} 
+ 2 \log \frac{\bar \mu^2}{K^2} + \frac{11}2 \, \Big) \, ,
\label{VI 20 div}
\ee
which we subtract for plotting purposes.
(For the related master integral with $m=0$ and $n=2\,$, 
one replaces $s_2 \to s_1$ and $s_5 \to s_4$ in the above.)
Figure~\ref{fig: diagram 6 20} depicts  
$\Im\ {\cal I}^{(2)}_{11111}$ as a function of energy.
Although not shown, 
this quantity also seems to be discontinuous across the light cone.
The figure uses an average three momentum-squared, 
defined in Eq.~\eq{k2ave}, and the energy $k_0 = \rt{M^2 + k_{\rm ave}^2}\,$.

\section{Results and conclusions \label{end}}

We have computed a list of spectral functions that originate from 
self-energy diagrams with two loops,
generalising the results of Refs.~\cite{laine1,laine2}
 to below the light cone and considering a larger set of master integrals 
with $m,n >0\,$.
In so doing, we have separated the ultraviolet divergence where appropriate 
and shown how to determine the finite remainder numerically.
Validity of the KLN theorem was explicitly demonstrated in 
Secs.~\ref{sec: 3} and \ref{sec: 4}, by careful analysis of the collinear phase space.
We considered any
arrangement of propagating bosons or fermions allowed by the diagram's topology.
The code used for our numerical evaluation is publicly available
at Ref.~\cite{code}.

\bigskip

Returning at last to the QCD corrections for the photon spectral function, 
which was our original motivation, 
the imaginary parts of Eqs.~\eq{masters, mn} and \eq{masters, 00} can now be evaluated.
Firstly, all the temperature dependent divergences that were individually
isolated end up cancelling and the vacuum NLO result is recovered.
Since the thermal parts carry no ultraviolet difficulties, zero-temperature
counterterms suffice for renormalisation.
Some integrals give zero because they have no 
$k_0$-dependence prior to taking the imaginary
part, specifically
$$\Im\,{\cal I}_{01020}^{(0)} = \Im\,{\cal I}_{00120}^{(0)} = 
\Im\,{\cal I}_{01110}^{(0)} = 0 \ .$$
Other terms in \eq{masters, mn} and \eq{masters, 00}
also do not contribute, in particular those 
proportional to $\epsilon$ without a compensating divergence
in the master integral.
It is important to keep some of these terms so
that the vacuum result is recovered, but
$\Im\,{\cal I}_{11100}^{(0)}$ and $\Im\,{\cal I}_{11111}^{(1,1)}$
end up not contributing at all.
And \eq{masters, 00} can be simplified thanks to exact relations like
\bea
 K^2\, \Im\,{\cal I}_{11020}^{(0)} 
- 4k_0\, \Im\,{\cal I}_{11020}^{(1)} 
+ 4\, \Im\,{\cal I}_{11020}^{(2)} 
 &=& 0 \, . 
\eea

\begin{figure}[h]
  \includegraphics[scale=\figscale]{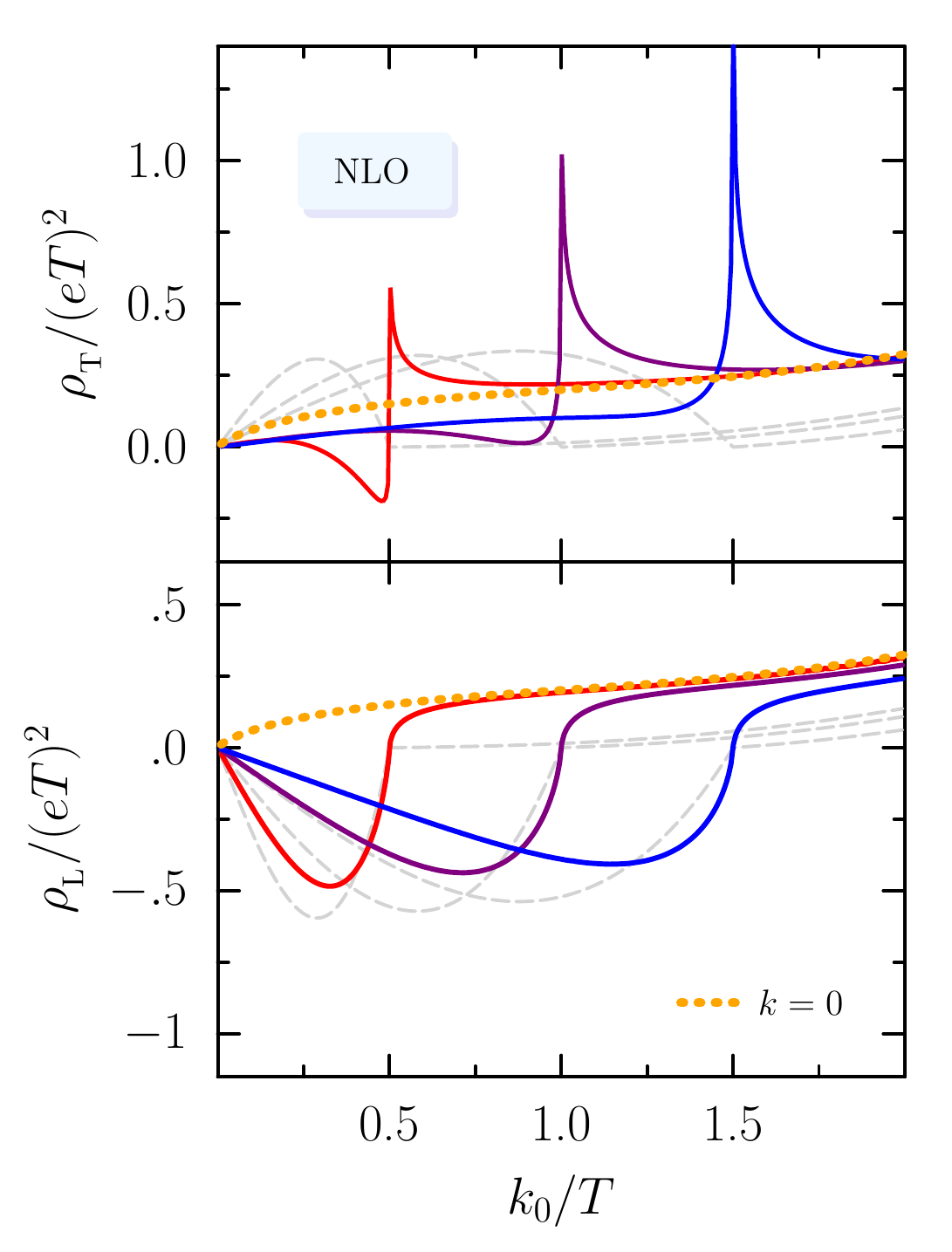}
  \caption{
    Transverse (upper panel) and 
    longitudinal (lower panel) spectral functions for 
    $k/T=0.5,1,1.5$\,.
    Since $\rho_{_{\rm T}} = \rho_{_{\rm L}}$ for a photon at rest,
    the curves with $k=0$ are identical.
    As defined in Eq.~\eq{expansion def}, $e^2$ was factored out and
    we used a fixed strong coupling $g=3$ for illustration.
    The thin dashed lines are the LO result (\ie $g=0$).
  }
  \label{fig: rhoV}
\end{figure}

At finite temperature the spectral function for the current-current 
correlator 
is specified by two scalar functions,
$\rho_{_{\rm T,L}} = \Im [ \Pi_{_{\rm T,L}} ]$ according to \eq{TL}.
Figure~\ref{fig: rhoV} shows the energy dependence of
the NLO spectral functions for several momenta.
The behaviour in \eq{log div} prevails near the light cone 
for the transverse polarisation, while
the factor of $K^2$ is enough to ensure that $\rho_{_{\rm L}}$ is 
zero there.
Both functions approaches zero for $k_0 \to 0\,$, as they should.
The large $K^2$ behaviour (of the NLO parts) 
can be found in Eq.~\eq{large K photon} of the Appendix.

In heavy-ion collisions \cite{McLerran1985,Weldon1990iw},
the {\em observable} photon and dilepton rates are proportional to 
$-\Im\, \Pi^\mu_{\ \mu} = 2\rho_{_{\rm T}} + \rho_{_{\rm L}}$ 
for $K^2 =0$ and $K^2 \geq 4m^2_{l}$ respectively ($m_l$ is the lepton's mass),
where perturbative studies have hitherto focused\footnote{%
   Including the special case $m_l \to 0\,$.}.
The complementary region, $K^2<0\,$, is not merely academic:
It provides an opportunity to cross-check the weak coupling framework, 
\eg with non-perturbative 
Euclidean correlators (of either polarisation) 
provided by lattice QCD \cite{Brandt2017}.

\bigskip
As noted in the Introduction, our results are relevant for 
deploying (fixed-order) perturbation theory in contexts 
where simplifying kinematic assumptions are not justified.
It is nevertheless worthwhile, \eg if analytic expressions are available, to
check that these numerical results reproduce the correct behaviour in those 
limits.
We have done this using the OPEs in Appendix~\ref{app: D} for $K^2 \gg T^2$
in each integral studied.
Our results are also compatible with recent HTL self-energies at NLO,
in the limit $k_0,\, k \ll T$ \cite{Carignano2019}.

It is worth recalling that the individual masters are {\em not} (usually) themselves
physical, rather they supply a convenient `basis' from which observables can be built.
That ubiquity makes a dedicated study constructive because of the
valuable resource it provides for future NLO developments at finite-$T\,$.
Although by no means all-inclusive, the list of loop integrals compiled
here should satisfy a wide variety of needs or take little generalising to do so.

\begin{acknowledgments}

I thank Mikko Laine for helpful discussions and indispensable guidance,
as well as Andr\'{e} Peshier for useful comments on the manuscript.
This work was partly supported by the 
Swiss National Science Foundation (SNF) under Grant No. 200020B-188712.

\end{acknowledgments}

\appendix
\renewcommand{\thesubsection}{\Alph{section}.\arabic{subsection}}
\renewcommand{\theequation}{\Alph{section}.\arabic{equation}}
\makeatletter 
\def\p@section{}
\def\p@subsection{}
\makeatother

\section{Thermal Phase Space \label{app: A}}

In this appendix, we discuss two important phase space integrals
which were used in the main text.
The following results are valid both above and below the lightcone:
$k_0 \lg |\bm k|\,$.
Without loss of generality, we assume $k_0$ is positive.
Energy conservation is imposed by cutting the diagram, handled with the replacement
$$\Im \,\frac1{k_0 +i0^+ - A} \,\to\, - \pi \delta(k_0-A)\, .$$

\subsection{Two-particle decay \label{A1}}

The production rate due to binary encounters, e.g. $q\bar q \to \g^{*}$
(the asterisk indicates a virtual photon),
is given by a phase-space integration over the momenta of  
interacting partons.
For a given external momentum $K=(k_0, \bm k)$, we simplify
the integration measure
\bea
\int_{[1,2]}  &\equiv& 
\sum_v \int_{\bm p, \bm q} \frac{ (2\pi)^{d+1} }{4E_1E_2} \d^{(d+1)} (K-v_1P-v_2Q) \, ,
\label{A1 - 1}
\eea
with the (on-shell) energies $E_1 = |\bm p|$, $E_2=|\bm q |$ .
The $v_i=\pm 1$ are  summed over, for $i=\{1,2\}$, so that 
in \eq{A1 - 1} each channel is represented.
Here we also used 
dimensional regularisation for
the terms ${\cal O}(\epsilon)$ needed later.
Integrating over $d^d\bm q$ is trivial; momentum conservation
fixes $\bm q = v_2 ( \bm k - v_1 \bm p)$ .
The on-shell energy is therefore 
\bea
E_2 &=& |\bm k -v_1 \bm p| \ = \ 
\rt{ k^2 + p^2 - 2v_1 p \, k \, \cos \theta_{kp}} \, ,
\label{A1 - 2}
\eea
where $\theta_{kp}$ is the angle between $\bm k$ and $\bm p$ .

\begin{figure}[h]
  \includegraphics[scale=\figscale]{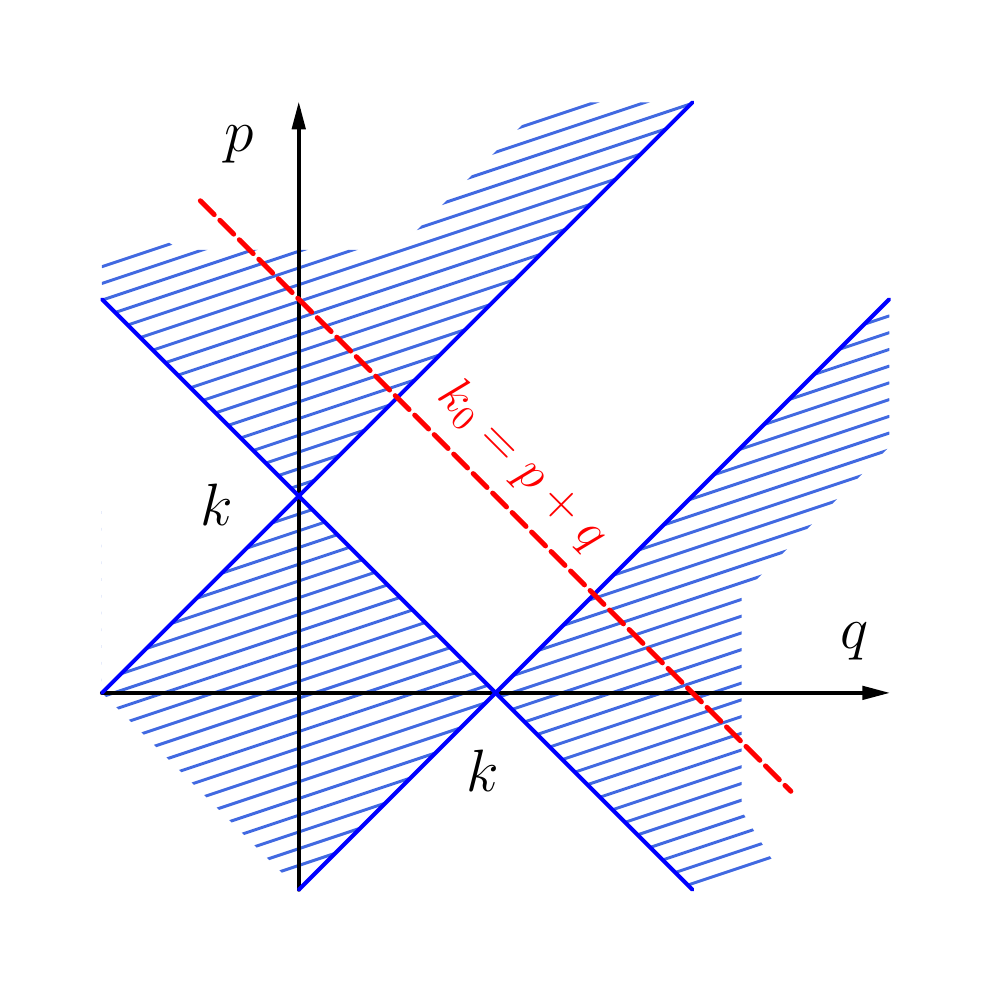}
  \caption{
    Allowable integration region in \eq{A1 - 1}, 
    where $p=v_1E_1$ and $q = v_2 E_2$.
    The dashed line is the energy constraint $k_0=p+q$, 
    while the blue shading is
    forbidden by the on-shell condition for $P$ and $Q=K-P$. (Here $K^2>0$.)
  }
  \label{fig: 1-loop}
\end{figure}

The external vector $\bm k$ distinguishes an orientation with which
to organise the remaining $d^d \bm p$-integration.
We choose to align the $p_z$ axis with $\bm k$, so that the
azimuthal integration is also trivial.
And instead of a polar integration over $\theta_{kp}$, we integrate
over the magnitude $q=E_2$ given by \eq{A1 - 2}.
The angular limits accordingly translate into a kinematic restriction; 
$|k-v_1p| < q < |k+v_1p|$ (see Fig.~\ref{fig: 1-loop}).
Let us therefore express the angular integration by
\bean
 \bar \mu^{2\epsilon} \int\displaylimits_{0}^{\infty}
\!\! dp \, p^{2-2\epsilon} \int\displaylimits_{0}^{\pi} \!\! d \theta_{kp} 
\ \sin^{1-2\epsilon} \theta_{kp} 
&=&
\int\displaylimits_{0}^{\infty} \!\! d p 
\int\displaylimits_{|k-v_1p|}^{|k+v_1p|} \!\! d q \,  \frac{pq}{k} 
\Big( \frac{4k^2\, \bar \mu^2}{4p^2q^2 - (k^2-p^2-q^2)^2} \Big)^\epsilon \, .
\eean
All the necessary scalar products can be formed in terms of these variables.

The combination $v_1 p$ and $v_2 q$ appears repeatedly above\footnote{%
The very same combinations will appear in the relevant integrands.
},
suggesting that we
formally extend the magnitudes $p$ and $q$ to negative values in lieu
of summing over $\{ v_i \}\,$.
Thus the subsequent $(p,q)$-integration covers all channels, see Fig.~\ref{fig: 1-loop},
and reads
\bea
\sum_v \int \!\!dp \, dq &=&
\int_{-\infty}^{+\infty}  \! dp \, \Big[\,
\int_{-|k+p|}^{-|k-p|}    \! dq +
\int_{|k-p|}^{|k+p|}      \! dq
\,\Big] \, . 
\label{A1 - 2b}
\eea
When the internal lines are on-shell, energy conservation implies that 
{$k_0 = p+q$ allowing the $q$ integration in \eq{A1 - 2b} to be carried out.
Therefore the final result is
supported only along dashed line shown in Fig.~\ref{fig: 1-loop},
and reads
\bea
\int_{[1,2]} &=&
\frac{1}{8\pi \, k}
\int^\prime \!\!\! dp \, 
\Big(\, 
\label{A1 - 3} 
   1 + \epsilon \log \frac{\bar\mu^2 \, k^2}{K^2 (p-k_-)(p-k_+)}  
+ {\cal O}(\epsilon^2) \,\Big) \, .
\eea
The $p$-integral was flagged with a prime as it depends on whether
$K^2$ is positive or negative.
In the former case (with $k_0>0$), only the channel having $v_1=v_2=+1$
contributes in \eq{A1 - 2b}.
Precisely one of $v_1$ and $v_2$ equals $-1$ in the latter case.

Altogether, one has
\bea
\int^\prime \!\! dp 
&\equiv& 
  \Theta (k_-) \int_{k_-}^{k_+} \!\! d p  
\label{A1 - 3b} 
  - \Theta (-k_-) \Big[\
\int_{-\infty}^{k_-} \!\! d p  +
\int_{k_+}^{\infty} \!\! d p  \ \Big] \, . 
\eea
The discontinuity at $k_0 = k$ is thus given by an integral
over all real $p\,$, in the principal valued sense.
In the rest frame, 
$\bm k = 0\,$, \eq{A1 - 3} has the net effect of setting $p=k_0/2$\,.

\bigskip

As an application of the foregoing discussion, define
\bea
\psi_{s_1,s_2}^{(\nu)} &=&
\frac{n_{s_0}^{-1}}{k} \int^\prime \!\! dp \, n_{s_1} (p) n_{s_2} (k_0-p) 
\Big(  \frac{p}{k_0}  \Big)^\nu  \ , \nonu\\
  \label{psi}
\eea
and consider it for $\nu = \{0,1,2\}$ in particular.
This function, which is needed for the one-loop discontinuity,
was also used in the virtual parts of our two-loop calculations.

Letting the exponentials be abbreviated by 
\bean
{\cal E}_\pm \equiv \exp [ \,- \b | k_\pm |\, ] \, ,
\eean
we find that
\bea
\psi_{s_1,s_2}^{(0)} &=&
\Theta (k_-) \ +\ \frac{T}{k} \log 
\frac{(1-s_1 {\cal E}_+)(1-s_2 {\cal E}_+)}
     {(1-s_1 {\cal E}_-)(1-s_2 {\cal E}_- )}
\ , \label{psi0} \\
\psi_{s_1,s_2}^{(1)} &=&
\frac12 \Theta (k_-) \ +\ \frac{T}{k k_0} \,\bigg[\
               k_0   \log \frac{1-s_2 {\cal E}_+}{1-s_2 {\cal E}_-}
+ k_+ \log \frac{1-s_1 {\cal E}_+}{1-s_2 {\cal E}_+}
- k_- \log \frac{1-s_1 {\cal E}_-}{1-s_2 {\cal E}_-}   \label{psi1} \\
& & \ -\  T \Big( \, 
\Li_2(s_1 {\cal E}_+)
+ \sgn(k_-) \Li_2(s_1 {\cal E}_-)
-\Li_2(s_2 {\cal E}_+)
- \sgn(k_-) \Li_2(s_2 {\cal E}_-) \, \Big)
\ \bigg] \ ,\nonu \\
\psi_{s_1,s_2}^{(2)} &=&
\frac1{4 k_0^2} \Big(\, k_0^2 + \frac{k^2}{3}\, \Big) \Theta (k_-)
\ +\ \frac{T}{k k_0^2} \, \bigg[\
 k_+^2  \log \frac{1-s_1 {\cal E}_+}{1-s_2 {\cal E}_-}
-k_-^2  \log \frac{1-s_1 {\cal E}_-}{1-s_2 {\cal E}_+} \nonu \\
& & \ -\  2 T k_+ 
\Big(\, \Li_2 (s_1 {\cal E}_+) + \sgn (k_-) \Li_2 (s_2 {\cal E}_-) \, \Big)
+ 2 T k_- 
\Big(\, \Li_2 (s_2 {\cal E}_+) + \sgn (k_-) \Li_2 (s_1 {\cal E}_-) \, \Big) 
\nonu \\ & & \ - \ 2T^2 \Big(\,
  \Li_3 ( s_1 {\cal E}_+)
- \Li_3 ( s_1 {\cal E}_-)
+ \Li_3 ( s_2 {\cal E}_+)
- \Li_3 ( s_2 {\cal E}_-)
\,\Big) \ \bigg]  \ . \label{psi2}
\eea
These quantities are discontinuous on the light cone, explicitly with
\bean
\renewcommand{\tabcolsep}{4mm}
\begin{tabular}{c || c | c | c }
  $\nu$ & $0$ & $1$ & $2$ \\ \hline
  & & & \\[-4mm]
  $\displaystyle \psi^{(\nu)}_{s_1,s_2}\big|_{k_0=k+0^+}
                -\psi^{(\nu)}_{s_1,s_2}\big|_{k_0=k-0^+}$
  & $1$ 
  & $\displaystyle 
     \frac12 + 2 \frac{T^2}{k^2} \big( \Li_2(s_1) - \Li_2(s_2) \big)$
  & $\displaystyle \frac13 - 4 \frac{T^2}{k^2} \Li_2 (s_2)$ \\[2mm]
\end{tabular} \ .
\eean

\subsection{Three-particle decay \label{A2}}

It is sufficient to work in $d=3$ dimensions for
phase space integrations of the {\em real} two-loop corrections 
(those where three propagating particles are put on-shell).
Extending \eq{A1 - 1}, we define
\bea
\int_{[1,2,3]} &\equiv&
\sum_{v} \int_{\bm p, \bm q, \bm r} \frac{(2\pi)^4}{8E_1 E_2 E_3} 
\d^{(4)} (K - v_1 P -v_2 Q - v_3 R) \, ,
\label{dGamma}
\eea
with the (on-shell) energies $E_1 = |\bm p|$, $E_2=|\bm q|$ and $E_3 = \rt{\bm r^2 +\lam^2}$.
(The regulator $\lam$ is necessary to observe the cancelling divergence between real and
virtual corrections.)
The $v_i=\pm 1$ are summed over, for $i=\{1,2,3\}$ .
One of the integrals may be simplified using energy and momentum conservation;
we choose the $r$-integral, and complete it by writing
\bean
\int_{\bm r} \frac{1}{2E_3} &=& 
2\pi \int_{\bm r} \int_{-\infty}^{+\infty} \! \frac{d r_0}{2\pi} \d(R^2 - \lam^2) \Th(r_0) \, .
\eean
Hence, by fixing $\underline{R} = v_3 (K-v_1 P - v_2 Q)$, we are left with
\bea
\int_{[1,2,3]} &=&
2\pi \sum_{v} \int_{\bm p,\bm q} 
\frac{v_3}{4E_1E_2}\,\Th(\underline{r}_0) \d \big( \underline{R}^2 - \lam^2 \big) \, .
\label{R-int}
\eea

Let us now specify a coordinate system in order to proceed.
We choose, following Ref.~\cite{Peigne2007},
to align the $z$-axis along the direction of $\bm k$
and orient the $zy$-plane to contain $\bm p$, viz.
\bea
\bm k &=& k(0,0,1) \, , \nonu \\
\bm p &=& p(0,\sin \theta_1,\cos\theta_1) \, , \nonu \\
\bm q &=& q(\sin\theta_2 \sin\phi,\sin\theta_2 \cos\phi,\cos\theta_2) \, .
\eea
The integral over the azimuthal angle $\phi$ can then be performed in \eq{R-int}.
To do so, express the argument of the $\delta$-function by
$\underline{R}^2-\lam^2 = A + B \cos \phi$, where
\bea
A &=& K^2 - \lam^2 + 
      2\big[ v_1 p\, v_2 q(1-x_1x_2) 
      - v_1 p(k_0 - x_1 k) - v_2 q(k_0 - x_2 k) \big] 
      \, , \nonu \\
B &=& -2 \Big(\, v_1 p \rt{1-x_1^2} \, \Big) 
         \Big(\, v_2 q \rt{1-x_2^2} \, \Big) 
      \, . \label{A and B}
\eea
Here we have abbreviated $x_i=\cos \theta_i\,$.
Consider then the integral of a function $g(\phi)\,$,
$$ \int_0^{2\pi} \! d \phi \ \d \big(\underline{R}^2 -\lam^2\big)\, g(\phi)
= \frac{\Th(h)}{\rt{h}} \sum_{\pm} g(\phi_\pm) \, , $$
where $h=B^2-A^2$ and the angles $\phi_\pm = \pi \pm {\rm acos}(A/B)$.
For our purposes, the function $g$ depends on $\phi$ via $\cos \phi$ and
hence $\sum_\pm g(\phi_\pm) \to 2 g(\phi_\pm)$ .

We have accomplished five of the nine integrals in \eq{dGamma}, and are left with
\bea
  \int_{[1,2,3]} &=&
\frac1{4(2\pi)^4}  \sum_v 
\int \!\! dp \, dx_1 
\int \!\! dq \, dx_2  
\sgn(k_0-p-q)
\, pq \frac{\Th(h)}{\rt{h}} \, ,
\label{4 dim integral}
\eea
which cannot be simplified further in general.
The combination $v_1 p$ and $v_2 q$ in \eq{A and B} once again suggests that we
formally extend the magnitudes $p$ and $q$ to negative values.
The factor of $\Th(\underline{r}_0)$ in \eq{R-int} has been dropped because
$\underline{r}_0 = v_3 (k_0 -v_1 p -v_3 q)$ and so exactly one term in the sum over 
$v_3 =\pm1$ will contribute.
(The relevant value for $v_3$ is determined for a given $p$ and $q$ --
extended to take negative values.)
And then, as before, we can just forget the sum over $\{ v_i \}$.

The $h$-function of \eq{4 dim integral} is quadratic in each of its arguments
$p$, $x_1$, $q$, and $x_2$.
Requiring $h\geq 0$ (equivalent to taking the real part of $\rt{h}$) 
summarises the allowable phase space.


\section{One-loop Auxiliary Functions\label{app: B}}

\subsection{$F_m$ for $m=0,1,2$\label{B5}}

The formula for $F_0$ was already given in the main text: see \eq{F0}.
Here we derive that result after discussing some simple properties.
The expansion of the function $F_0$ about the light cone energy, $k_0=k\,$, is
\bea
F_0 (K; s_1, s_2) &=&
\frac{-1}{32\pi k (k_0-k)} \bigg[\,
T\frac{1+s_2}{(k_0-k)} 
\label{B5 - 3}  
+ \big(\,\tfrac12 + s_1 n_{s_1}(k)\, \big)
\, +\, {\cal O}(k_0-k) \, \bigg] \, . 
\eea
Whether the leading term has a double-pole depends upon $s_2 = +1\,$, implying the
propagator that appears twice is bosonic. 
If rather $s_2=-1\,$, then the pole is only simple.
Equation \eq{F0} includes the vacuum result $-1/(16\pi K^2)\,$, which we 
subtract in Fig.~\ref{fig: F0}.
Therefore the vertical intercept (at $k_0=0$) in this figure is merely $k^2/T^2\,$.

\begin{figure}[h]
  \includegraphics[scale=\figscale]{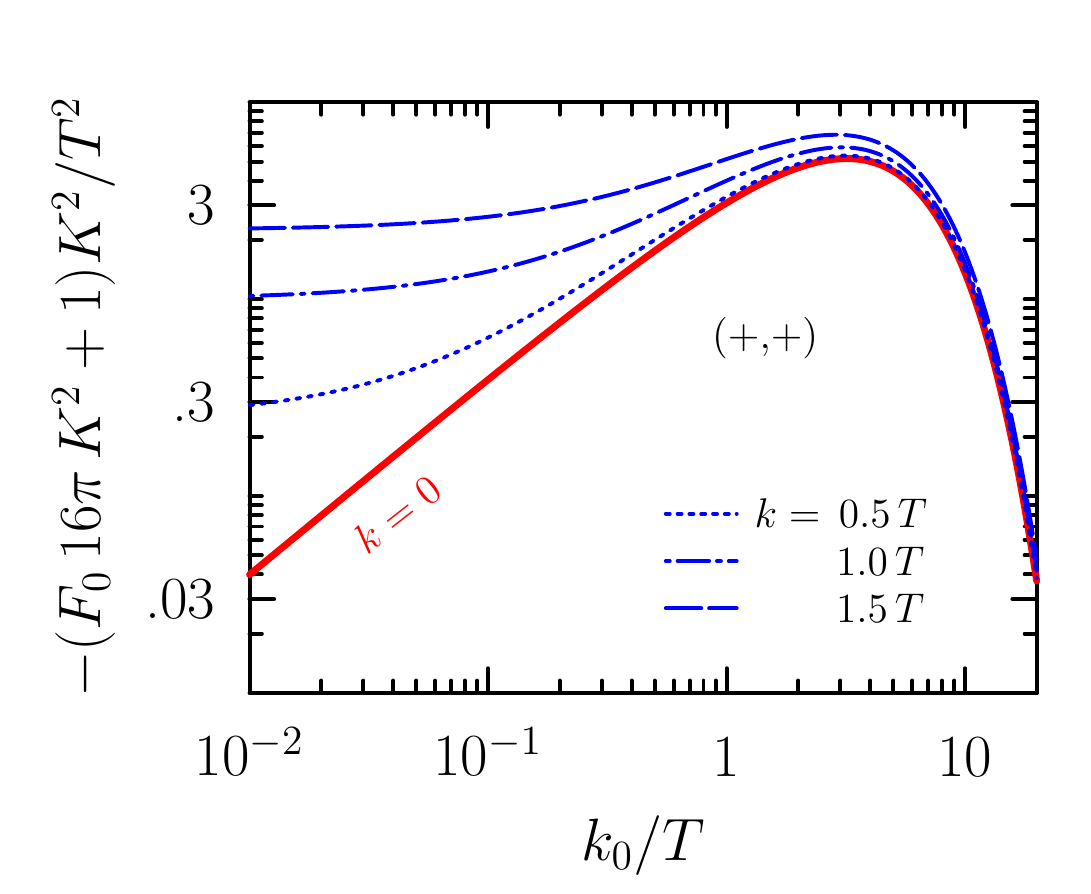}
  \caption{
    The thermal part (i.e.~we subtracted the $T\to 0$ limit) of the function $F_0(K;+,+)$, 
     for several values of the three momentum.
    The continuity in the function at $k_0=k$ for $k>0$
    is an aspect of the repeated pole in \eq{B5 - 3}.
  }
  \label{fig: F0}
\end{figure}

A mass regulator $\lam$ helps to evaluate $F_0\,$,
by writing the repeated propagator as
$$
\D^2_{s_2} (L) \ = \ 
 \pd_{\lam^2} \bm( \D_{s_2} (L_\lam) \bm)  \big\vert_{\lam\to 0} \, ,
$$
where $L_\lam = (\ell_0,\, \bm \ell_\lam)$ is a four momentum
with $|\bm \ell_\lam| = \rt{\ell^2 + \lam^2}\,$.
This calls for the one-loop results, provided in Appendix~\ref{app: A},
 to be endowed with a mass on one of the propagators.

The integration methods
are only slightly modified by the presence of a mass term above.
On-shell energies are then
$E_1 = | \bm p|$ and $E_2 = |\bm \ell_\lam|\,$,
and the integration limits are 
determined by energy conservation.
We find, by generalising definition \eq{A1 - 1} to massive particles,
\bea
\Im 
\,\sumint{\, L} \D_{s_1}(K-L) \D_{s_2} (L_\lam)  
&=&
\! - \int_{[1,2]} \big(
f_{s_1}^{v_1}f_{s_2}^{v_2}
-f_{s_1}^{-v_1}f_{s_2}^{-v_2} \big)
\label{B5 - 1} \\
&=&
\! - \frac{n_{s_0}^{-1}}{8\pi k} \int \!\! dp \, d\ell \frac{p\,\ell}{E_1E_2} 
\d (k_0 - e_1 - e_2 ) \, 
n_{s_1}(e_1) n_{s_2}(e_2) \, .
\nonu
\eea
The signed energies introduced are $e_1 = \sgn(p) E_1$ and $e_2 = \sgn(q)E_2\,$.
It is necessary to take the derivative of 
\eq{B5 - 1} with respect to $\lam^2$ (and evaluate it at $\lam=0$).
For this purpose the integration variables are changed to $e_1$ and $e_2$
so that $\lam$-dependence is swept into the limits of integration.
The $\d$-function then imposes the following limits on the $p$-integration:
\be
\l\{
\begin{array}{l}
  \displaystyle 
  p > k_- + \frac{\lam^2}{4k_-} \quad {\rm and}\quad p<k_++\frac{\lam^2}{4k_+} \\[.4cm]
  \displaystyle
  p < k_- + \frac{\lam^2}{4k_-} \quad \ {\rm or}\ \quad p>k_++\frac{\lam^2}{4k_+} \\
\end{array} \r. \quad {\rm for} \quad k \lg k_0 \, . \nonu
\ee
We only need to evaluate the integrand at the appropriate boundary values (with $\lam=0$),
to obtain $F\,$.
The result is stated in Eq.~\eq{F0}, in a way that is valid for either sign of $K^2$.

\bigskip
Using the approach stated in Sec.~\ref{strategy} to include powers
of $p_0$ into the above derivation, one obtains
expressions for $F_1$ and $F_2\,$.
Casting them altogether gives \eq{F0}.
The resulting functions are plotted in Fig.~\ref{fig: Fm},
omitting the large-$k_0$ 
behaviour of \eq{F0}, namely
[see also Eq.~\eq{ope: factorisable}]
\be
& F_0 = -1/(16\pi \, K^2) \, ,
\qquad
F_1 = -k_0/(16\pi \, K^2 ) \, , \nonu \\
& F_2 = -3k_0^2/(64\pi \, K^2 )  \ ;
\quad k_0 \to \infty \, . \nonu
\ee
We note that in this limit, the coefficient of the 
$T^2$ thermal contribution is zero. 
(Which is also the case for the statistical combinations not plotted.)

Some relations can be derived for these functions. 
For the special case that $s_1 = s_2\,$, one has $F_1 = k_0 \, F_0\,$.
In general, $F_2$ is actually expressible by $F_1$ and $F_0$ due
to the identity:
$$ K^2 \, F_0  - 4( k_0 \, F_1 - F_2 ) = 0 \, ,$$
which can be used in \eq{V and VI} for the first of the two master integrals when $m=2\,$.

\begin{figure}[h]
  \includegraphics[scale=\figscale]{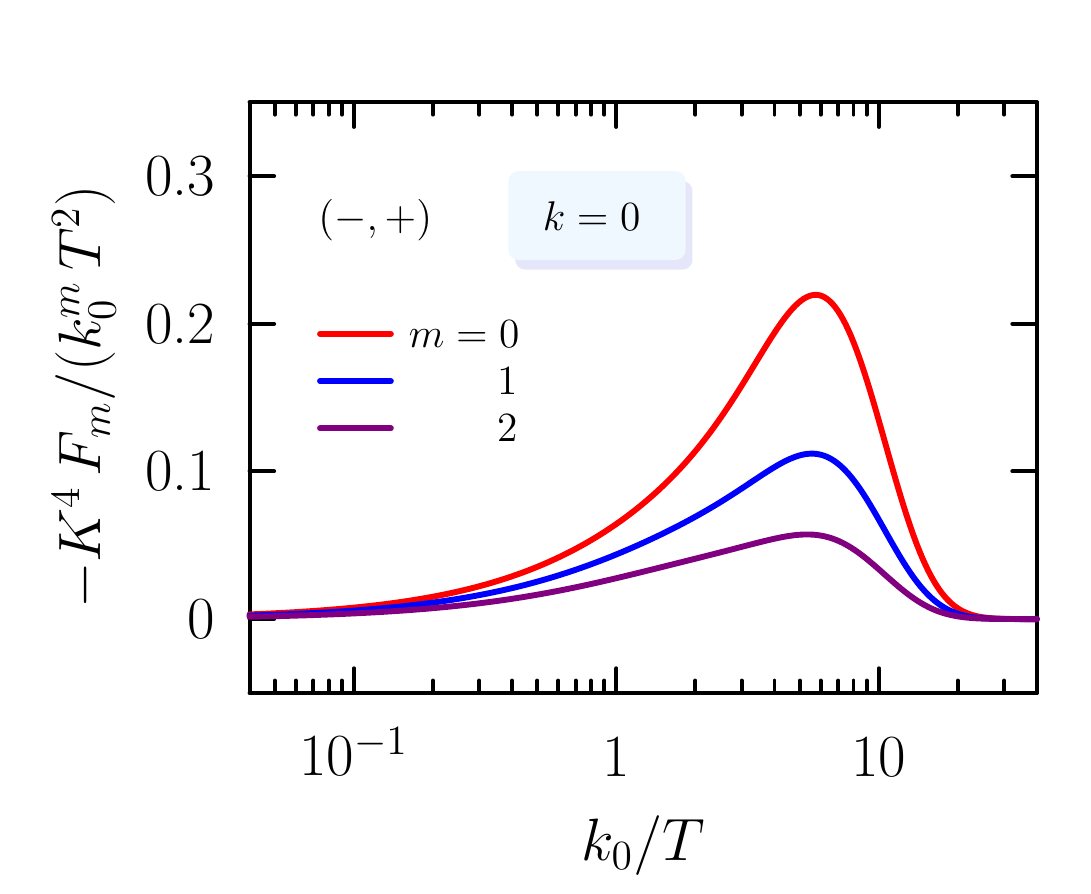}
  \caption{
    The energy dependence of the family of functions $F_m$
    for $m=\{0,1,2\}$ at $k=0\,$.
    We show here the functions for $s_1=-1$ and $s_2=+1\,$, with 
    their $k_0\to \infty$ limits subtracted out.
    The $m=0$ curve was also shown in Fig.~\ref{fig: F0}.
  }
  \label{fig: Fm}
\end{figure}

\subsection{$H_m$ and its ${\cal O}(\epsilon)$ contribution\label{Hn}}

In this section, we explain the functions $H_m^{[0,1]}$ as 
defined by Eq.~\eq{H order eps} of the main text.
To first establish $H_m$ from \eq{Hm int},
it can be written as 
\bea
H_m(K; s_1,s_2) &=&
- \int_{[1,2]} \big(
f_{s_1}^{v_1}f_{s_2}^{v_2}
-f_{s_1}^{-v_1}f_{s_2}^{-v_2} \big) (v_1 E_1)^m \, ,
\label{B5 - 5} 
\eea
for $m=0,1,2$ with the special notation of Appendix~\ref{A1}.
In the special case that $s_1=s_2\,$, one may show that
$H_1 = \tfrac12 k_0 H_0$ by a change of integration variables.
But in general, one must adhere to \eq{A1 - 3} for the integration.
According to that formulation we can set $v_1=v_2=+1$
enabling the distribution functions to be written
$$
 f_{s_1}^{+} f_{s_2}^{+} -
 f_{s_1}^{-} f_{s_2}^{-} 
 \ =\ 
 \bm( e^{k_0/T} - s_1s_2 \bm)\,  n_{s_1} n_{s_2} \ ,
$$
where the arguments were omitted.

The results are immediate:
[The primed integral is defined in \eq{A1 - 3b}.]
\bea
 H_m^{[0]} &=& k_0^m \, \psi^{(m)}_{s_1,s_2} \  , \\
 H_m^{[1]} 
&=&
\frac{ n_0^{-1} }{k} \int^\prime \!\! dp\ p^m \, n_1 n_2 
\log \frac{k^2}{(p-k_-)(p-k_+)} 
\ .\nonu
\label{B6 - 1} 
\eea

\subsection{Finite part of $G_m$ \label{Gm}}

We now consider $G_m^{[0]}$, defined in 
Eq.~\eq{H order eps} of the main text, for $m=\{0,1\}$.

The whole function $G_0$ is equal to the real part of
$$
- \sum_v \int_{\bm p} \frac1{4E_1E_2}
\frac{ v_1(\tfrac12 + s_2 n_{s_2} (E_2)) + v_2(\tfrac12 + s_1 n_{s_1}(E_1)) }{
  k_0 - v_1 E_1 - v_2 E_2} \ .
$$
In the sum above, where $v_1=v_2$ the `$\tfrac12$'-terms combine into
what becomes the vacuum result.
Anything proportional to the distribution functions gives thermal effects, 
and we use a symmetry to write $n_{s_2}$ with the argument $E_1\,$.
Making use of \eq{A1 - 2b}, without imposing the energy constraint, it is necessary to 
also add the contribution coming from ${k_0 \to -k_0}\,$.
We can complete all but one of the integrations to obtain
\bea
G_0^{[0]} (K; s_1, s_2) &=& 
\frac1{k} \int_0^\infty \! dp 
\big( s_1 n_{s_1} (p) + s_2 n_{s_2}(p) \big) 
\G_0 ( p, k_0,k) \ , \nonu \\
\G_0 (p, k_0, k) &\equiv& 
\log \frac
{(k_+ + p)(k_- - p)}
{(k_+ - p)(k_- + p)} \, .
\label{B6 - 2} 
\eea
For small $p$ relative to the external momentum, $\G_0 \simeq -8p\, k/K^2\,$
and therefore the integral is finite even if one of the distribution functions
is bosonic.
We have decided to work with all logarithms taking the absolute values of their
arguments, but if we were careful to keep the correct sign one could use the
imaginary part of \eq{B6 - 2} to determine $H_0\,$.


The function $G_1^{[0]}$ reads
\bea
G_1^{[0]} (K; s_1, s_2) &=& 
 \frac{k_0}{k} \int_0^\infty \!\! dp \, \Big\{ \,
 s_2 n_{s_2} (p) \G_0 (p,k_0,k)  
+ 
 \big( s_1 n_{s_1} (p) - s_2 n_{s_2}(p) \big) \G_1 ( p, k_0,k)  \, \Big\}
 \ , \nonu \\
\G_1 (p, k_0, k) &\equiv& 
\frac{p}{k_0}
\log \frac
{k_+^2(k_-^2 - p^2)}
{k_-^2(k_+^2 - p^2)} \, ,
\label{B6 - 4} 
\eea
and $\G_0$ is as before, see \eq{B6 - 2}.
We note that $G^{[0]}_1 = \tfrac12 G^{[0]}_0$ if $s_1=s_2\,$,
as can be derived from the integral expression for $G_1\,$.



\section{Two-loop Kernels \label{app: C}}

In the diagrams labelled IV, V and VI, index $c$ is nonzero and therefore the 
momenta running in the loop do not decouple.
After carrying out the sums over $p_0$ and $q_0$ in \eq{I def},
following \eg Ref.~\cite{BP}, the terms can be sensibly collected according
to which energies are on-shell.
The main virtue of our strategy for $m,n\geq0$ (discussed in Sec.~\ref{list})
is that it means we can perform the frequency sums assuming $m=n=0\,$.

For IV, the outcome is \eq{im phi} with all intermediate particles on-shell.
  That is generally the form the `real' corrections will take.
Having more propagators (via $d,e\neq 0$) will facilitate other
permutations of cuts.
The type-V master integral has more,
which are represented by the two
terms in the following expression for the imaginary part:
\bea
\Im\ {\cal I}_{11110}^{(0)}
  \!&=&  
 \int_{[1,4]} 
 \big(\, f_{s_1}^{v_1} f_{s_4}^{v_4} -  f_{s_1}^{-v_1} f_{s_4}^{-v_4} \, \big) 
\label{C2 - 1}\\
 &\times& \Big\{\,
\sum_{v_2} \int_{\bm q} \frac{\frac12 + s_2 n_2(E_2)}{2E_2(R^2 - \lam^2)}
\,+ \sum_{v_3} \int_{\bm r} \frac{\frac12 + s_3 n_3(E_3)}{2E_3(v_4L-v_3R)^2}
 \, \Big\}     \nonu \\
 &-&  \int_{[1,2,3]}
 \frac{
 f_{s_1}^{ v_1} f_{s_2}^{ v_2} f_{s_3}^{ v_3} -
 f_{s_1}^{-v_1} f_{s_2}^{-v_2} f_{s_3}^{-v_3} \  }{(K-v_1 P)^2}   \ .
 \nonu
\eea
(The notation of the outermost integrals was defined in Appendix \ref{app: A}.)
Three momenta are put on-shell in the third line, now with
an internal propagator that was not needed before.
The virtual correction [equal to the part of \eq{C2 - 1}]
has factored into a binary decay amplitude multiplying 
a one-loop vertex amplitude.
Note that the `mass' $\lam^2$ has been introduced as a regulator,
so that $R$ has on-shell energy $E_3 = \rt{r^2 + \lam^2}$.
Later we will take the limit $\lam \to 0$ and find that the
real and virtual pieces dovetail together, leaving a result
that is both finite and $\lam$-independent.
The other energies are as before: $E_1=|\bm p|$ and $E_2=|\bm q|$,
and we also denoted $L=K-v_1P$ and $R = K-v_1P-v_2Q$.

The imaginary part of the special integral in \eq{star} is given explicitly by
\bea
& & \!\!\!\!\!\!
\Im\ {\cal I}_{11110}^\text{\,\large $\star$} 
=
  \int_{[1,4]} 
 \big(\, f_{s_1}^{v_1} f_{s_4}^{v_4} -  f_{s_1}^{-v_1} f_{s_4}^{-v_4} \, \big) 
\label{C2 - 1b}   \\
 &\times& \bigg\{\,
\sum_{v_2} \int_{\bm q} 
\frac{\frac12 + s_2 n_2(E_2)}{2E_2(R^2 - \lam^2)} (2v_2\, K\cdot Q) 
\,+
\sum_{v_3} \int_{\bm r} 
\frac{\frac12 + s_3 n_3(E_3)}{2E_3(v_4L-v_3R)^2}(2v_4\,K\cdot L-2v_3\, K\cdot R)
 \, \bigg\}  \nonu\\
 &-&\int_{[1,2,3]}
 \big( \, f_{s_1}^{ v_1} f_{s_2}^{ v_2} f_{s_3}^{ v_3} -
 f_{s_1}^{-v_1} f_{s_2}^{-v_2} f_{s_3}^{-v_3} \, \big) 
 \frac{2v_2\, K\cdot Q  }{(K-v_1 P)^2}  \, .\nonu
\eea

The most intricate two-loop topology, yet benefiting from the most symmetry,
as four different cuts contribute to the discontinuity,
given by the following expression:
\bea
\Im & & \!\!\!\!\!\!\!\!\!\!
{\cal I}^{(0)}_{11111}  
 \,=\,
 \bigg[ \int_{[1,4]}
 \big(\, f_{s_1}^{v_1} f_{s_4}^{v_4} -  f_{s_1}^{-v_1} f_{s_4}^{-v_4} \, \big) 
\label{C3 - 1} \\
  &\times& \bigg\{\,
\sum_{v_2}\! \int_{\bm q} 
\frac{
  \frac12 + s_2 n_{s_2}(E_2)
}{2E_2(R^2- \lam^2) V^2} 
+ 
\sum_{v_5}\! \int_{\bm v} 
\frac{
  \frac12 + s_5 n_{s_5}(E_5)
}{2E_5(R^2- \lam^2) Q^2} 
  +
\sum_{v_3}\! \int_{\bm r} 
\frac{\frac12 + s_3s_{s_3}(E_3)}{2E_3 Q^2  V^2}
 \, \bigg\} 
 \nonu \\
&-&
 \int_{[1,2,3]}
 \frac{
 f_{s_1}^{ v_1} f_{s_2}^{ v_2} f_{s_3}^{ v_3} -
 f_{s_1}^{-v_1} f_{s_2}^{-v_2} f_{s_3}^{-v_3} \  }{(K-v_1 P)^2(K-v_2Q)^2} \, \bigg] 
  + \bigg[\quad s_1\leftrightarrow s_5\, ,\  s_2 \leftrightarrow s_4   \quad \bigg] \  . \nonu
\eea
The meaning of the third line should be clear, the second line is one of the virtual corrections. 
It possess three terms  which ought to be clarified.
The momenta $P$ and $L=v_4(K-v_1P)$ are on-shell with
$E_1 = |\bm p|$ and $E_4 = |\bm \ell|\,$.
Inside the 
curly braces we have defined
\be
 R \ =&   v_4L - v_2Q \ &=\  v_5V - v_1 P \  , \nonu \\
 V \ =&   K-v_2Q      \ &=\  v_3 R -v_1 P \  , \nonu \\
 Q \ =&   K-v_5V      \ &=\  v_4L  -v_3 R \  ,
\label{C3 - 2}
\ee
together with 
the energies $E_2 = |\bm q|$, $E_3 = \rt{\bm r^2 + \lam^2}$ and $E_5 = |\bm v|\,$.
(The necessary form to use should be clear from the $v$-sum and spatial integrals.)

\subsection{Real corrections, $W(p,q)$}

Here we derive the expression given in \eq{W formula},
by explicitly carrying out the angular integrals from \eq{im phi}.
The result of Appendix~\ref{app: A}, and specifically Eq.~\eq{4 dim integral},
shows that we can identify
\bea
W_{\rm I\!V}(p,q) &=& 
\sgn(k_0 - p - q) \frac{pq}{\pi} \int \! d x_2 \, d x_1 \frac{\Th (h)}{\rt{h}} \, . 
\label{C1 - 1}
\eea
The meaning of $h$ and the boundaries on the integrals were given there.
One may safely set $\lam = 0$ 
(it does not generate the collinear logarithms), 
which simplifies the kinematical constraints
from $h\geq 0$.
Starting with the angular integration over $x_1 = \cos \theta_{kp}$ in \eq{4 dim integral},
we write the function $h=h(x_1)$ as a quadratic: $ax_1^2 + bx_1 +c$ \cite{Peigne2007}.
The coefficients, which can be calculated from \eq{A and B}, are
\bea
a\! &=&\! -4p^2(k^2+2k q x_2 + q^2) \  , \nonu\\
b\! &=&\! 4p(k-q x_2) \big[ k_0^2 + k^2 -2(k_0-p)(k_0-q) -2kq x_2 \big] \, , \nonu\\
c\! &=&\! 4p^2 q^2 (1-x_2^2) - 
\big( K^2 + 2[pq-k_0p-q(k_0-kx_2)]  \big)^2 \, . 
\label{C1 - 2}
\eea
Because $a<0\,$, the $\Theta$-function in \eq{C1 - 1} 
dictates the upper and lower limits on the $x_1$ integration.
We can parametrise $x_1$ in terms of the `angle' $\xi\in(0,\pi)$ with
\bea
x_1(\xi) &=& \frac{ -b + \cos \xi \rt{\Delta}}{2a} \, ,
\label{C1 - 3}
\eea
where $\Delta = b^2 - 4ac$ is the discriminant.
Changing the integration variable from $x_1$ to $\xi$ thus
removes the $\Theta$-function and yields
\bean
\int_{-1}^{+1} \! dx_1 \frac{\Theta(h)}{\rt{h}} F(x_1) 
&=& |a|^{-\nicefrac{1}{2}} \! \int_0^\pi \! d\xi F\bm(x_1(\xi)\bm) \, .
\eean
It turns out that $\Delta \geq 0$ summarises the allowable phase space,
which we have elaborated previously (in the momenta $p$ and $q$).
We already assumed a permissible $(p,q)$ configuration
by writing out the $x_1$-integral.

Returning to \eq{C1 - 1}, this gives
\bea
W_{\rm I\!V}(p,q) &=&
\sgn \Big( p(k_0-p-q) \Big) 
\label{C1 - 4}
\frac{q}2\int_{x_2^{\rm min}}^{x_2^{\rm max}} \!\! \frac{d x_2  }{\rt{k^2 - 2kp x_2 + p^2}} \, . 
\eea
The limits above follow from requiring $\D \geq 0$ in \eq{C1 - 3}, so that 
the integral has non-zero support.
They are, for sanctioned $p$ and $q$ values,
\bea
 x_2^{\rm max}  &=& 
\, \min \Big[ +1, \max \big[ 
X,Y \big] \Big] \ , \nonu\\
 x_2^{\rm min} &=&
\max \Big[ -1, \,\min \big[ 
X,Y \big] \Big]  \ . 
\label{C1 - 5}
\eea
where 
$$
X =  \frac{2k_0q-K^2}{2kq} \quad
{\rm and} \quad Y = X+ \frac2{kq} p (k_0-p-q) \, .
$$
One may check that this reproduces the formula in \eq{W formula}.

The same approach works for calculating $W_{\rm V}$ and $W_{\rm VI}\,$,
however $\lam \neq 0$ must be kept wherever a log-divergence may occur.
Determining $h$ and the limits $x_2^{\rm min}$ and $x_2^{\rm max}$
in each of the regions defined by Figs.~\ref{k<k0} and \ref{k>k0} leads
to the expressions \eq{C2 - 2} and \eq{C3 - 2b}.

Let us note that the angular limits in \eq{C1 - 5} apply
to {\em all} the real two-loop contributions.
Harking back to Figs.~\ref{k<k0} and \ref{k>k0},
these limits are specifiable in each region of the $(p,q)$-plane.
Considering them individually reveals how for $k_0 \simeq k\,$,
the {\em prima facie} incompatible regions interchange 
when moving from above to below the light cone:
If one takes the limit $k_0 \to k+0^+$ 
(see Fig.~\ref{k=k0}),
the $x_2$-limits coincide thus giving zero
precisely where $p$ and $q$ are kinematically forbidden 
in the spacelike region.
A similar argument works out for $k_0 \to k - 0^+\,$.
Moreover, the angular limits are well-defined for $k_0=k$
in the nontrivial regions that remain (i.e.~`2', `3' and `4' from
Fig~\ref{k=0}).
That implies\footnote{%
  We skipped over the technicality of setting $\lam=0\,$.
  However the conclusion is still true: $W$ is continuous for 
  finite $\lam\,$, as are the virtual corrections.
  Therefore once combined, and the limit $\lam\to 0$ is taken,
  they remain continuous at $k_0=k\,$.
}
the functions $W(p,q)$ that we calculate
are continuous at $k_0=k\,$.

\begin{figure}[h]
  \includegraphics[scale=\figscale]{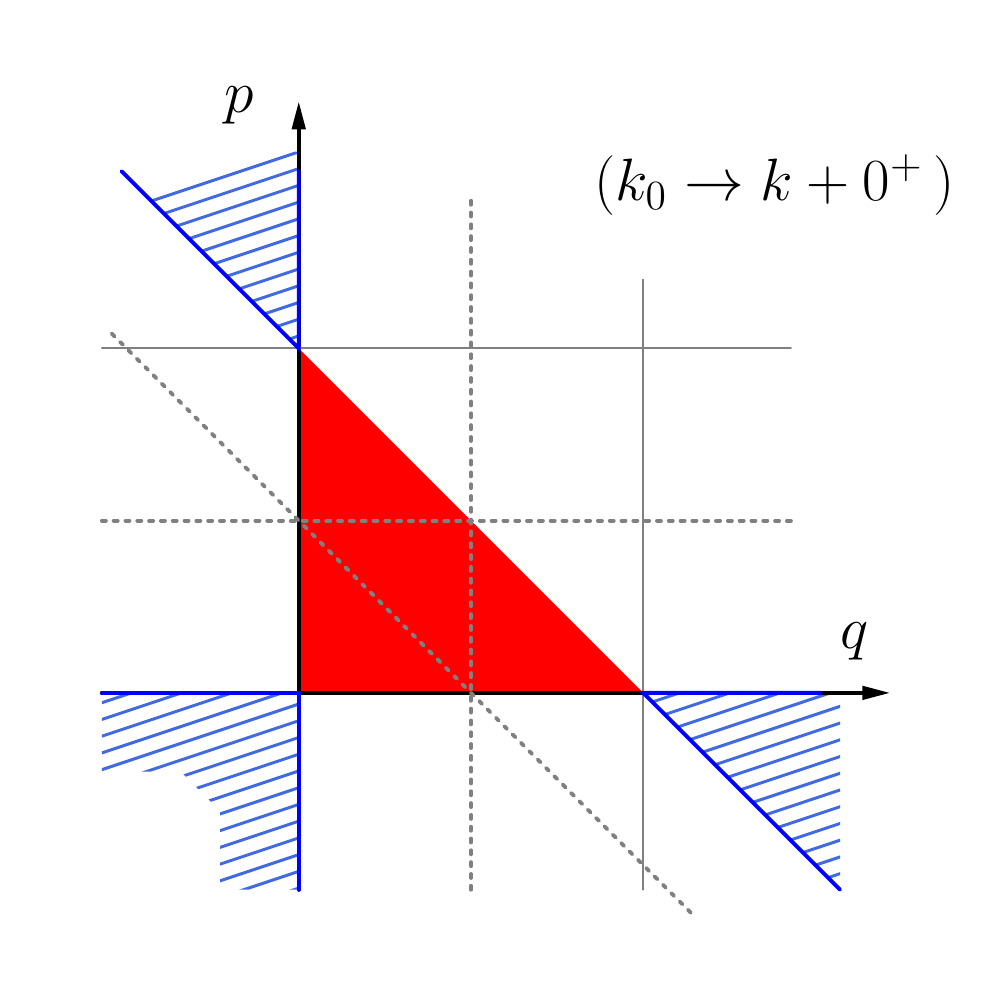}
  \caption{
    The allowable integration region in variables $p$ and $q\,$,
    as $k_0$ approaches $k$ from above.
    Region `1' from Fig.~\ref{k=0} has been coloured (red)
    and is excluded by observing that the upper and lower limits 
    in the $x_2$-integration coincide.
  }
  \label{k=k0}
\end{figure}

This does {\em not} preclude the final master integrals from
being infinite on the light cone.
However, for those that are, the singularity must be the same whether
approached from above or below.
Some of the master integrals are finite at $k_0=k\,$,
others diverge either logarithmically or due to a pole.

\subsection{Virtual corrections, 
            $U_{\rm V}(p\,;\lam)$ 
            \label{B2}}

The two terms in the second line of \eq{C2 - 1} 
may be calculated by a standard Feynman 
parametrisation. Let us manipulate these terms 
assuming $v_1=v_4=+1$ (it does not lose any generality).
We find 
\bea
\sum_{v_2} & & \!\!\!\!\!
\int_{\bm q} \frac{\frac12 + s_2 n_2(E_2)}{2E_2(R^2 - \lam^2)}
\,=\,
\frac1{(4\pi)^2\ell}
\label{C2 - 3}
\int_{-\infty}^{+\infty} \!\! dq \big(\,\tfrac12 + s_2 n_{s_2}(q)\,\big)
\log \frac{\lam^2}{4\ell q + \lam^2} \, , 
\eea
where $L=(\ell_0,\bm\ell) =K-P$ and $R = L - v_2 Q\,$. (Here, as before, we
absorb the sum over $v_2$ into the sign of $q$.)
Similarly, 
\bea
\sum_{v_3} & & \!\!\!\!\!\!\!\!\!
\int_{\bm r} \frac{\frac12 + s_3 n_3(E_3)}{2E_3(L-v_3R)^2}
\,=\,
\frac1{(4\pi)^2\ell}
\label{C2 - 4}\\
&\times&\int_{-\infty}^{+\infty} \!\! dr \frac{r}{e_3} \big(\,\tfrac12 + s_3 n_{s_3}(e_3)\,\big)
\log \frac{\lam^2 - 2 \ell (e_3-r)}{\lam^2 - 2\ell(e_3+r)} \, , \nonu 
\eea
with a signed energy $e_3 = \sgn (r) \rt{\lam^2+r^2}$.
In this term, we change integration variables from $r$
to $q=k_0-p-e_3$ so that \eq{C2 - 4} and \eq{C2 - 3} are ready
to be combined.
But before doing that, 
a divergent contribution needs to be salvaged from this vertex correction.
It is handily isolated 
(and underlined in what follows)
by writing
$$
  \big(\,\tfrac12 + s_2 n_{s_2}(q) \,\big) 
  \ = \
  \u{\sgn (q) \tfrac12} + \sgn(q) s_2 n_{s_2}\big( |q| \big) \ .
$$
in \eq{C2 - 3}, and similarly a term like $\sgn(r)\frac12$ in
\eq{C2 - 4}.
The other parts lead to momentum integrals that are rendered finite by 
the thermal weights.
But the divergent vacuum result, using dimensional regularisation, is equal to
\bea
\int_{Q} \ \frac1{Q^2[(L-Q)^2-\lam^2]} 
&=&
\frac1{(4\pi)^2} \Big(\
  \frac1{\epsilon} + \log \frac{\bar \mu^2}{\lam^2} + 1
\ \Big) \, , \nonu\\
\label{C2 - 6}
\eea
where $L^2=(K-P)^2 = 0$ was needed.
This is what the curly braces in \eq{C2 - 1}
produce in vacuum.
With a large cut-off $\Lambda$ 
on the magnitude of $\bm q\,,$
it is easy to show how the
same type of divergence arises.
The restricted integral, 
\bean
& & \int_{-\Lambda}^{+\Lambda} \frac{dq}2 \,
\Big[ \,
\sgn(q) \log \frac{\lam^2}{4\ell\,q} +
\sgn(\ell -q) \log \frac{\lam^2 q}{4\ell\, (\ell -q)^2}  \, \Big] \\
& &  =  - \ell \, \Big( \, 
\log \frac{4\Lambda^2}{\lam^2} + 1 + {\cal O}(\ell^2 / \Lambda^2 ) \, \Big) \, ,
\eean
where we assumed $\lam \to 0\,$. 
In this limit  $e_3 = r + \frac{\lam^2}{2r} + \ldots$\ , which enabled the arguments of the 
logarithms to be simplified.
Hence using $\log 4 \Lambda^2 = \epsilon^{-1} + \log \bar \mu^2 $ is
the choice consistent with the two-point function in \eq{C2 - 6}.

Returning to Eqs.~\eq{C2 - 3} and \eq{C2 - 4}, these two contributions 
can be drawn up against their real counterparts, by substituting
$$   \big(\,\tfrac12 + s_2 n_{s_2} \,\big)
\ =\ -\big(\,\tfrac12 + s_3 n_{s_3} \,\big)
+  \frac{n_{s_2} n_{s_3}}{n_{s_4}} 
 \, ,
$$
where we identified $\ell = p+e_3$
and omitted arguments of the distribution functions.
($\ell$ is the argument of $n_{s_4}$ and $s_4=s_0\, s_1$.)
For small $\lam$, the expression in curly 
braces from \eq{C2 - 1} thus simplifies,
\bea
\lim_{\lam \to 0} \big\{\ \cdots\ \big\} &=& 
\frac1{(4\pi)^2\ell} 
\int \!\! d q\ \Big[\
 \frac{n_{s_2} n_{s_3}}{n_{s_4}} \log \frac{\lam^2}{4\ell q} 
+ \big(\,\tfrac12 + s_3 n_{s_3} \,\big) \log \frac{q^2}{r^2}
\ \Big] \, .
\label{C2 - 5} 
\eea
The first summand combines with the real corrections to
remove any dependence on $\lam$ overall.
In the second, we again single out the divergent
part and calculate it with a cut-off regulator.
Namely, from
\bean
& & \int_{-\Lambda}^{+\Lambda} \! dq \  \sgn(\ell -q) \log \frac{q^2}{(\ell - q)^2} 
= - 2 \ell \Big( \, \frac1{\epsilon}
+ \log\frac{\bar \mu^2}{4\ell^2} 
    + 2 + {\cal O}(\ell^2/\Lambda^2)\, \Big) \, ,
\eean
which was converted to dimensional regularisation.
Now ${\cal O}(\epsilon)$ terms in
Eq.~\eq{A1 - 3} 
 must be kept for the outer integration.
The sequestered part of \eq{C2 - 5} therefore generates
\bea
\frac{-1}{(4\pi)^2}
  \Big[ \,
  \frac1{\epsilon} + 2 \log \frac{\bar \mu^2}{K^2} + 
  \log \frac{K^2 \, k^2}{4\ell^2(p-k_-)(p-k_+)} + 2 \, \Big] \, .
\label{C2 - 7}
\eea
Equation \eq{C2 - 7} contains an ultraviolet divergence, which will end up being 
multiplied by a function of the temperature in the final expression.
The first term in \eq{C2 - 1} can be written, 
after using signed momenta to incorporate the sum over the $\{ v_i\}$,
\bea
  & & \frac{1}{4(4\pi)^3\,k} 
  \int^\prime \! dp \, 
 \big(\, f_{s_1}^{+} f_{s_4}^{+} -  f_{s_1}^{-} f_{s_4}^{-} \, \big) 
\big\{\ \cdots\ \big\} 
\ = \
\frac{n_{s_0}^{-1}}{(4\pi)^3}
\int \! dp \, U^{(0)}_{\rm V}(p\,;\lam) \, n_{s_1} n_{s_4} \, .\nonu 
\eea
The formula for $U^{(0)}_{\rm V}$ was given in \eq{U_V} of the main text.

Including an extra power of $q=v_2 E_2$ [from $n=1$ in \eq{11110}]
is a simple matter.
None of the Feynman parametrisations or angular integrals are modified.
The $\lam$-singularities are also unchanged in 
\eq{C2 - 5}, apart from an extra factor of $q\,$.
The vacuum result\footnote{%
Again making use of $L^2=(K-P)^2 = 0$ to simplify.
  }
\bea
\int_{Q} \ \frac{q_0}{Q^2[(L-Q)^2-\lam^2]} 
&=&
\frac{\ell_0}{4(4\pi)^2} \Big(\
  \frac1{\epsilon} + \log \frac{\bar \mu^2}{\lam^2} + \frac12
\ \Big) \, , 
\label{U1 vac}
\eea
must be recovered by our regularisation procedure.
This can be checked by redoing the calculation with
a cut-off, as before.
But now one finds that
$\log 4 \Lambda^2 = \epsilon^{-1} + \log \bar \mu^2 +2$
is needed to consistently convert 
 the restricted integral,
\bean
& & \int_{-\Lambda}^{+\Lambda} \! dq \ 
q\, \sgn(\ell -q) \log \frac{q^2}{(\ell - q)^2}  
 = - \ell^2\, \Big( \, \log\frac{\Lambda^2}{\ell^2} 
    - 1 + {\cal O}(\ell^2/\Lambda^2)\, \Big) \, ,
\eean
to dimensional regularisation.
The argument is otherwise unchanged, and
altogether leads to the formula for $U^{(1)}_{\rm V}(p\,;\lam)$
that was given in \eq{U1_V}.

\bigskip

The special master integral, defined in \eq{star}, also follows 
along the lines above.
Here we give some more details:
The curly braces of Eq.~\eq{C2 - 1b} give, for $\lam \to 0\,$,
\bea
\big\{ \ \cdots \ \big\}  \ = \ 
\frac1{(4\pi)^2\ell^2}
\int \! d q \, &\bigg[&
 \label{U star 1} 
 q\, \big(\,\tfrac12 + s_2 n_{s_2} \,\big) 
 \Big(\, 2(K^2-2\ell k_0) + K^2 \log \frac{\lam^2}{4\ell q} \, \Big)
 \\
 &-& \!\!  \big(\,\tfrac12 + s_3 n_{s_3} \,\big) 
 \Big(\, 2r(K^2-2\ell k_0) - qK^2 \log \frac{\lam^2 q}{4\ell r^2} \, \Big)
\ \bigg] \, . \nonu 
\eea
Anything in the integrand that is proportional to a distribution function
$n_{s_2}(|q|)$ or $n_{s_3}(|r|)$ (evaluated at {\em positive} arguments)
will be finite.
The divergent terms are easily isolated using the same approach 
as for $U_{\rm V}^{(0)}$ and $U_{\rm V}^{(1)}$.
They can be calculated by a cut-off regulator, and altogether must give
\bean
\int_{Q} \ \frac{2\, K\cdot Q}{Q^2[(L-Q)^2-\lam^2]} 
&=&
\frac{-K^2}{2(4\pi)^2} \Big(\
  \frac1{\epsilon} + \log \frac{\bar \mu^2}{\lam^2} + \frac12
\ \Big) \, .
\eean
Once again, the hard cut-off integral can be given explicitly:
\bean
& & \int_{-\Lambda}^{+\Lambda} \! dq \,
\Big[ \,
 \sgn (q) 
 \Big(\, 2q(K^2-2\ell k_0) + qK^2 \log \frac{\lam^2}{4\ell q} \, \Big)\\
& & \qquad\quad -\, \sgn(r)
 \Big(\, 2r(K^2-2\ell k_0) - qK^2 \log \frac{\lam^2 q}{4\ell r^2} \, \Big)
\, \Big] \\
& & =
- \ell^2 \, 
\Big( \, K^2 \log \frac{4\Lambda^2}{\lam^2} + \frac12K^2 - 4k_0\ell 
   + {\cal O}(\ell^2/\Lambda^2)\, \Big) \, .
\eean
This part of \eq{U star 1} should give the complete vacuum result
for the vertex correction and consequently, 
should be converted to dimensional regularisation using 
$\log 4 \Lambda^2 = \epsilon^{-1} + \log \bar\mu^2 + 4k_0\ell/K^2\,$.
Going back to \eq{U star 1} and rearranging the distribution functions,
the quantity in curly braces becomes, for $\lam \to 0\,$,
\bea
\big\{\ \cdots\ \big\} \ =\ 
\frac1{(4\pi)^2\ell^2}
\int \! d q  &\bigg[ &
 \label{U star 2} 
   q \, \frac{n_{s_2} n_{s_3}}{n_{s_4}} 
 \Big(\, 2(K^2-2\ell k_0) + K^2 \log \frac{\lam^2}{4\ell q} \, \Big)
 \\
 &+& \big(\,\tfrac12 + s_3 n_{s_3} \,\big) 
 \Big(\, -2\ell \, (K^2-2\ell k_0) + qK^2 \log \frac{q^2}{r^2} \, \Big)
\ \bigg] \, .\nonu 
\eea
The first line above will combine with the real corrections;
cf. Eq.~\eq{C2 - 2}.
A remnant, which is proportional to $\sgn(r)\frac12$ in the second line,
diverges and can now be calculated with a regulator,
\bean
 \int_{-\Lambda}^{+\Lambda} \! &\displaystyle dq& \!  \sgn(r) \, \Big[ \,
2 \ell \, \Big(\, \frac{2\ell k_0}{K^2} - 1 \,\Big) + q \log \frac{q^2}{r^2} 
\, \Big] 
\ =\ - \ell^2 \, \Big( \, \frac1{\epsilon}
+ \log \frac{\bar \mu^2}{4\ell^2} + 3 - \frac{4\ell k_0}{K^2} \, \Big)
\, .
\eean
With that, inserted into \eq{U star 2}, we are able to single out a fragment
that is the same as $U_{\rm V}^{(1)}$ up to a factor of $K^2/\ell$.
We thus arrive at Eq.~\eq{Ustar} of the main text, after
using the principal valued integral
$$
{\cal P} \! \int_{-\infty}^{+\infty} \! dq  \
\sgn(\ell-q) \, n_{s_3}(|\ell-q|) = 0 \, ,
$$
to drop some polynomial parts of the thermal proportion.

\bigskip
\subsection{Virtual corrections, 
            $U_{\rm VI}(p\,;\lam)$}

Moving along to the function $U_{\rm VI}^{(n)}(p\,;\lam)$ needed in \eq{11111},
we derive it from \eq{C3 - 1}.
Let us focus on the three terms in curly braces and consider them individually.
The first, for $\lam \to 0\,$, is
\bea
\sum_{v_2} \int_{\bm q} 
\frac{
  \bm( \frac12 + s_2 n_{s_2}(E_2) \bm) \, q^n
}{2E_2(R^2- \lam^2) V^2} 
&=& 
\label{C3 - 3}
\frac1{(4\pi)^2 K^2} \int_{-\infty}^{+\infty} \!\! \frac{d q\, q^n}{k_0-p-q} \\
&\times& 
  \big( \, \tfrac12 + n_2 \, \big)
  \log \frac{\lam^2(k_0-p) (k_--q)(k_+-q)}{K^2\, q(k_0-p-q)^2} \ .
\nonu
\eea
To carry out the $v_i$-sums,
the momenta $p$ and $q$ were extended to negative values.
[Here only the $q$ integration is made plain, the impending 
outer $p$-integration takes the form
of \eq{A1 - 3}.]
Similarly we find, for $\lam \to 0\,$,
\bea
\sum_{v_5} \int_{\bm v} 
\frac{
  \bm( \frac12 + s_5 n_{s_5}(E_5) \bm) q^n
}{2E_5(R^2- \lam^2) Q^2} 
&=&
\frac1{(4\pi)^2 K^2} \int_{-\infty}^{+\infty} \!
 \frac{d v (k_0-v)^n}{v-p} 
\label{C3 - 4}
 \\
&\times& 
  \big( \, \tfrac12 + n_5 \, \big)
  \log \frac{K^2\, v (v-p)^2}{\lam^2 p(k_--v)(k_+-v)} \ , 
\nonu
\eea
where the integration variable $v$ has been extended to negative values. 
For the term where `particle-3' is on-shell, we use the integration variable
$r = v_3E_3$.
Then, for $\lam \to 0\,$,
\bea
\sum_{v_3} \int_{\bm r} 
\frac{ \bm( \frac12 + s_3s_{s_3}(E_3) \bm)\,q^n}{2E_3 \, Q^2 \, V^2}
&=&
\frac{1}{(4\pi)^2 K^2} 
\bigg(\! \int_{-\infty}^{-\lam} \! dr +\!
  \int_{+\lam}^{+\infty} \! dr  \bigg)\ 
  \frac{(\ell-r)^n}{r}
\label{C3 - 5} \\
&\times& \!\!
\big(\, \tfrac12 + n_3 \, \big)
\log \frac{\lam^4\, p\,(k_0-p-r)\, (k_0-p)(r+p)}{K^4\,  r^4 } \ .
\nonu
\eea
The restriction $E_3 \geq \lam$ assists to explicitly regulate the $r\approx 0$ singularity.
Unlike in \eq{C3 - 3} and \eq{C3 - 4}, where the corresponding integral may be understood in
the principle valued sense, the risk that $s_3=+1$ in \eq{C3 - 5} would make this futile. 
Therefore we keep $\lam$ to control the exclusion of this integration interval.

Let us change integration variables to $q=k_0-v$ instead of $v$
and $q=k_0-p-r$ instead of $r$
in \eq{C3 - 4} and \eq{C3 - 5} respectively.
The
distribution functions can be rewritten in a way that
the singular terms coalesce with their real counterparts.
\begin{widetext}
We do this by using an identity to rewrite the integrand of 
\eq{C3 - 5},
\bean
\big( \tfrac12 + n_3 \big)
\log \frac{\lam^4\, pq\ell v}{K^4\,  r^4 }
&=&
\Big[  - \big( \tfrac12 + n_2\big) 
+ \frac{n_2 n_3}{n_4} \, \Big] \log \frac{\lam^2 \ell v}{K^2 r^2} \\
  &+& \Big[  + \big( \tfrac12 + n_5\big) 
- \frac{n_3 n_5}{n_1} \, \Big] \log \frac{\lam^2 pq}{K^2 r^2} \  .
\eean
The two lines above are related by the exchange $p \leftrightarrow \ell$ 
and $q\leftrightarrow v$ (together with same swap in statistics).
Combining the result with \eq{C3 - 4} and \eq{C3 - 5} 
 gives, for the curly braces
in \eq{C3 - 1}, 
\bea
\big\{ \, \cdots \, \big\}
\ \to\ 
\frac{1}{(4\pi)^2 K^2} 
\int \! \frac{d q \, q^n}{r}  \!\! &\Big[& 
\big(\, n_2 - n_5 \, \big)
\log \frac{(k_--q)(k_+-q)}{ q v }  \nonu \\
 &+& 
\frac{n_2 n_3}{n_4}
\log \frac{\lam^2 \ell v}{K^2 r^2}
\ -\ \frac{n_3 n_5}{n_1}
\log \frac{\lam^2 p q}{K^2 r^2}
\  \Big] \, . 
\label{C3 - 6}
\eea
And therefore, taking into account the outer limits of integration 
[e.g.~by using \eq{A1 - 3}], it is clear that \eq{C3 - 6} contributes
to a subregion of the available $(p,q)$-plane in the real corrections.
This is exactly where $o_p=1$ if $k_0>k$, and the dependence on $\lam^2$
in the second line of \eq{C3 - 6} disappears when combined with
the two real corrections in Eq.~\eq{11111}.
(The same occurs if $k<k_0$, but
then the cancellation happens when $\bar o_p =1$.)
Note that the second term in the second line of \eq{11111} is to be
included in the whole result.

The case $n=2$ needs to be handled with some care, as the subsequent
$q$-integral diverges.
[Equation~\eq{11111_20} of Appendix~\ref{app: D} also exposes this fact.]
That can be seen from the ultraviolet behaviour of the 
first line in \eq{C3 - 6}, 
since  for $q\to \pm \infty$ the difference $(n_2 - n_5)$
becomes $\pm 1\,$, and the logarithm
$$
\log \frac{(k_--q)(k_+-q)}{ q v } 
\ \simeq \ 
\frac{K^2}{4q^2} \, \Big[ \,
1 + \frac{k_0}{q} + \frac{7 k_0^2 + k^2}{8 q^2} + \ldots \, \Big] \, .
$$
Hence the integration in \eq{C3 - 6} will be log-divergent for $n=2\,$, due
to the $1/r = 1/(k_0-p-q)$ left over.
It can be attributed to the vertex correction, \ie the three-point function
studied in Ref.~\cite{PV}.
The entry of the rank-2 tensor that we need, is given by
\bea
\lim_{\lambda\to 0} \int_{Q} \ \frac{q_0^2}{Q^2\, V^2(R^2-\lam^2)} 
&=&
\frac1{(4\pi)^2\, K^2} \Big[\,
\frac{K^2}4 
\Big(\, \frac1{\epsilon} + 4  + \frac{k^2-6p\ell}{K^2}+ \log \frac{\bar{\mu}^2}{K^2} \, \Big)
\label{C3 - 7} \\
&+& 
\big(\, \tfrac32 k_0^2 - 5k_0 p + 3p^2 \, \big)
\Big(\, 1 + \log \frac{\lam^2}{K^2} \, \Big)
\ -\ 
 \frac{ \ell^2 }2 \, \Big(\, 
  \frac{\pi^2}{3} -
 \log^2 \frac{\lam^2}{K^2}
  \, \Big)
\,\Big] \, . \nonu
\eea
\end{widetext}
The `$\tfrac12$'-terms next to the distribution functions in 
Eqs.~\eq{C3 - 3}-\eq{C3 - 5} concomitantly {\em ought} to reproduce 
this vacuum result.
Explicitely taking these equations together, with an imposed cut-off for large-$q$ and
an infrared regulator for $q \simeq \ell\,$, we find, for $\lam \to 0\,$,
\bean
  &\bigg(&\! \int^{\ell-\lam}_{-\Lambda} \!dq +\!
  \int_{\ell+\lam}^{+\Lambda} \! dq  \bigg) 
\frac{q^2}{2(\ell - q)} \Big[ 
\  \sgn(q) \log \frac{ \lam^2 \ell (k_+-q)(k_--q)}{K^2 q r^2}\\
& & \quad \qquad +\ \sgn(v) \log \frac{K^2 v r^2}{ \lam^2 p (k_+-q)(k_--q)} 
  +\ \sgn(r) \log \frac{ \lam^4 p q \ell v}{K^4 r^4}
  \quad  \Big] \\[.4cm]
  &=& 
-\, \frac{ K^2 }4
  \Big(\, \log \frac{4\Lambda^2}{K^2} +
  2+ \frac{k^2-6p\,\ell}{K^2} \, \Big)  \\
   & & -\, \big(\, \tfrac32 k_0^2 - 5 k_0 p + 3p^2 \,\big) 
  \Big(\, 1 + \log \frac{\lam^2}{K^2} \Big) 
   +\, \frac{\ell^2}2 \, \Big(\, 
  \frac{\pi^2}{3} -
 \log^2 \frac{\lam^2}{K^2}
 \, \Big) \ .
\eean
Terms that vanish as $\Lambda \to \infty$ were omitted.
This integral appears in $U_{\rm VI}^{(2)}(p\,;\lam)\,$, to give 
\eq{C3 - 7} once the dust has settled and all factors are collected.
This means that the cut-off regulator should be replaced by
$$\log 4 \Lambda^2 \to \epsilon^{-1} + 2\log \bar\mu^2 + 2\, .$$


\section{Large-$K^2$ expansions\label{app: D}}

For external energies $k_0\,$, that are much larger than both the temperature
$T$ and momentum $\bm k\,$, spectral functions can be studied by OPE 
techniques \cite{CaronHuot2009ns}.
The resulting approximations are applicable in the 
deeply virtual regime $K^2 \gg T^2\,$ and may also be obtained 
systematically from the master sum-integrals themselves \cite{Laine2010}.

Carrying out the Matsubara sums of \eq{I def}
produces terms, besides the vacuum result, 
with different loop momenta put `on-shell' and weighted by 
a thermal distribution.
These thermal contributions are multiplied by coefficients
that resemble $T=0$ amplitudes of a simpler kind.
In general, we can relabel and shift integration variables until
the result is in the form: (omitting $a,b,c,d,e$ and $m,n$)
\bea
{\cal I} &=&
\lim_{T\to 0}\, \Big( \, {\cal I} \, \Big) \label{OPE 1} \\
&+&
{\textstyle\sum}_i 
\int_{\bm p} 
n_{s_i}(p) 
\Big[\, {\cal A}_i(P) \,\Big]_{p_0=\pm p} 
+
{\textstyle\sum}_{i<j}
\int_{\bm p,\, \bm q} 
n_{s_i}(p) n_{s_j}(q)
\Big[\, {\cal B}_{ij}(P,Q) \,\Big]_{p_0=\pm p, q_0=\pm q} \ . \nonu 
\eea
Above, within the first square bracket one may set $P^2=0$ and in the second one
may set both $P^2=0$ and $Q^2=0\,$.
The presentation in \eq{OPE 1} folds together all the physical
reactions described by the Boltzmann equation, thus 
only linear and quadratic terms in the distribution functions are present.
The former, proportional to ${\cal A}_i$ (given below), contain 
leading thermal corrections of the OPE.

For general $a$, $b$, $d$, $c$ and $d$,
\bean
{\cal A}_1 &=&
    \phi_a(p) \, p_0^m \D_{K-P}^d \int_Q q_0^n \D_Q^b \D_{K-P-Q}^c \D_{K-Q}^e \, ,\\
{\cal A}_2 &=&
    \phi_b(p) \, p_0^n \D_{K-P}^e \int_Q q_0^m  \D_Q^a \D_{K-P-Q}^c \D_{K-Q}^d \, ,\\
{\cal A}_3 &=&
    \phi_c(p) 
    \int_Q (k_0-p_0-q_0)^m q_0^n \D_Q^b \D_{K-P-Q}^a \D_{P+Q}^d \D_{K-Q}^e  \, ,\\
{\cal A}_4 &=&
    \phi_d(p) \, (k_0-p_0)^m \D_{K-P}^a 
    \int_Q (k_0-q_0)^n \D_Q^e \D_{K-P-Q}^c \D_{K-Q}^b \, ,\\
{\cal A}_5 &=&
    \phi_e(p) \, (k_0-p_0)^n \D_{K-P}^b 
    \int_Q (k_0-q_0)^m \D_Q^d \D_{K-P-Q}^c \D_{K-Q}^a \,
\eean
These five terms are compatible with the original symmetry of the diagram, e.g.
${\cal A}_2$ with $P \leftrightarrow Q$ gives a result with $Q$ (associated with 
$\D_Q^b$ in the original labelling) being the on-shell momenta.
In ${\cal A}_1\,$, $\phi_a(p)$ denotes the residue of $\D_P^a$
at its positive pole $p_0 = p\,$, viewing the scalar propagator 
$\D(P) = (p_0^2-p^2)^{-1}$ as a function the complex  energy.
It may be expressed using the gamma function as
\bea
\phi_a(p) &\equiv&
\frac{(-1)^{a-1} }{\rt{\pi}}
\frac{\G (a+\tfrac12)}{\G(a) p^{2a-1} } \, .
\label{def phi_a}
\eea

For the vacuum-like $Q$ integrations  it is safe to expand the integrand
in $P=(p_0,p)\,$.
Doing just that, for what we need\footnote{%
  We write four products $X\cdot Y = x_0 y_0 - \bm x \cdot \bm y\,$.
  }
 (dropping any terms we can, thanks to $P^2=0$)
\bea
\D_{K-P}^a 
&\simeq &
\frac1{K^{2a}} \Big\{ \, 
  1 + 2a \frac{K\cdot P}{K^2} 
  + 2a(a+1) \frac{(K\cdot P)^2}{K^4} + \, \ldots
\, \Big\} \, ,\nonu \\
\D_{K-P-Q}^a 
&\simeq &
\frac1{(K-Q)^{2a}} \Big\{ \,
   1 + 2a \frac{(K-Q)\cdot P}{(K-Q)^2} 
   +  2a(a+1) \frac{\big((K-Q)\cdot P\big)^2}{(K-Q)^4} + \, \ldots
\, \Big\} \, ,\nonu \\
\D_{P+Q}^a 
&\simeq &
\frac1{Q^{2a}} \Big\{ \, 
  1 - 2a \frac{Q\cdot P}{Q^2} 
  + 2a(a+1) \frac{(Q\cdot P)^2}{Q^4} + \, \ldots
\, \Big\} \, . \label{OPE 2}
\eea
After the necessary expansion is inserted into the definitions of each 
${\cal A}_i\,$,
one finds a variety of ordinary vacuum integrals.
These types of integrals are all derivable from a class of 1-loop tensors 
\cite{PV}, viz. (not the same $m$, $n$ as before)
$$
{\cal J}^{\mu_1 \mu_2 \, \cdots} _{m,n}\equiv
\int_Q \frac{Q^{\mu_1} Q^{\mu_2}\, \cdots }{Q^{2m} (K-Q)^{2n}} \ .
$$
In particular, Lorentz invariance implies that 
they are each linear combinations of independent tensor
(of appropriate rank) that can be constructed from just $K^\mu$ 
and $g\omn\,$.
We need only those with up to four indices, denoted
\bean
{\cal J}^\mu_{m,n} &=&
A_{m,n} \, K^\mu \, ,\\
{\cal J}\omn_{m,n} &=&
C_{m,n} \, g\omn + B_{m,n} \, K^\mu K^\nu \, ,\\
{\cal J}^{\mu\nu\rho}_{m,n} &=& 
E_{m,n}\, \big(K_m\, g^{\nu\rho} + \,{\rm sym.}\, \big) 
+ D_{m,n}\, K^\mu K^\nu K^\rho \, ,\\
{\cal J}^{\mu\nu\rho\s}_{m,n} &=&
H_{m,n} \, \big(\, g\omn g^{\rho\s} + \,{\rm sym.}\, \big) 
+
G_{m,n} \, \big( \, K^\mu K^\nu g^{\rho\s} +\, {\rm sym.}\,\big) 
+F_{m,n} \, K^\mu K^\nu K^\rho K^\s \, .
\eean
The coefficients $A,B,C,D,E,F,G$ and $H$ can all be related to
the fundamental scalar integral ${\cal J}_{m,n}$ 
(it has no powers of $q_0$ in the numerator).

Without loss of generality, we assume $m \geq n$ so that 
the pairs $(m,n)$ of interest here are
$(0,0)$, $(1,0)$, $(1,1)$ and $(2,0)\,$.
Therefore the contractions needed are as follows.
\bea
P_\mu {\cal J}^\mu_{m,n} &=& (K\cdot P) A_{m,n} \nonu \\
{\cal J}^0_{m,n} &=& k_0 A_{m,n} \\[.25cm]
P_\mu P_\nu {\cal J}\omn_{m,n} &=&
(K\cdot P)^2
B_{m,n}  \nonu \\
P_\mu {\cal J}^{\mu 0}_{m,n} &=&
p_0 C_{m,n}  +  k_0 (K\cdot P) B_{m,n} \nonu \\
{\cal J}^{00}_{m,n} &=&
C_{m,n} + k_0^2 B_{m,n} \\[.25cm]
P_\mu P_\nu {\cal J}^{\mu\nu 0}_{m,n} &=&
2 p_0 (K\cdot P) E_{m,n}  + k_0 (K\cdot P)^2 D_{m,n} \nonu \\
P_\mu {\cal J}^{\mu 00}_{m,n} &=&
\big[\, (K\cdot P) + 2k_0p_0 \,\big] E_{m,n} 
    + k_0 (K\cdot P)^2 D_{m,n} \\[.25cm]
P_\mu P_\nu {\cal J}^{\mu\nu 00}_{m,n} &=&
2 p_0^2 H_{m,n} 
+
(K \cdot P) \big[\, (K\cdot P) + 4k_0 p_0 \, \big] G_{m,n} 
+ k_0^2 (K\cdot P)^2 \, F_{m,n}
\eea
These expressions still depend on the relative angle between $\bm k$ and 
$\bm p\,$.
That is to be taken into account when performing the integrals in \eq{OPE 1},
which we carry out using \eg
\bean
\int_{\bm p} \, f(p) \Big[\, (K\cdot P)^2 \,\Big]_{p_0=\pm p}
\!\!\!\! &=& 2 \int_{\bm p} \, f(p) p^2\, 
 \Big(\, k_0^2 + \frac{k^2}{3-2\epsilon} \, \Big) \, , \\
\int_{\bm p} \, f(p) \Big[ \, p_0 (K\cdot P) \, \Big]_{p_0=\pm p}
\!\!\!\! &=& 2 \int_{\bm p} \, f(p) p^2\, k_0  \, ,
\eean
and any other angular averaging useful for 
simplifying what comes from implementing  Eqs.~\eq{OPE 2}.
(For example, terms in the square bracket that are odd in $p_0$
will vanish after summing over $p_0 = \pm p\,$.)
Such manipulations will put the spectral function\footnote{%
  The approach here {\em could} be used for the whole master integral, but
   presently we focus only on the imaginary part.}
from \eq{OPE 1} into the form
$$
\Im \  {\cal I} \ \sim\ 
\omega_0 \,  K^2
+ \omega_2 \, T^2 + \omega_4 \, \frac{T^4}{K^2} + \, \ldots \,  ,
\ (K^2/T^2 \to \infty )
$$
where $\omega_i$ will depend on details of the master integral.
The first coefficient, $\omega_0\,$, is exactly the vacuum result and
therefore might contain a term $(\epsilon^{-1} + 2 \log \bar \mu^2/K^2)\,$.
Coefficients of powers of $T^2\,$, which are all finite, arise from
moments of equilibrium distribution functions. So, for instance, 
$\omega_4$ stems from $\int_{\bm p} p n_{s_i}(p) = {\cal O}(T^{4})\,$.
Only $\omega_0$ is independent of the statistical nature (via $s_i=\pm1$)
of the propagators.

\bigskip

With the abbreviation $n_i \equiv s_i n_{s_i}(p)\,$, we list some of 
those integrals now:

The masters that factor into a product with a tadpole diagram
are zero in vacuum: they
only start at ${\cal O}(T^2)\,$.
It turns out that they also have no $T^4$-term,
\bea
\Im \ {\cal I}_{10110}^{(0)} &=&
+ \int_{\bm p} 
\frac{ n_3 }{16\pi p} \, + \, {\cal O} \Big( \frac{T^6}{K^4} \Big) \, ,\\
K^2\, \Im\ {\cal I}_{10120}^{(0)} &=&
- \int_{\bm p} 
\frac{ n_3 }{16\pi p} 
\, + \, {\cal O} \Big( \frac{T^6}{K^4} \Big) \, .
\label{ope: factorisable}
\eea
For the cases with $m\geq 1$ they can be expressed in terms 
of the results above.
The following equalities are only valid to ${\cal O}(T^6/K^4)\,$, although
the first is true in general if $s_1=s_4\,$,
\bean
\Im \ {\cal I}_{10110}^{(1)} &=& \frac12\, k_0 \, \Im \ {\cal I}_{10110}^{(0)} 
\, , \\
\Im \ {\cal I}_{10120}^{(1)} &=& {k_0}\, \Im \ {\cal I}_{10120}^{(0)}  
\, , \\
\Im \ {\cal I}_{10120}^{(2)} &=&
\frac34 \Big(\, k_0^2 + \frac{k^2}3 \, \Big)\,
\Im \ {\cal I}_{10120}^{(0)} 
\, .
\eean
One may also obtain the masters ${\cal I}_{11010}$ 
and ${\cal I}_{11020}$ by replacing $s_3$ with $s_2$ in the above.

The sunset integral has the expansion
\bea
\Im \ {\cal I}_{11100}^{(0)} &=& 
\frac{K^2}{8(4\pi)^3} \, + \,
\int_{\bm p} \frac{n_1+n_2+n_3}{16\pi p} 
\, + \, {\cal O} \Big( \frac{T^6}{K^4} \Big)  \, , 
\eea
which is evidently symmetric in $\{ s_i \}$ for the indices $i=1, 2$ and $3\,$.
It also has no $T^4$-term, and the $T^2$-term can equal zero if only
one of the particles is bosonic, the rest fermionic.

For the spectacle diagram, which bears an ultraviolet divergence (from
the 1-loop factor) that persists after taking the imaginary part,
we find
\bea
 K^2 \, \Im \ {\cal I}_{11011}^{(0)} &=&
- \frac{K^2}{2(4\pi)^3} \Big( \, \frac1{\epsilon}
 + 2 \log \frac{\bar \mu^2}{K^2} + 4 \, \Big) \\
&+& \int_{\bm p} \frac{n_1+n_2+n_4+n_5}{16\pi p} 
+
\int_{\bm p} \frac{p( n_1+n_2+n_4+n_5) }{4\pi \, K^4 } 
\Big(\, k_0^2 + \frac{k^2}3 \, \Big) 
+ {\cal O} \Big( \frac{T^6}{K^4} \Big)  \, .  \nonu
\eea
The result is symmetric in $\{s_i\}$ for $i=1,2,3$ and $4\,$.
Moreover, given any combination of statistics,
the $T^2$- and $T^4$-order corrections
are not zero.

Considering next the squint two-loop diagrams (with 
$a=b=c=d=1$ and $e=0$), the simplest yields
\bea
K^2 \, \Im\ {\cal I}_{11110}^{(0)} &=&
 - \frac{K^2}{4(4\pi)^3} \Big( \, \frac1{\epsilon}
 + 2 \log \frac{\bar \mu^2}{K^2} + 5 \, \Big) 
+ \int_{\bm p} \frac{n_1-(n_2+n_3)}{16\pi p}  \nonu\\
&+&
\int_{\bm p} \frac{p\,
\bm( 3n_1-(n_2+n_3)\bm) 
}{12\pi \, K^4 } 
\Big(\, k_0^2 + \frac{k^2}3 \, \Big) 
+ {\cal O} \Big( \frac{T^6}{K^4} \Big)  \, ,
\label{ope: V 00}
\eea
which is symmetric in $s_2$ and $s_3\,$.
The master with $m=1$ ($n=0$) has the same symmetry:
\bea
\frac{K^2}{k_0} \, \Im\ {\cal I}_{11110}^{(1)} &=& 
 - \frac{K^2}{8(4\pi)^3} \Big( \, \frac1{\epsilon}
 + 2 \log \frac{\bar \mu^2}{K^2} + \frac{11}2 \, \Big) 
 \, - \, \int_{\bm p} \frac{n_2+n_3}{16\pi p}  \\ 
&-&
\int_{\bm p} \frac{p}{12\pi \, K^4} \bigg[ 
(n_2+n_3) \Big(\, k_0^2 + \frac{k^2}3 \, \Big)
- \frac12 ( 3 n_1 + n_2 +n_3)K^2 
\bigg]
\, + \, {\cal O} \Big( \frac{T^6}{K^4} \Big)  \, . \nonu 
\eea
So do all ${\cal I}_{11110}^{(m)}$ (with $n=0\,$, for any $m$), but 
those with $n\geq 1$ have no such symmetry in the 
statistical factors.
Indeed consider $m=0$ and $n=1\,$, which has the expansion
\bea
\frac{K^2}{k_0} \, \Im \ {\cal I}_{11110}^{(0,1)} &=& 
 - \frac{K^2}{16(4\pi)^3} \Big( \, \frac1{\epsilon}
 + 2 \log \frac{\bar \mu^2}{K^2} + \frac92 \, \Big) 
 \, + \, \int_{\bm p} \frac{n_1}{32\pi p}  \\ 
&+&
\int_{\bm p} \frac{p}{24\pi \, K^4} \bigg[ 
3 n_1 \Big(\, k_0^2 + \frac{k^2}3 \, \Big)
- \frac12 \bm( 3 (n_1 + n_2) -n_3\bm)K^2 
\bigg]
\, + \, {\cal O} \Big( \frac{T^6}{K^4} \Big)  \, . \nonu 
\eea
Within the same class of master integrals, having $a=b=c=d=1\,$, 
it is also useful to cater for $e=-1\,$.
Then let us consider the particular combination
\bea
\Im\ {\cal I}_{11110}^{\text{\,\large $\star$}} &\equiv& 
\Im \big[\, {\cal I}_{10110}^{(0)} + K^2 \, {\cal I}_{11110}^{(0)} - 
{\cal I}_{1111(-1)}^{(0)} \,\big] \label{11110_01}\\
&=&
 - \frac{K^2}{8(4\pi)^3} \Big( \, \frac1{\epsilon}
 + 2 \log \frac{\bar \mu^2}{K^2} + \frac92 \, \Big) 
 \, + \,  \int_{\bm p} \frac{n_1}{16\pi p}  
  \nonu \\ &+&
\int_{\bm p} \frac{p}{24\pi \, K^4 } 
\bm( 3(n_1-n_2)+n_3\bm) 
\Big(\, k_0^2 + \frac{k^2}3 \, \Big)
\, + \, {\cal O} \Big( \frac{T^6}{K^4} \Big)  \, . \nonu 
\eea
Finally, it remains to discuss the cats-eye topology.
The simplest case ($m=n=0$) is actually zero in vacuum,
and the expansion starts at ${\cal O}(T^2)\,$:
\bea
K^4
\, \Im \ {\cal I}_{11111}^{(0)} &=& 
- \int_{\bm p} \frac{n_1+n_2+2n_3+n_4+n_5}{16\pi p}  \\ 
&-&
\int_{\bm p} \frac{p}{24\pi \, K^4} 
\bm( 11 (n_1 + n_2 + n_4+n_5)  +6 n_3 \bm) 
\Big(\, k_0^2 + \frac{k^2}3 \, \Big)
\, + \, {\cal O} \Big( \frac{T^6}{K^4} \Big)  \, , \nonu 
\eea
which has, as it should, a total
 symmetry in $\{ s_i \}$ for $i=1,2,4$ and $5\,$.
We may assume $m \geq n$ without loss
of generality, due to the symmetry in $s_1$ and $s_2\,$.
Those integrals with $m<n$ follow from those with $m>n$
under this exchange. 
We give the first such case, (with $m=1$ and $n=0\,$)
\bea
\frac{K^4}{k_0}
\, \Im \ {\cal I}_{11111}^{(1)} &=& 
 \, - \, \int_{\bm p} \frac{n_2+n_3+n_4}{16\pi p}   
- \int_{\bm p} \frac{p}{24\pi\, K^4 } \bigg[  \label{11111_10}\\
& & \!\!\!\!\!\!\! 
\bm( 11(n_2+n_4)+3n_3 \bm) \Big(\, k_0^2 + \frac{k^2}3 \, \Big) 
+ \frac12 \bm( 9(n_1-n_4) -5(n_2-n_5) \bm)K^2
\bigg]
\, + \, {\cal O} \Big( \frac{T^6}{K^4} \Big)  \, . \nonu 
\eea
And the closely related ${\cal I}_{11111}^{(0,1)}$ 
may be obtained by simultaneously swapping
$s_1$ with $s_2$ and $s_4$ with $s_5\,$. (The latter is
automatic if we enforce $s_4 = s_0 s_1$ and $s_5 = s_0 s_2\,$.)
For some integrals with higher powers of energies, we obtain
(with $m=n=1$)
\bea
K^2
\, \Im \ {\cal I}_{11111}^{(1,1)} &=& 
 + \frac{K^2}{16(4\pi)^3} 
 \, + \, 
\int_{\bm p} \frac{p}{48\pi \, K^4 }  \bigg[ 
\frac12 \bm( 3 (n_1 + n_2 + n_4+n_5)  -2 n_3 \bm) K^2 
\nonu\\
&-&
\bm( 9(n_1+n_2)+2n_3+5(n_4+n_5) \bm) k_0^2
\bigg]
\, + \, {\cal O} \Big( \frac{T^6}{K^4} \Big)  \, , 
\label{11111_11}
\eea
and (with $m=2$ and $n=0$)
\bea
K^2
\, \Im \ {\cal I}_{11111}^{(2)} &=& 
 - \frac{K^2}{16(4\pi)^3} \Big( \, \frac1{\epsilon}
 + 2 \log \frac{\bar \mu^2}{K^2} + \frac{11}2 \, \Big)  
- \int_{\bm p} \frac1{16\pi p \, K^2}  
  \Big(\, (n_2+n_3+n_4)k_0^2 \label{11111_20}\\
  &-&
\frac14 (n_2+n_5) K^2  \, \Big)  
- \int_{\bm p} \frac{p}{24\pi \, K^4 } \bigg[ 
 \frac1{2} \bm( 3(n_1+n_4) + 2(n_2 + n_3 + n_5) \bm) K^2 \nonu \\
&-& \bm( 7n_2 + n_3 + 9n_4 + 2n_5 \bm) k_0^2
+ \bm( 11 (n_2+n_4) + 3n_3 \bm) \frac{k_0^2}{K^2} 
\Big( k_0^2 + \frac{k^2}3  \Big)
\bigg]
\, + \, {\cal O} \Big( \frac{T^6}{K^4} \Big)  \, . \nonu 
\eea
In all the explicit expansions above, we have left
the momentum integrals (over $\bm p\,$) in the
coefficients $\omega_2$ and $\omega_4$
 as is.
But they are trivial to carry out for given 
$s_0\,$, $s_1$ and $s_2\,$.
They are all of the form 
\bea
\int_{\bm p} p^{\nu-2} \, n_i
&=& 
\frac{s_i T^{\nu+1}}{2\pi^2}  
\Big[\, 1 - \frac{\Theta (-s_i)}{2^\nu} \, \Big] \,
\G (\nu+1) \zeta (\nu+1) \, , 
\label{p integrals}
\eea
where $\nu = \{ 1,3 \}\,$.
\bigskip

These expansions can be used for the photon spectral funcion 
in a QCD medium, given in
Eqs.~\eq{masters, mn} and \eq{masters, 00}.
The expansions for large $K^2$ are
\bea
\Im \big[ \, g\mn \Pi_{(1)}\omn \, ] \,=  
&-& N \cf \, \Big\{ \,
  \frac{3K^2}{(4\pi)^3}
  + \frac{\pi\,T^4}{9 K^4}
  \Big( \, k_0^2 + \frac{k^2}3 \, \Big) 
  \, \Big\}  
  + \, \ldots \ , 
\nonu \\
\Im \big[ \, \Pi_{(1)}^{00} \, ] \,= & &
 \!\!\!\!\!\!\!\! N \cf \, k^2\, \Big\{ \,
  \frac{1}{(4\pi)^3}
  + \frac{\pi\,T^4 }{27 K^4}
  \, \Big\} + \, \ldots \ .
  \label{large K photon}
\eea
The leading term is the vacuum result and 
thermal corrections would start at ${\cal O}(T^2)$, but this term is absent
in accordance with \cite{CaronHuot2009ns}.

\bibliographystyle{JHEP}
\bibliography{refs}

\end{document}